\documentclass[a4paper, 11pt]{article}
\pdfoutput=1 
 
\usepackage{jheppub} 
           
\usepackage[T1]{fontenc} 
           
\usepackage{amsmath}  
\usepackage{amssymb}  
\usepackage{latexsym}
\usepackage{graphicx}
\usepackage{subfig} 
\usepackage{empheq}
\usepackage{upgreek} 
\usepackage{cancel}

\usepackage{braket}
\usepackage{amssymb}
\usepackage{amsmath}
\usepackage{amsthm}
\usepackage{mathrsfs}
\usepackage{skak}
\usepackage{framed}
\usepackage{hyperref}
\usepackage{slashed}
\usepackage{array}

\hypersetup{colorlinks=true}
\hypersetup{linkcolor=violet}
\hypersetup{citecolor=violet}
\hypersetup{urlcolor=violet}

\numberwithin{equation}{section}

\usepackage
{datetime}
\usepackage{fancyhdr}
\usepackage{tikz, pgf}
\usetikzlibrary{shapes.misc}
\usetikzlibrary{shapes}
\usetikzlibrary{fit}
\usetikzlibrary{arrows}

\usetikzlibrary{decorations.pathmorphing}	
\usetikzlibrary{decorations.markings}
 \usetikzlibrary{patterns}

\tikzset{
    sugra/.style={decorate, decoration={snake}, draw=black},
    scalarphi/.style={dashed,draw=black, postaction={decorate},
        },
    scalarchi/.style={draw=brown}, 
    hwbou/.style={draw=blue, postaction={decorate}, ultra thick
        },
    vector/.style={draw=blue,decorate, decoration={snake}, draw},
	provector/.style={decorate, decoration={snake,amplitude=2.5pt}, draw},
	antivector/.style={decorate, decoration={snake,amplitude=-2.5pt}, draw},
   	 fermion/.style={draw=cyan, postaction={decorate},
        decoration={markings,mark=at position .55 with {\arrow[draw=black]{>}}}},
    fermionbar/.style={draw=cyan, postaction={decorate},
        decoration={markings,mark=at position .55 with {\arrow[draw=black]{<}}}},
    fermionnoarrow/.style={draw=black},
    gluon/.style={decorate, draw=red,
        decoration={coil, amplitude=4pt, segment length=5pt}},
    scalar/.style={dashed,draw=black, postaction={decorate},
        decoration={markings,mark=at position .55 with {\arrow[draw=black]{>}}}},
    scalarbar/.style={dashed,draw=black, postaction={decorate},
        decoration={markings,mark=at position .55 with {\arrow[draw=black]{<}}}},
    electron/.style={draw=black, postaction={decorate},
        decoration={markings,mark=at position .55 with {\arrow[draw=black]{>}}}},
    scalarnoarrow/.style={dashed, draw=black},
    electron/.style={draw=black, postaction={decorate},
        decoration={markings, mark=at position .55 with {\arrow[draw=black]{>}}}},
	bigvector/.style={decorate, decoration={snake, amplitude=4pt}, draw},
    photon/.style={draw=violet, decorate, decoration={snake}, draw},
    higgs/.style={dashed, draw=black, postaction={decorate},
        },	
        goldstone/.style={draw=brown, postaction={decorate},
        },    
          ghost/.style={dashed, draw=magenta, postaction={decorate},
        decoration={markings, mark=at position .55 with {\arrow[draw=black]{>}}}
        },  
          antighost/.style={dashed, draw=magenta, postaction={decorate},
        decoration={markings, mark=at position .55 with {\arrow[draw=black]{<}}}
        }, 
            scalartwo/.style={dashed,draw=brown, postaction={decorate},
        decoration={markings,mark=at position .55 with {\arrow[draw=black]{>}}}},
    scalarbartwo/.style={dashed,draw=brown, postaction={decorate},
        decoration={markings,mark=at position .55 with {\arrow[draw=black]{<}}}}, 
    fermiontwo/.style={draw=purple, postaction={decorate},
        decoration={markings,mark=at position .55 with {\arrow[draw=black]{>}}}},
    fermionbartwo/.style={draw=purple, postaction={decorate},
        decoration={markings,mark=at position .55 with {\arrow[draw=black]{<}}}},    
        mphoton/.style={decorate, decoration={snake}, draw=violet},
        realscalar/.style={draw=black}, 
        fakerealscalar/.style={draw=white}, 
        realscalarone/.style={ draw=black},
    	realscalartwo/.style={draw=brown},    	    pseudoscalar/.style={draw=brown},
        mgluon/.style={decorate, draw=blue,
        	decoration={coil,amplitude=4pt, segment length=5pt}},
         weylfermion/.style={draw=orange, postaction={decorate},
        decoration={markings,mark=at position .55 with {\arrow[draw=black]{>}}}},
         weylfermionbar/.style={draw=orange, postaction={decorate},
        decoration={markings,mark=at position .55 with {\arrow[draw=black]{<}}}}, 
    majorana/.style={draw=cyan, postaction={decorate},
        decoration={markings,mark=at position .55 with {\arrow[draw=black]{><}}}},
    majoranabar/.style={draw=cyan, postaction={decorate},
        decoration={markings,mark=at position .55 with {\arrow[draw=black]{><}}}},    
   	wboson/.style={draw=blue,decorate, decoration={snake,amplitude=4pt}, draw},  
    zboson/.style={draw=violet, decorate, decoration={snake}, draw},   
    lepton/.style={draw=black, postaction={decorate},
        decoration={markings, mark=at position .55 with {\arrow[draw=black]{>}}}},
    leptonbar/.style={draw=black, postaction={decorate},
        decoration={markings, mark=at position .55 with {\arrow[draw=black]{<}}}}, 
    clepton/.style={draw=cyan, postaction={decorate},
        decoration={markings, mark=at position .55 with {\arrow[draw= black]{>}}}},
    cleptonbar/.style={draw=cyan, postaction={decorate},
        decoration={markings, mark=at position .55 with {\arrow[draw=black]{<}}}},   
   nlepton/.style={draw=orange, postaction={decorate},
        decoration={markings, mark=at position .55 with {\arrow[draw=black]{>}}}},
    nleptonbar/.style={draw=orange, postaction={decorate},
        decoration={markings, mark=at position .55 with {\arrow[draw=black]{<}}}},              
        graviton/.style={draw=blue, decorate, decoration={snake, amplitude=4pt}, draw},  
        spinj/.style={draw=red, decorate, decoration={snake, amplitude=4pt}, draw},  
        bgraviton/.style={draw=blue, decorate, decoration={snake, amplitude=4pt}, draw},  
        gravitino/.style={draw=red, postaction={decorate}, 
        decoration={snake,  markings, mark=at position .55 with {\arrow[draw=black]{><}}}},
    	gravitinobar/.style={draw=red, postaction={decorate},
        decoration={snake, markings, mark=at position .55 with {\arrow[draw=black]{><}}} },  
    phir/.style={draw=blue, postaction={decorate},},
   phil/.style={dashed,draw=blue,},
     phiav/.style={draw=cyan, postaction={decorate},},
   phidif/.style={dashed,draw=cyan,},  
    chir/.style={draw=red, postaction={decorate},},
   chil/.style={dashed,draw=red,},  
}

\newcommand{\treelevelthreepointskel}[4]{
\begin{scope}[shift={(0,0)}, rotate=#4]
	\draw[#1] (0,0)--(1.5,0);    
	\draw[#2][rotate=120] (0,0)--(1.5,0);  
	\draw[#3][rotate=-120](0,0)--(1.5,0);  

\end{scope}	 
}

\newcommand{\labeltreelevelthreepoint}[4]{
\begin{scope}[shift={(0,0)}, rotate=#4]

\node at (2,0) {#1}; 
\begin{scope}[shift={(0,0)},rotate=120]	
\node at (2,0) {#2};
\end{scope}	
\begin{scope}[shift={(0,0)},rotate=240]	
	\node at (2,0) {#3};
\end{scope}	
	
\end{scope}	 
} 

\newcommand{\treefourpointexchange}[6]{
\begin{scope}[rotate=#6]

\begin{scope}[shift={(-.75,0)} ]

\treelevelthreepointskel{fakerealscalar}{#1}{#2}{0}	
\begin{scope}[shift={(1.5,0)}]	
\treelevelthreepointskel{#5}{#3}{#4}{180}	
\end{scope}	

\end{scope}	 
\end{scope}	 
}

\newcommand{\labeltreefourpointexchange}[6]{
\begin{scope}[rotate=#6]

\begin{scope}[shift={(-.75,0)} ]

\node at (.75,-.5) {#5}; 

\labeltreelevelthreepoint{}{#1}{#2}{0} 
\begin{scope}[shift={(1.5,0)}]	
\labeltreelevelthreepoint{}{#3}{#4}{180}
\end{scope} 
	
\end{scope}	 
\end{scope}	 
}

\tikzstyle{block} = [draw, rectangle, 
    minimum height=3em, minimum width=6em]

\usepackage{comment}

 \specialcomment{rudradetails}{%
  \begingroup \color{blue}}{%
  \endgroup}

\excludecomment{rudradetails}

\def\amom{J}

\def\Tr{\textrm{Tr}}

\def\iimg{ {\bf i}}

\def\diag{\textrm{diag}}

\def\invc{\textrm{in}}
\def\outvc{\textrm{out}}

\hbadness=\maxdimen
\vbadness=\maxdimen

\allowdisplaybreaks[1]

\newcommand{\gegen}{\mathcal{G} }

\newcommand{\legendre}{P} 

\newcommand{\jacobi}{\mathcal{J} }

\title{\boldmath Spinning amplitudes from scalar amplitudes}

\vspace{10mm}

\author[ \symbishop ]{ Mahesh KN Balasubramanian }
\author[ \symrook]{, Raj Patil }
\author[ \symbishop ]{, Arnab Rudra }
\author[]{\\ }
\vspace*{4.0ex} 
\vspace{10mm}
\affiliation[\symbishop]{Indian Institute of Science Education and Research Bhopal,\\
 Bhopal Bypass Rd, Bhauri, Madhya Pradesh 462066, India.\\ }
\vspace*{4.0ex} 
\affiliation[\symrook]{Indian Institute of Science Education and Research Pune,\\
Dr Homi Bhabha Rd, Ward No. 8, NCL Colony, Pashan, Pune, Maharashtra 411008, India.\\ }
\vspace*{4.0ex} 
\vspace*{2.0ex}
\emailAdd{mahesh16@iiserb.ac.in}
\emailAdd{patil.raj@students.iiserpune.ac.in}  
\emailAdd{rudra@iiserb.ac.in .} 

\abstract{ We provide a systematic method to compute tree-level scattering amplitudes with spinning external states from amplitudes with scalar external states in arbitrary spacetime dimensions. We write down analytic answers for various scattering amplitudes, including the four graviton amplitude due to the massive spin $J$ exchange. We verify the results by computing angular distributions in $3+1$ dimensions using various identities involving Jacobi polynomials. }

\begin{document} 
\maketitle

\newpage

\section{Introduction}
A relativistic particle with a spin more than two is commonly known as a Higher spin particle. These particles are necessarily massive since it is not possible to construct interacting, unitary, Poincare invariant quantum field theory for massless higher spin particles \footnote{To be more precise, it is not possible to have a theory of massless higher spin particle(s) with long range interaction.}. Such particles don't play any role in constructing the standard model of particle physics. However, there has been a lot of interest in understanding theories with massive higher spin particles in recent times. There are primarily two motivations. The first motivation comes from the theoretical quest to understand the structure of gravitational theories. The general relativity provides a consistent \footnote{One can contest this claim in the strong gravity regime. Hawking-Penrose singularity theorem predicts singularity for generic initial conditions. However, in such cases, the singularity is inside a horizon and hence not observable, at least classically.} classical dynamics for  gravity. But is this theory unique, or is it just a candidate amongst many possible classical theories of gravity? One way to approach this question is to understand the allowed space of higher-derivative corrections to the Einstein gravity. Any higher derivative modification is dimension-full, and hence they become relevant only at (and above) a particular energy scale. There are at least two possibilities for choosing the energy scale: All the higher derivative corrections may become important at the Planck scale $M_p$. To understand this class of theories, it is equally important to understand quantum corrections. Considering Einstein gravity (and all its supersymmetric cousins) is non-renormalizable\footnote{Though it is one-loop finite \cite{tHooft:1974toh}.} in 3+1 dimensions, it is not possible to understand them using our current knowledge of Quantum field theory. Another possibility is that when the higher-derivative corrections become relevant at a scale (say $\alpha$) which is parametrically smaller than Planck scale $M_p$ \footnote{String theory is a model quantum gravity, and it falls into this category. But from the low energy considerations, there is no compelling reason for the existence of such a scale. One can only draw an analogy to the Fermi theory of beta decay. The theory is non-renormalizable, and the quantum corrections become extremely important at 300 GeV. This problem is resolved in the Electroweak theory, where new physics/higher derivative corrections become important around 80 GeV. }. In such theories, it is possible to study perturbative corrections due to higher derivative operators without worrying about quantum/loop corrections. In \cite{Camanho:2014apa}, the authors used argument involving causality  to show that any quadratic or cubic higher derivative correction to Einstein gravity must be accompanied by an infinite tower of higher spin particles. This result was further strengthened in \cite{Chowdhury:2019kaq}, where the authors considered quartic corrections (= quartic contact interactions). The authors showed that there are no possible corrections to the graviton four-point contact interaction in $D\leq 6$. The question that one would like to understand is that what is the most general tree-level $S$-matrix for four gravitons such that it is consistent with Causality, Unitarity, and Lorentz invariance.   

In \cite{Chowdhury:2019kaq}, the authors made a series of conjectures on tree-level four gravitons $S$-matrix (the CRG conjectures). The tree-level four graviton amplitude in various String theories is an inspiration behind this series of conjectures. The work of \cite{Camanho:2014apa} establishes that in a model with particles spin, not more than 2, the consistent tree-level four gravitons $S$ matrix is unique. One would like to test these conjectures in the presence of higher-spin particles. For that purpose, it is necessary to find the expression for the $S$ matrices due to the massive higher spin exchanges. We wish to address that in this work.

The second motivation behind studying the massive higher spin particles is due to the recent discovery of gravitational waves. There has been a vast amount of work to understand the dynamics of two black holes or the scattering of gravitational waves from a black hole. Such physics can be understood by considering the spinning black holes as massive Higher spin particles \cite{Holstein:2008sx, Vaidya:2014kza, Guevara:2018wpp, Guevara:2019fsj}. A simple way to understand this connection \cite{Arkani-Hamed:2019ymq} is to note that the black holes in 3+1 dimensions are uniquely characterized mass, angular momentum, and charge (No-hair theorem). These are the same three quantum numbers that describe a quantum particle. A Kerr black hole can be considered a limit of highly spinning particle (to be more precise $|\vec S|\rightarrow \infty$, $\hbar \rightarrow 0$, keeping $|\hbar \vec S|$  fixed). Given these connections, the scattering of Higher spin particles can be used to compute the effective dynamics of spinning black hole system(s) \cite{Chung:2018kqs,Chung:2019duq, Arkani-Hamed:2019ymq,Bern:2020buy, Antonelli:2020ybz, Antonelli:2020aeb, Khalil:2020mmr, Siemonsen:2019dsu,Kosmopoulos:2021zoq}.

Scattering amplitudes with massive higher spin particles  got a new boost from the work Arkani-Hamed et. al. \cite{Arkani-Hamed:2017jhn}. The authors extended the formalism of spinor helicity variables to massive particles in $3+1$ dimensions and used them to compute the three-point function of massive higher spin particles. The authors also computed four-point amplitude due to massive exchanges.  In \cite{Chakraborty:2020rxf, Chowdhury:2020ddc}, the authors wrote down the photon-photon-massive spin $J$ and graviton-graviton-spin $J$ three-point in arbitrary dimensions. In spirit, our work is closer to the work of \cite{Chakraborty:2020rxf, Chowdhury:2020ddc}.

\subsection{Main results }

This paper considers amplitudes due to the exchange of massive higher spin particles, which transform in the symmetric, traceless representations of the little group ($SO(D-1)$). The main results of the papers are the following
\begin{enumerate}
	\item We showed that the amplitudes with the spinning external state could be obtained from {\it applying certain differential/multiplicative operators on the four scalar amplitude}. These operators can be constructed from the information of on-shell three-point functions. This is explained in sec. \ref{sec:ramderivativemethod}.  

A similar method for spinning CFT correlators can be found in \cite{Costa:2011dw}.

	\item Using this method, we derived an explicit expression for the following amplitudes in arbitrary spacetime dimensions

\begin{center}

\texttt{psss}, \texttt{psps}, \texttt{ppss}, \texttt{pppp}, \texttt{gsss}, \texttt{gsgs}, \texttt{ggss}, \texttt{ggpp}, and \texttt{gggg}.
	
\end{center}
Here $\texttt{s}$ stands for scalar, $\texttt{p}$ stands for photon, and $\texttt{g}$ stands for the graviton. These explicit expressions can be used in $S$-matrix bootstrap, bounding the co-efficient of EFTs and, in general, understanding the structure of higher spin theories. 

The operator approach can be used to construct amplitudes of any spinning external state. 

	\item In 4 spacetime dimensions, using various properties of Legendre polynomials and Jacobi polynomials, we were able to show that these amplitudes produce correct angular distribution in the center of mass frame. This provides an independent verification for the derivative method. 
\end{enumerate}

\paragraph{Organisation of the paper }

The organization of the paper is the following: In section \ref{sec:ramsetup}, we review various basic things that are essential throughout the paper. This section and the appendix \ref{app:ramnotation} also serve the purpose to set up the notation and convention of this paper. The key insight of this work is explained in section \ref{sec:ramderivativemethod}. In this section, we provide a method to construct spinning amplitudes from the action of certain operators on the scalar amplitudes. These operators can be constructed from the knowledge of the on-shell three-point functions. We use this derivative method to compute various amplitudes in sections \ref{sec:ramclassIamplitudes} and \ref{sec:ramclassIIamplitudes}. In all these sections, we also verify our answer by analytically computing the angular distributions in the center of the mass frame of a 3+1 dimensional theory. In section \ref{sec:ramconclusion}, we conclude with future directions . In appendix \ref{app:rampartialwave}, we summarize partial wave expansion of spinning amplitudes in $3+1$ dimensions. The appendix \ref{app:orthogonalpoly} summarizes all the essential features of Gegenbauer polynomials, Jacobi polynomials, Wigner matrices, and various mathematical identities.

\section{Set-up}
\label{sec:ramsetup}
A quantum particle is an irreducible representation of the Poincare group \cite{Weinberg:1995mt, Kessel:2017mxa, Bekaert:2006py}. The unitary representations of the Poincare group can be classified by the unitary representation of the Little group. In this work, we are primarily interested in massive particles. In $D$ spacetime dimensions, the little group for massive particles is $SO(D-1)$ . The representations can be denoted using the Young tableau. We focus only on the symmetric traceless representations of $SO(D-1)$; they have only one row in the young diagram. We denote it as $(m_a,J_a)$ - $m_a$ is the mass and $J_a$ is the number of boxes in young diagrams. In 3+1 dimensions, $J_a$ is the spin of the particle. In $3+1$ dimensions, a massive spin $J$ representation has $2J+1$ degrees of freedom; $2J$ is a non-negative integer. A massive spin $J$ representation for integer $J$ can be represented by a symmetric, traceless, transverse tensor of rank $J$. Mathematically, these conditions can be written as  
\begin{eqnarray}
\textrm{eom}\qquad&:&\qquad  (k^2+m^2)\Phi_{\mu_1\cdots\mu_J}=0 \qquad ,
\label{ramsetup1}
\\
\textrm{Symmetric}\qquad&:&\qquad  \Phi_{\mu_1\cdots\mu_J}=\Phi_{(\mu_1\cdots\mu_J)} \qquad ,
\label{ramsetup2}
\\
\textrm{Traceless}\qquad&:&\qquad    
\eta^{\mu \nu}
\Phi_{\mu\nu\mu_1\cdots\mu_{J-2}}=0 \qquad \text{and}
\label{ramsetup3}
\\
\textrm{Transverse}\qquad&:&\qquad    
k^{\mu}
\Phi_{\mu\mu_1\cdots\mu_{J-1}}=0 \qquad,
\label{ramsetup4}
\end{eqnarray}
where $k^\mu $ is the momentum. One can show that a rank $J$ tensor in $3+1$ dimensions, satisfying all the above conditions, has $2J+1$ degrees of freedom in $3+1$ dimensions. This rank $J$ tensor is known as the polarisation. The polarisation can also be written in-term simple auxiliary variable $\epsilon$
\begin{eqnarray}
  \Phi_{\mu_1\cdots\mu_J}= \epsilon_{\mu_1} \cdots \, \epsilon_{\mu_J}
~.   
\label{ramsetup5}
\end{eqnarray}
In terms of the auxiliary variables, the transverse and the traceless conditions simply means
\begin{equation}
k\cdot \epsilon =0 \qquad \text{and}  \qquad \epsilon \cdot \epsilon =0
~.  
\label{ramsetup11}
\end{equation}
Poincare group also has massless representations which can classified by representations of $ISO(D-2)$. If a particle is massless, then it enjoys gauge invariance
\begin{equation}
	\epsilon_\mu(k) \rightarrow \epsilon_\mu(k) +\alpha(k)\, k_\mu 
~. 	
\label{ramsetup12}
\end{equation}
In an unitary quantum field theory, the amplitude of massless particles must be gauge invariant. As a result, the amplitudes with external massless particles can always be written in terms of gauge invariant objects. For massless spin 1 particles, the amplitude can be written in terms of linearised field strength
\begin{equation}
	\mathcal{B}_{\mu \nu}^{(a)}= k_\mu^{(a)} \epsilon_\nu^{(a)} -k_\nu^{(a)} \epsilon_\mu^{(a)} 
~. 	
\label{ramsetup13}
\end{equation} 
We use $a,b$ to label particles. For future purposes, we define the following quantities 
\begin{equation}
	\mathcal{W}_{\mu \nu}^{(ab)}= \eta^{\rho\sigma}\mathcal{B}_{\mu \rho }^{(a)}\, \mathcal{B}_{\sigma \nu }^{(b)}	\quad\quad\text{and}\quad\quad \mathcal{W}^{(ab)}=\mathcal{W}_{\mu \nu}^{(ab)}\eta^{\mu\nu}~.
\label{ramsetup14}
\end{equation}
This will be useful later to write down the three point function of two photons and one spin $J$ particle. A massless spin 2 particle is represented by a symmetric, traceless tensor $\zeta_{\nu \sigma }$ with the following gauge invariance 
\begin{equation}
	\zeta_{\mu\nu }(k)\longrightarrow \zeta_{\mu\nu }(k)+k_\mu\, \alpha_\nu+k_\nu\, \alpha_\mu ~.
\label{ramsetup15}
\end{equation}
The gauge-invariant amplitude can be written in term of the linearized Riemann tensor 
\begin{equation}
	\mathcal{R}_{\mu \nu \rho \sigma }= k_\mu k_{\rho}\zeta_{\nu \sigma }+k_{\nu}k_{\sigma }\zeta_{\mu \rho }-k_\nu k_{\rho}\zeta_{\mu \sigma }-k_\mu k_{\sigma}\zeta_{\nu \rho}~.
\label{ramsetup21}
\end{equation}
Note that the linearized Ricci tensor (and Ricci scalar) vanishes due to Linearized Einstein's equations 
\begin{equation}
	\eta^{\mu \rho}\mathcal{R}_{\mu \nu \rho \sigma }= k^2\,\zeta_{\nu \sigma }+k_{\nu}k_{\sigma }\eta^{\mu \rho}\zeta_{\mu \rho }-k_\nu k^{\mu}\zeta_{\mu \sigma }-k^{\rho} k_{\sigma}\zeta_{\nu \rho}=0~.
\label{ramsetup22}
\end{equation}
For future convenience we define the following three tensors :  
\begin{equation}
\begin{split}
&	\mathcal{Y}^{(ab)}=\mathcal{R}_{\mu \nu \rho \sigma }^{(a)}(\mathcal{R}^{(b)})^{\mu \nu \rho \sigma }
\qquad,\qquad 	
\mathcal{Y}^{(ab)}_{\sigma_1\sigma_2}=	\mathcal{R}_{\mu \nu \rho \sigma_1 }^{(a)}{(\mathcal{R}^{(b)})^{\mu \nu\rho }}_ { \sigma_2 } \qquad \text{and}
\\
&	\mathcal{Y}^{(ab)}_{\mu_1\rho_1\mu_2\rho_2}=\mathcal{R}_{\mu_1 \nu \rho_1 \sigma }^{(a)}
	{{{(\mathcal{R}^{(b)})_{\mu_2}}^{\nu}}_{\rho_2}}^\sigma ~.
\end{split} 
\label{ramsetup23}
\end{equation}

\paragraph{Note on double copy} An extremely useful technique to compute massless amplitudes in $3+1$ dimensions is the double copy relations \cite{Kawai:1985xq, Bern:2008qj, Bern:2010ue}. These relations relate the gravity amplitudes to the gauge theory amplitudes. One key observation behind these relations is to note that the graviton polarization can be written in terms of two copies of photon polarizations
\begin{equation}
	\zeta_{\mu \nu}[2h]=\epsilon_{\mu }[h]\,\tilde \epsilon_{\nu }[h]
\label{ramsetup24}
\end{equation}
where, $h=\pm1$. This observation leads to the following facts. We have already defined linearized field strength of the photon field in \eqref{ramsetup13}. Then one can show that the Linearized Riemann tensor, given in \eqref{ramsetup21}, can be written as a product of two Maxwell field strengths 
\begin{equation}
\mathcal{R}_{\mu \nu \rho \sigma }^{(a)}=\mathcal{B}_{\mu \nu}^{(a)}\, \widetilde {\mathcal{B}}_{\rho \sigma }^{(a)}	~.
\label{ramsetup25}
\end{equation}
As a result, the tensor factors written in \eqref{ramsetup23} can also be written as the double copy of photon tensor structures  
\begin{equation}
\mathcal{Y}^{(ab)}= \mathcal{W}^{(ab)}\, 	\widetilde{\mathcal{W}}^{(ab)}
\qquad,\qquad 
\mathcal{Y}^{(ab)}_{\mu\nu}=\mathcal{W}^{(ab)}\,	\widetilde{\mathcal{W}}^{(ab)}_{\mu \nu }
\qquad\text{and}\qquad 
\mathcal{Y}^{(ab)}_{\mu\nu\rho\sigma}=\mathcal{W}^{(ab)}_{\mu \nu }\, 	\widetilde{\mathcal{W}}^{(ab)}_{\rho\sigma}~.
\label{ramsetup31}
\end{equation}
We use these relations while computing gravitational amplitudes.

\subsection{Three point function and propagators of higher spin particles}
The purpose of this paper is to compute four point function due to spin $J$ exchange. In order to do that first we write down the three point functions. The three-point function of two photons (labeled by 1 and 2) with a massive spin $J$ (labeled by 3) has two possible structures \cite{Arkani-Hamed:2017jhn, Chowdhury:2019kaq}
\begin{equation}
\Big[\mathcal{W}_{(12)}^{\mu \nu }(\epsilon_3)_{\mu }(\epsilon_3)_{\nu }\Big] (\epsilon_3 \cdot k_{12})^{J-2}
\qquad\quad\text{and}\quad\qquad 
\Big[\mathcal{W}_{(12)}^{\mu \nu }\eta_{\mu\nu }\Big] (\epsilon_3 \cdot k_{12})^{J}~
\label{ramsetup32}
\end{equation}
where, $k_{ab}^\mu=k_a^\mu-k_b^\mu$. 
\footnote{The tensors $\mathcal{W}_{(12)}^{\mu \nu }$ satisfy the following identities 
\begin{equation}
	\mathcal{W}_{(12)}^{\mu \nu} k_{1~\mu} = 0 = \mathcal{W}_{(12)}^{\mu \nu} k_{2~\nu} \quad,\quad \mathcal{W}_{(12)}^{\mu \nu} k_{2~\mu} = -\frac{1}{2} k_{2}^\nu \mathcal{W}_{(12)} \quad\text{and}\quad \mathcal{W}_{(12)}^{\mu \nu} k_{1~\nu} = -\frac{1}{2} k_{1}^\mu \mathcal{W}_{(12)}. \nonumber
\end{equation}
Using these equations we can show any other Lorentz scalar (for example, $\mathcal{W}_{(12)}^{\mu \nu} k_{12~\mu} k_{12~\nu}$) can be written in terms of the two Lorentz scalars given in equation (\ref{ramsetup32}).
}
The second one has more factors of momentum(/derivative) compared to the first one. So, the first one is the minimal coupling, and the second one is the non-minimal coupling. Similarly, the three-point functions of two gravitons and one spin $J$ particle are given by \cite{Arkani-Hamed:2017jhn, Chowdhury:2019kaq}
\begin{equation}
\begin{split}
&	\mathcal{Y}^{(12)}(k_{12}\cdot\epsilon_3 )^{J}\qquad ,
\\	
&	(\mathcal{Y}^{(12)}_{\sigma_1\sigma_2}\epsilon_3^{\sigma_1}\epsilon_3^{\sigma_2})(k_{12}\cdot\epsilon_3 )^{J-2}\qquad \text{and}
\\	
&	(\mathcal{Y}^{(12)}_{\mu_1\rho_1\mu_2\rho_2}\epsilon_3^{\mu_1}\epsilon_3^{\rho_1}\epsilon_3^{\mu_2}\epsilon_3^{\rho_2})(k_{12}\cdot\epsilon_3 )^{J-4}~.
\\	
\end{split}	
\label{ramsetup33}
\end{equation}
Given a three point function of two massless and one massive spin $J$ particle: $\mathcal{M}$, we define the following quantity 
\begin{equation}
	\mathcal{M} = 
M^{\mu_1\cdots\mu_J}
	\epsilon_{\mu_1}^{(3)} \cdots \epsilon_{\mu_J}^{(3)}
\label{ramsetup34}
\end{equation} 
where $M^{\mu_1\cdots\mu_J}$ is the stripped three-point function; here we have written the polarization for spin $J$ particles explicitly. It will be useful when we try to compute four point function. 
The other thing that we need to derive the expression of various tree level amplitudes is the higher spin propagators. The expression for the propagator of higher spin particles, transforming in the symmetric traceless representation, is \cite{Singh:1974qz, Ingraham:1974un} \footnote{A derivation of the expression for propagators in $D$ dimensions can be found in section 4.1 of \cite{Chandrasekaran:2018qmx}; see also eqn L.2 of \cite{Chowdhury:2019kaq}. }
\begin{equation}
\langle\Phi_{\mu_1\cdots \mu_J}(p)\, \Phi_{\nu_1\cdots \nu_J}(-p) \rangle =	\frac{-\iimg }{p^2+m_J^2}\mathcal{P}^{(J)}_{\mu_{1}\ldots\mu_{J};\nu_{1}\ldots\nu_{J}}(p)~.
\label{hspinexchange1}
\end{equation}
The tensor $\mathcal{P}^{(J)}_{\mu_{1}\ldots\mu_{J};\nu_{1}\ldots\nu_{J}}(p)$ appearing in the numerator of the propagator is a projector. Given a momentum $p^\mu$, it projects to symmetric traceless representations of $SO(D-1)$, which leaves $p^\mu$ invariant (i.e. the little group of $p^\mu$). One can also construct similar projectors other representations of $SO(D-1)$ \cite{Costa:2016hju}. $\mathcal{P}^{(J)}_{\mu_{1}\ldots\mu_{J};\nu_{1}\ldots\nu_{J}}(p)$ is given by, \footnote{The propagator $\mathcal{P}^{(J)}$ can be determined from the fact that it is the spin $J$ projector
	\begin{equation}
	\begin{split}
	{\mathcal{P}^{(J)}}_{\mu_{1}\ldots\mu_{J}\,;\,\nu_{1}\ldots\nu_{J}}&=
	{\mathcal{P}^{(J)}}_{(\mu_{1}\ldots\mu_{J})\,;\,\nu_{1}\ldots\nu_{J}}
	={\mathcal{P}^{(J)}}_{\mu_{1}\ldots\mu_{J}\,;\,(\nu_{1}\ldots\nu_{J})}
	\\	
	p^{\mu_1}{\mathcal{P}^{(J)}}_{\mu_{1}\ldots\mu_{J}\,;\,\nu_{1}\ldots\nu_{J}}&=0=p^{\nu_1}{\mathcal{P}^{(J)}}_{\mu_{1}\ldots\mu_{J}\,;\,\nu_{1}\ldots\nu_{J}}
	\\	
	\Theta^{\mu_1\mu_2}{\mathcal{P}^{(J)}}_{\mu_{1}\ldots\mu_{J}\,;\,\nu_{1}\ldots\nu_{J}}&=0=\Theta^{\nu_1\nu_2}{\mathcal{P}^{(J)}}_{\mu_{1}\ldots\mu_{J}\,;\,\nu_{1}\ldots\nu_{J}}
	\\
	{\mathcal{P}^{(J)}}_{\mu_{1}\ldots\mu_{J}\,;\,\rho_{1}\ldots\rho_{J}} {{\mathcal{P}^{(J)}}^{\rho_{1}\ldots\rho_{J}\,}}_{\,\nu_{1}\ldots\nu_{J}} &=\mathcal{P}^{(J)}_{\mu_{1}\ldots\mu_{J}\,;\,\nu_{1}\ldots\nu_{J}}
	\nonumber
	\end{split}
	\end{equation}
}
\begin{equation}
\mathcal{P}^{(J)}_{\mu_{1}\ldots\mu_{J}\,;\,\nu_{1}\ldots\nu_{J}}=\sum_{a=0}^{\lfloor\frac{J}{2}\rfloor}~A(J,a,D)~\Bigg\{\Theta_{\mu_1\mu_2}\Theta_{\nu_1\nu_2}\ldots\Theta_{\mu_{2a-1}\mu_{2a}}\Theta_{\nu_{2a-1}\nu_{2a}}~\Theta_{\mu_{2a+1}\nu_{2a+1}}\ldots\Theta_{\mu_{J}\nu_{J}}\Bigg\}_\text{sym($\mu$)\, sym($\nu$)}
\label{hspinexchange2}
\end{equation}
where $D$ is the number of space-time dimensions. $\lfloor n\rfloor$ denotes greatest integer which is less than or equal to $n$. . The sum has $\lfloor\frac{J}{2}\rfloor+1$ terms; for each term, we have to symmetrize over all the $\mu$ and $\nu$ indices. After symmetrization, the number of terms in the spin $J$ propagator is $	\frac{(2J)!}{2^J\cdot J!}$. $A(J,a,D)$ is given by 
\begin{equation}
 A(J,a,D) = \Bigg[\frac{(-1)^a J!(2J+D-2a-5)!!}{2^a a!(J-2a)!(2J+D-5)!!}\Bigg]~.
\label{hspinexchange3}
\end{equation}
Here $\Theta_{\mu\nu}$ is defined as
\begin{equation}
\Theta_{\mu\nu} =\eta_{\mu \nu}-\frac{p_\mu\, p_\nu}{p^2}~.
\label{hspinexchange4}
\end{equation}
$\Theta_{\mu\nu}$ is essentially the flat (euclidean) metric defined on the plane perpendicular to $p^\mu$. 
$\Theta_{\mu\nu}$ will play an important role in our analysis. From the definition it follows that  
\begin{eqnarray}
	\Theta^{\mu \nu}&=&\eta^{\mu\nu}-\frac{p^\mu p^\nu}{p^2}\implies {\Theta^\mu}_{\nu}={\eta^\mu}_\nu-\frac{p^\mu p_\nu}{p^2}\quad,
\label{hspinexchange5}
\\
	\Theta^{\mu \nu}\, \Theta_{ \nu\rho}&=& {\Theta^{\mu}}_\rho\qquad\text{and}\qquad {\Theta^\mu}_ \nu\, {\Theta^\nu}_\rho= {\Theta^{\mu}}_\rho~.
\label{hspinexchange11} 	
\end{eqnarray}
For future purposes, we set this convention: We define {\it vectors with only lower indices} and we use two kinds of dot product: 
\begin{equation}
	A\cdot B = \eta^{\mu \nu}A_{\mu }B_{\nu}
\qquad\qquad\text{and}\qquad\qquad	
	A\odot B = \Theta^{\mu \nu}A_{\mu }B_{\nu}~.
\label{hspinexchange12}
\end{equation}
For two arbitrary vectors these two dot products don't agree.

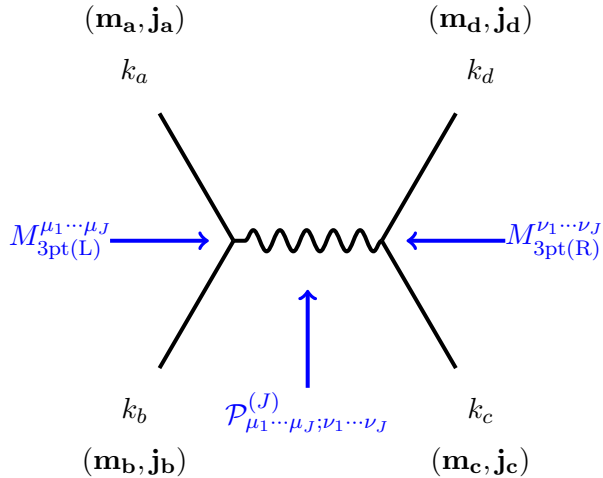
\begin{figure}[h]
\begin{center}
\begin{tikzpicture}[line width=1.5 pt, scale=1.3]
 
\begin{scope}[shift={(0,0)}]	
		
\treefourpointexchange{realscalar}{realscalar}{realscalar}{realscalar}{graviton, black}{0}	

\labeltreefourpointexchange{ $k_a$ }{ $k_b$ }{ $k_c$ }{ $k_d$ }{}{0}

 \node at (-1.75,2.25) {$(\bf m_a,j_a)$};	
 \node at (-1.75,-2.25) {$(\bf m_b,j_b)$};	  
 \node at (1.75,-2.25) {$(\bf m_c,j_c)$};	  
 \node at (1.75,2.25) {$(\bf m_d,j_d)$};

\begin{scope}[shift={(-1,0)}]		
	\draw[realscalar, blue][<-] (0,0)--(-1,0);
 \node at (-1.5,0) {${\color{blue}M_{\text{3pt(L)}}^{\mu_1\cdots\mu_J}	}$};	  
\end{scope}

\begin{scope}[shift={(1,0)}]		
	\draw[realscalar, blue][<-] (0,0)--(1,0);
 \node at (1.5,0) {${\color{blue}M_{\text{3pt(R)}}^{\nu_1\cdots\nu_J}	}$};	  
\end{scope}

\begin{scope}[shift={(0,-.5)}]		
	\draw[realscalar, blue][<-] (0,0)--(0,-1);
 \node at (0,-1.25) {${\color{blue}\mathcal{P}^{(J)}_{\mu_1\cdots\mu_J;\nu_1\cdots\nu_J}	}$};	  
\end{scope}

\end{scope}

\end{tikzpicture} 
\end{center}
\caption{Schematic Feynman diagram for a massive spinning particle exchange}
\label{fig:fourpointdiag}
\end{figure}

Given all these ingredients, a general four point function (see fig. \ref{fig:fourpointdiag}) for a particular channel is given by 
\begin{equation}
\Big(M_{\text{3pt(L)}}^\texttt{(i)}\Big)^{\mu_1\cdots\mu_J} ~\mathcal{P}^{(J)}_{\mu_1\cdots\mu_J,\nu_1\cdots\nu_J}~\Big(M_{\text{3pt(R)}}^\texttt{(j)}\Big)^{\nu_1\cdots\nu_J} \left[ 	\frac{-\iimg }{p^2+m_J^2}\right]
\label{hspinexchange13}
\end{equation}
where $M_{\text{3pt(L)}}$ and $M_{\text{3pt(R)}}$ are the stripped three point functions, defined in eqn \eqref{ramsetup34} and the value of $\texttt{i}$ denotes $n^\texttt{i}m$ three point function on the left of the propagating particle, and similarly for the right side.

The non-trivial part is to compute various Lorentz contractions for a general higher spin particle. One can always do the computation in a brute force way; the computations are algebraically intensive. In this paper, we find the expressions of the numerators by the action of certain differential/multiplicative operators on scalar amplitudes. We describe this method in section \ref{sec:ramderivativemethod}.

\paragraph{Notation}
In our convention, all the particles are outgoing. We write the tree-level non-analytic piece of the full four point amplitude in the following way 
\begin{align}
\mathcal{A}_{\texttt{xyzw}}~.
\end{align}
This notation denotes the fact that it is scattering amplitude of four particles: \texttt{x, y, z, w}. We are focussing on tree level four point amplitude due to exchange of massive particle(s). So these amplitudes depend on the three point function. In general the three point function can have more than one independent Lorentz invariant structure. Every independent structure comes with independent coupling constant. So the full amplitude given by 
\begin{align}
\mathcal{A}_{\texttt{xyzw}} = \sum_{\texttt{i,j}}\,g^\texttt{(i)}_{\texttt{xyJ}} g^\texttt{(j)}_{\texttt{Jzw}}\frac{\iimg}{s - m_\text{J}^2} ~\texttt{A}^\texttt{(i|j)}_{\texttt{xyzw}}
+
\textrm{other two channels}~.
\label{hspinexchange14}
\end{align}
The sum denotes summation over various contribution that is obtained from independent Lorentz invariant three point functions. Here the value of $\texttt{i}$ denotes $n^\texttt{i}m$ three point function on the left of the propagating particle, and similarly for the right side. For example $\texttt{i}=0$ denotes the minimal coupling, $\texttt{i}=1$ denotes the non-minimal coupling and so on. $g^\texttt{(i)}_{\texttt{xyJ}}$ and $g^\texttt{(j)}_{\texttt{Jzw}}$ are corresponding coupling constants. They are real numbers. The subscript of $\texttt{A}^\texttt{(i|j)}_{\texttt{xyzw}}$ mean that the particle \texttt{x} and the particle \texttt{y} meet at vertex and the other particles \texttt{z} and \texttt{w} meet at the other vertex. $\texttt{A}^\texttt{(i|j)}_{\texttt{xyzw}}$ is the part obtained from contracting $\mathcal{P}^{(J)}_{\mu_1\cdots\mu_J\,;\,\nu_1\cdots\nu_J}$ of spin $J$ vertices with (stripped) three point functions; the super-script $^\texttt{(i|j)}$ denotes what three points goes into this contribution. 

For example, consider the four photon amplitude due to massive spin $J$ exchange. The photon-photon spin $J$ has two independent structures. So, the full amplitude is given by  
\begin{align}
\label{ram_equation_to_refer_later}
\mathcal{A}_{\texttt{pppp}} = \sum_{\texttt{i}=0}^{1}\sum_{\texttt{j}=0}^{1} g_{\texttt{ppJ}}^\texttt{(i)} ~g_\texttt{Jpp}^\texttt{(j)}~\frac{\iimg}{s - m_\texttt{J}^2} ~\texttt{A}_\texttt{pppp}^\texttt{(i|j)}~
+
\textrm{other two channels}~.
\end{align}
Sometimes, we drop the notation for the minimal and the non-minimal couplings and instead use the superscript for denoting the total helicity of the incoming and outgoing particles. For example, the minimally coupled \texttt{pppp} amplitude (just the analytic part) for helicity exchange $2^+ \rightarrow 2^+$ is denoted as $ \texttt{A}_{\text{\texttt{pppp}}}^{2^+\rightarrow 2^+}$.
In the rest of the paper, we only compute the numerator of the s-channel amplitude $\texttt{A}_{\texttt{xyzw}}$ for arbitrary dimensions. Amplitudes for the other channels can be obtained from s-channel by changing the labels of the incoming and outgoing particles, subjected to the conservation laws.
	For example, consider the following amplitude involving four scalars
\begin{equation}
	\Phi_{q_1}(k_1)+\Psi_{q_2}(k_2)\longrightarrow \Phi_{q_1}(k_3)+\Psi_{q_2}(k_4)~.
\label{hspinexchange16}
\end{equation}
where, $q_i$ is the charge of the corresponding particle due to some internal symmetries.
For simplicity, let's assume $\Phi$ and $\Psi$ have charge $q_1$ and $q_2$ ($q_1\ne q_2$). Then numerator for the $s$ channel exchange and $t$ channel exchange vanishes simply because of charge conservation.

\subsection{Angular distribution and connection to the Wigner matrices}
For these scattering processes, one can always go to the center of mass frame. 
In the center of mass frame these amplitudes admit a partial wave decomposition \cite{Jacob:1959at}. In $3+1$ dimensions, for on-shell massless spinning states it is possible to define a Lorentz invariant quantity, namely the helicity; helicity is the projection of the spin along the spatial component of the 4-momentum
\begin{equation}
	\frac{\vec p\cdot \vec J}{|\vec p|}
\label{angdist1}
\end{equation} 
where $\vec p$ is the spatial part of the 4-momentum; $\vec J$ is the angular momentum operator. We choose the center of mass frame in such a way that the scattering process happen in $x-z$ plane (see appendix \ref{subsec:comconvention}). The incoming particle are travelling along the $z$ axis and the scattering angle is $\theta$. For s-channel tree-level scattering of four massless spinning particles, the helicity amplitude in the center of mass is given by 
\begin{equation}
\mathcal{M}(1^{h_1},2^{h_2};3^{h_3},4^{h_4} )=	 \widetilde{\mathcal{N}}_{J;h,h^\prime} \left[\frac{s}{2}\right]^{\frac{|h_1+h_2|+|h_3+h_4|}{2}} \frac{1}{s-m_J^2+\iimg \varepsilon } s^{J} \, d^{(J)}_{hh^\prime }(\theta)
\label{angdist2}
\end{equation} 
where $h=h_1-h_2$ and $h^\prime =h_3-h_4$. $m_J$ and  $J$  are the mass and the spin the of the exchange particle.  $d^{(J)}_{hh^\prime }(\theta)$ is the Wigner little-$d$ matrix (see \ref{subsec:wignermatrix} for a review). $\widetilde{\mathcal{N}}_{J;h,h^\prime}$ is given by 
\begin{equation}
\widetilde{\mathcal{N}}_{J;h,h^\prime}=	\mathcal{C}_{J,h}\, \mathcal{C}_{J,h^\prime }
\qquad;\qquad
 \mathcal{C}_{J,h}=	2^{\frac{J}{2}-h} \sqrt{ \frac{ \Gamma (J-h+1)\, \Gamma (J+h+1)}{ \Gamma (2 J+1)} }~.
\label{angdist3}
\end{equation}
This formula can be derived from the $SO(3)$ representation theory. (section 2 of \cite{Joglekar:1973hh}, eqn 6.47 in \cite{Arkani-Hamed:2020blm}, \cite{Hebbar:2020ukp}, We present it in the appendix \ref{app:rampartialwave}. We also provide a derivation for the normalization.)

One important ingredient in the expression for the above formula is the three-point function. In a generic quantum field theory, the 3-point functions are a linear combination of all possible Lorentz scalars that one can write down from the momenta and the polarization. In this paper, we require the three-point function of two massless particles and one massive spin $J$ particle. The structure of such a three-point function is highly constrained \cite{Arkani-Hamed:2017jhn, Chakraborty:2020rxf} due to gauge invariance of the massless legs. The three-point function of one massive and two massless particles can be studied in the rest frame of the massive particle. The massless particles travel along the positive and negative $z$ axis. We measure the spin of a massive particle along $z$ (This is just a convenient choice; we can measure the spin of the massive particle along any other axis. This can be obtained by rotating it with the spin $J$ Wigner matrix). Since $J_z$ is an additive quantum number, we can find $J_z$ of the massive particle from the helicity of the massless particles. We have shown one example in the table. \ref{tab:helicitytable}

\begin{table}[h]
 \begin{center}
\begin{tabular}{ |p{2.5cm}|p{2.5cm}||p{2.5cm}| }
   \hline
   \hline
   \multicolumn{3}{|c|}{{Minimal coupling} } \\
   \hline
   \hline
   $h_1$ & $h_2$ & $J_z^{(3)}$ \\
   \hline
    1& -1 & +2\\
   \hline
    -1 & 1 &-2 \\
   \hline
   \hline
   \hline
   \multicolumn{3}{|c|}{{Non-minimal coupling} } \\
   \hline
   \hline
    1& 1 & 0\\
   \hline
    -1 & -1 &0  \\
   \hline
     \end{tabular}
  \caption{Helicity decomposition of photon-photon-massive spin $J$ three point function}
  \label{tab:helicitytable}
 \end{center}
\end{table}

This simplicity of the three-point function simplifies our analysis drastically. For example, if we use the minimal coupling and construct the four-point function out of it, then the four-point function is non-zero for helicity configurations given in the table \ref{tab:helicitytable}; Otherwise, the amplitude has to be zero. It should be valid for any spin. This provides an important cross-check for our computation. For example, consider the angular distribution formula \eqref{angdist2} for the two photons and two scalar amplitude. The structure of minimal coupling implies that if set the non-minimal coupling to $0$, then the amplitude must be proportional to $d_{\pm20}^{(J)}(\theta)$. We show this explicitly in sec \ref{subsec:ppssamplitude}.

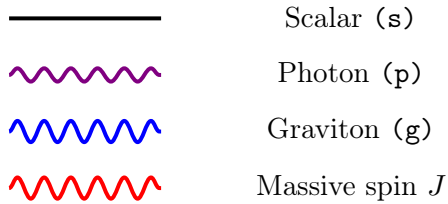
\begin{figure}[h]
\begin{center}
\begin{tikzpicture}[line width=1.5 pt, scale=1]

\begin{scope}[shift={(6,0)}]

\draw[realscalar] (-1,0)--(1,0); 
\node at (3.5,0) {Scalar \texttt{(s)}}; 

\draw[photon] (-1,-.75)--(1,-.75); 
\node at (3.5,-.75) {Photon \texttt{(p)}}; 
 
\draw[graviton] (-1,-1.5)--(1,-1.5); 
\node at (3.5,-1.5) {Graviton \texttt{(g)}}; 

\draw[spinj] (-1,-2.25)--(1,-2.25); 
\node at (3.5,-2.25) {Massive spin $J$}; 

 
\end{scope}

\end{tikzpicture}
\end{center}
\caption{Conventions for Feynman diagrams} 
\label{fig:ramfeyndiagconv}
\end{figure}

In this paper, $\texttt{s}$, $\texttt{p}$ and $\texttt{g}$ stands for scalar, photon and graviton respectively. We use black line to denote \texttt{scalar}, a violet wavy line for \texttt{photon}, a blue wavy line to denote \texttt{graviton} and a red wavy line to denote a \texttt{massive spin J} particle. These are shown in fig. \ref{fig:ramfeyndiagconv}.

\subsection{Review: Four scalar amplitude} 
We start with the simplest case: the scattering amplitude of four scalars due to massive higher spin-exchange \footnote{This result is well known; for example, appendix A.1 of \cite{Caron-Huot:2016icg}, eqn A.4 of \cite{Nayak:2017qru}, Eqn L.7 of \cite{Chowdhury:2019kaq}. }. The three-point function of two scalars and one spin $J$ is unique.

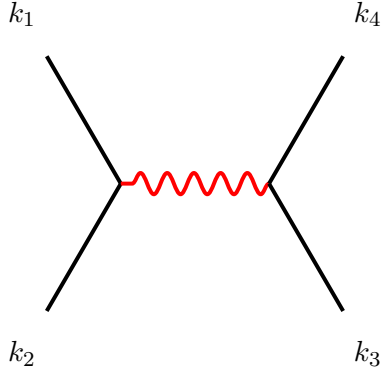
\begin{figure}[h]
\begin{center}. 
\begin{tikzpicture}[line width=1.5 pt, scale=1.3]
 
\begin{scope}[shift={(0,0)}]	
		
\treefourpointexchange{realscalar}{realscalar}{realscalar}{realscalar}{spinj}{0}	

\labeltreefourpointexchange{ $k_1$ }{ $k_2$ }{ $k_3$ }{ $k_4$ }{}{0} 
\end{scope}

\end{tikzpicture} 
\end{center}
\caption{Four scalar amplitude due to exchange of a massive spinning particle }
\label{fig:}
\end{figure}

The amplitude for four arbitrary scalars due to spin $J$ exchange is given by 
\begin{equation}
\texttt{A}_{\texttt{ssss}}= \sum_{a=0}^{\lfloor\frac{J}{2}\rfloor}~A(J,a,D) \Big[(k_{12}\odot k_{12})( k_{34}\odot k_{34} )\Big]^{a}(k_{12}\odot k_{34} )^{J-2a}
\label{sssshighspinex1}
\end{equation}
where $\odot$ is defined in \eqref{hspinexchange12}. If we choose the incoming particles to have same mass and the outgoing particles to have same masses, then we can replace the theta dot product $``\odot"$ by eta dot product $``\, \cdot\,"$ ($k_{12}\odot k_{12}\longrightarrow k_{12}\cdot k_{12}$ and $k_{34}\odot k_{34} \longrightarrow k_{34}\cdot k_{34}$). 

In the center of mass-frame and the in s-channel, the four scalar amplitude becomes Gegenbauer polynomial (see appendix \ref{subsec:gegenpoly} for a brief introduction to the Gegenbauer polynomial)
\begin{equation}
	\big(|k_{12}||k_{34}|\big)^{J}\frac{\Gamma(J+1)\Gamma\, \left( \beta\right)}{2^J\Gamma \left( \beta+J\right)} \gegen^{(\beta)}_J(\hat k_{12}\cdot \hat k_{34})
\qquad\text{where}\qquad
\beta= \frac{D-3}{2}	~.
\label{sssshighspinex3}
\end{equation}
If we put the mass of the external scalars to be zero then the scattering amplitude is
\begin{equation}
	\mathcal{N}_{J,D}\, s^J\gegen^{(\beta)}_J(\hat k_{12}\cdot \hat k_{34}) 
\qquad\text{where}\qquad
\mathcal{N}_{J,D\equiv2\beta+3}=\frac{\Gamma(J+1)\Gamma\, \left( \beta\right)}{2^J\Gamma \left( \beta+J\right)} ~.
\label{sssshighspinex4}
\end{equation}
The Gegenbauer polynomial satisfies the following property
\begin{equation}
	\gegen^{(\beta)}_J(z) >0 \qquad\text{for}\qquad z> 1 ~.
\label{sssshighspinex5}
\end{equation}
For physical scattering the variable $z$ is less than equal to $1$; $z> 1$ regime corresponds to unphysical scattering; this regime was considered in \cite{Caron-Huot:2016icg}.  
 
\paragraph{Angular distribution in $3+1$ dimensions}
In $3+1$ dimensions, let's write the answer in the special frame
\begin{equation}
	(k_{12})_\mu=(0,0,0,2p)
\qquad\text{and}\qquad
(k_{34})_\mu= (0,2\tilde p \sin \theta,0, 2\tilde p\cos\theta )	~.
\label{sssshighspinex11}
\end{equation}
In this frame the four-point amplitude is given by 
\begin{equation}
	g^2\mathcal{N}_{J,4}\frac{1}{s-m_J^2+\iimg \epsilon }(4p\tilde p)^{J}\, d^{(J)}_{00}(\theta) 
\qquad\text{where}\qquad 
	\mathcal{N}_{J,4}= \frac{\pi^{1/2}  \Gamma (J+1)}{2^{J}\, \Gamma \left(J+\frac{1}{2}\right)}~.
\label{sssshighspinex12}
\end{equation}

\subsection{Spinning amplitude(s), tensor factor(s) and form factor(s)}

We have discussed the scattering amplitude of four external scalars. But more generally the external states can have spin. We refer such amplitudes as spinning amplitude and this is the main interest in this paper. The external spinning states are characterised by momentum and polarization (see the discussion in sec \ref{sec:ramsetup}). Any flat space scattering amplitude is a Lorentz scalar which is constructed out of polarisation and momenta. The dependence on the polarization is always linear; however, the amplitude generally has a complicated dependence on momenta. The linear dependdance on the polarization allows us to write a spinning amplitude  in the following form 
\begin{eqnarray}
\sum_{I} \mathcal{T}_I(\{\epsilon_a,k_a\})\, \mathcal{F}_I(\{s_{ab}\})
\label{ramdecompositionI}
\end{eqnarray}
where $\mathcal{F}_I(\{s_{ab}\})$ is called {\it Form factor} - it does not depend on the polarisations, it only depends on the momenta. The polarization dependence is completely encoded in $\mathcal{T}_I(\{\epsilon_a,k_a\})$- they are called {\it Tensor/Polarization factor}. Generally, one can construct many independent Lorentz scalars out of polarisation and momenta; $I$ is the sum over all independent tensor factors. For massless particles, we choose a basis of the tensor factors such that it is gauge-invariant.

The non-analytic pieces of the form factors are fixed by Unitarity \cite{Cutkosky:1960sp, Veltman:1963th}, and the analytic piece is constrained by Causality \cite{Maldacena:2015waa, Camanho:2014apa, Chandorkar:2021viw}. 

In this paper, we only consider tree-level exchanges and thus write the form factors without the propagators. For the simple case of scalar amplitude, the tensor factor is simply identity. The form factor for spin $J$ exchange is
\begin{equation}
\sum_{a=0}^{\lfloor\frac{J}{2}\rfloor}~A(J,a,D) \Big[(k_{12}\cdot k_{12})( k_{34}\cdot k_{34} )\Big]^{a}(k_{12}\cdot k_{34} )^{J-2a}~.
\end{equation}

\subsection{A basis for the tensor structures}
\label{subsec:basisfortenstr}
In this paper, we write down expression for various spinning amplitudes. All such amplitudes can be written in the form \eqref{ramdecompositionI}. Here we list a basis for the Tensor structures that appears in various computation. We focus only on $s$-channel amplitudes since the expression for other channels can be obtained by $1\leftrightarrow 3$, $1\leftrightarrow 4$ exchange. We have already defined the linearized field strength $\mathcal{B}^{(a)}_{\mu \nu}$ in \eqref{ramsetup13} and $\mathcal{W}_{\mu \nu}^{(ab)}$ in \eqref{ramsetup14}. We define one more tensor
\begin{equation}
	\mathcal{X}^{(a;b)}_\mu = \mathcal{B}^{(a)}_{\mu \nu}\, (k_b)^\nu 	~.
\end{equation}
Using $\mathcal{W}_{\mu \nu}^{(ab)}$ and $\mathcal{X}^{(a;b)}_\mu$, we define the following 9 tensors which form a basis to write down the tensor structures for all the amplitudes in this paper.
\begin{equation}
\begin{split}
\widehat{\mathcal{T}}_{1}&=	\Big[\Tr\, \mathcal{W}^{(12)}\Big](k_{34})^2
\qquad,\qquad 
\widehat{\mathcal{T}}_{2}=	\Big[\mathcal{W}^{(12)}_{\mu\nu}(k_{34})^\mu(k_{34})^\nu\Big]
\qquad,
\\
\mathcal{T}_{3}&=\mathcal{W}^{(12)}_{\mu_1\mu_2}\Big[\eta^{\mu_1 \nu_1}\eta^{\mu_2 \nu_2}+\eta^{\mu_1 \nu_2}\eta^{\mu_2 \nu_1} \Big]\mathcal{W}^{(34)}_{\nu_1\nu_2}
\qquad\text{and}
\\
\mathcal{T}_{4}&=	\mathcal{W}^{(12)}_{\mu_1\mu_2}(k_{12})_{\mu_3}
\bigg[\eta^{\mu_1\nu_1}\eta^{\mu_2\nu_3}\eta^{\mu_3\nu_2}+\eta^{\mu_1\nu_2}\eta^{\mu_2\nu_3}\eta^{\mu_3\nu_1}\nonumber\\
 &~~~~~~~~~~~~~~~~~~~~~~~+\eta^{\mu_2\nu_1}\eta^{\mu_1\nu_3}\eta^{\mu_3\nu_2}+\eta^{\mu_2\nu_2}\eta^{\mu_1\nu_3}\eta^{\mu_1\nu_1}\bigg]
 \mathcal{W}^{(34)}_{\nu_1\nu_2}(k_{34})_{\nu_3} ~.
\end{split}
\end{equation}
These four structures (and various products of them) appear in \texttt{ppss}, \texttt{pppp}, \texttt{ggss}, \texttt{ggpp} and \texttt{gggg}. 
\begin{equation}
\begin{split}
\widehat{\mathcal{T}}_{5}=&
\mathcal{W}^{(12)}_{\mu_1\mu_2} \mathcal{W}^{(12)}_{\mu_3\mu_4} \mathcal{W}^{(34)}_{\nu_1\nu_2}(k_{34})_{\nu_3}(k_{34})_{\nu_4}
\\
&\Big[\eta^{\mu_1\nu_3}\eta^{\mu_2\nu_1}\eta^{\mu_3\nu_4}\eta^{\mu_4\nu_2}+(\mu_3\leftrightarrow \mu_4)
+(\mu_1\leftrightarrow \mu_2)
+ (\mu_3\leftrightarrow \mu_4)+(\mu_1\leftrightarrow \mu_2)
\Big]~.
\end{split}
\end{equation} 
$\mathcal{T}_{5}$ appear only for \texttt{ggpp} and \texttt{gggg}. 
\begin{equation}
\begin{split}
\mathcal{T}_{6}&=
(\mathcal{W}_{(12)})^{\mu_1\mu_2}(\mathcal{W}_{(12)})^{\mu_3\mu_4}\Bigg[(\mathcal{W}_{(34)})_{\,\mu_1\mu_4}(\mathcal{W}_{(34)})_{\,\mu_3\mu_2}
+(\mathcal{W}_{(34)})_{\,\mu_4\mu_1}(\mathcal{W}_{(34)})_{\,\mu_2\mu_3}
\\
 &+
2 (\mathcal{W}_{(34)})_{\,\mu_1\mu_4}(\mathcal{W}_{(34)})_{\,\mu_2\mu_3}+2
(\mathcal{W}_{(34)})_{\,\mu_1\mu_3}(\mathcal{W}_{(34)})_{\,\mu_2\mu_4}
+2
(\mathcal{W}_{(34)})_{\,\mu_1\mu_3}(\mathcal{W}_{(34)})_{\,\mu_4\mu_2}
\Bigg]\quad
\end{split}	
\end{equation}
\begin{equation}
\begin{split}
\mathcal{T}_{7}&=	\Bigg[
(k_{12})^{\mu_1} (k_{34})^{\mu_2} \mathcal{W}_{(12)\,\mu_2}^{\mu_3} \mathcal{W}_{(12)}^{\mu_4\mu_5} \mathcal{W}_{(34)\,\mu_1\mu_5} \mathcal{W}_{(34)\,\mu_3\mu_4} + (k_{12})^{\mu_1} (k_{34})^{\mu_2} \mathcal{W}_{(12)}^{\mu_3\,\mu_2} \mathcal{W}_{(12)}^{\mu_4\mu_5} \mathcal{W}_{(34)\,\mu_1\mu_5} \mathcal{W}_{(34)\,\mu_3\mu_4}\\
& + (k_{12})^{\mu_1} (k_{34})^{\mu_2} \mathcal{W}_{(12)\,\mu_2}^{\mu_3} \mathcal{W}_{(12)}^{\mu_4\mu_5} \mathcal{W}_{(34)\,\mu_1\mu_4} \mathcal{W}_{(34)\,\mu_3\mu_5} + (k_{12})^{\mu_1} (k_{34})^{\mu_2} \mathcal{W}_{(12)}^{\mu_3\,\mu_2} \mathcal{W}_{(12)}^{\mu_4\mu_5} \mathcal{W}_{(34)\,\mu_1\mu_4} \mathcal{W}_{(34)\,\mu_3\mu_5}\\
& + (k_{12})^{\mu_1} (k_{34})^{\mu_2} \mathcal{W}_{(12)\,\mu_2}^{\mu_3} \mathcal{W}_{(12)}^{\mu_4\mu_5} \mathcal{W}_{(34)\,\mu_3\mu_5} \mathcal{W}_{(34)\,\mu_4\mu_1} + (k_{12})^{\mu_1} (k_{34})^{\mu_2} \mathcal{W}_{(12)}^{\mu_3\,\mu_2} \mathcal{W}_{(12)}^{\mu_4\mu_5} \mathcal{W}_{(34)\,\mu_3\mu_5} \mathcal{W}_{(34)\,\mu_4\mu_1}\\
& + (k_{12})^{\mu_1} (k_{34})^{\mu_2} \mathcal{W}_{(12)\,\mu_2}^{\mu_3} \mathcal{W}_{(12)}^{\mu_4\mu_5} \mathcal{W}_{(34)\,\mu_1\mu_5} \mathcal{W}_{(34)\,\mu_4\mu_3} + (k_{12})^{\mu_1} (k_{34})^{\mu_2} \mathcal{W}_{(12)}^{\mu_3\,\mu_2} \mathcal{W}_{(12)}^{\mu_4\mu_5} \mathcal{W}_{(34)\,\mu_1\mu_5} \mathcal{W}_{(34)\,\mu_4\mu_3}\\
& + (k_{12})^{\mu_1} (k_{34})^{\mu_2} \mathcal{W}_{(12)\,\mu_2}^{\mu_3} \mathcal{W}_{(12)}^{\mu_4\mu_5} \mathcal{W}_{(34)\,\mu_3\mu_4} \mathcal{W}_{(34)\,\mu_5\mu_1} + (k_{12})^{\mu_1} (k_{34})^{\mu_2} \mathcal{W}_{(12)}^{\mu_3\,\mu_2} \mathcal{W}_{(12)}^{\mu_4\mu_5} \mathcal{W}_{(34)\,\mu_3\mu_4} \mathcal{W}_{(34)\,\mu_5\mu_1}\\
& + (k_{12})^{\mu_1} (k_{34})^{\mu_2} \mathcal{W}_{(12)\,\mu_2}^{\mu_3} \mathcal{W}_{(12)}^{\mu_4\mu_5} \mathcal{W}_{(34)\,\mu_4\mu_3} \mathcal{W}_{(34)\,\mu_5\mu_1}+ (k_{12})^{\mu_1} (k_{34})^{\mu_2} \mathcal{W}_{(12)}^{\mu_3\,\mu_2} \mathcal{W}_{(12)}^{\mu_4\mu_5} \mathcal{W}_{(34)\,\mu_4\mu_3} \mathcal{W}_{(34)\,\mu_5\mu_1}\\
& + (k_{12})^{\mu_1} (k_{34})^{\mu_2} \mathcal{W}_{(12)\,\mu_2}^{\mu_3} \mathcal{W}_{(12)}^{\mu_4\mu_5} \mathcal{W}_{(34)\,\mu_1\mu_4} \mathcal{W}_{(34)\,\mu_5\mu_3} + (k_{12})^{\mu_1} (k_{34})^{\mu_2} \mathcal{W}_{(12)}^{\mu_3\,\mu_2} \mathcal{W}_{(12)}^{\mu_4\mu_5} \mathcal{W}_{(34)\,\mu_1\mu_4} \mathcal{W}_{(34)\,\mu_5\mu_3}\\
& + (k_{12})^{\mu_1} (k_{34})^{\mu_2} \mathcal{W}_{(12)\,\mu_2}^{\mu_3} \mathcal{W}_{(12)}^{\mu_4\mu_5} \mathcal{W}_{(34)\,\mu_4\mu_1} \mathcal{W}_{(34)\,\mu_5\mu_3} + (k_{12})^{\mu_1} (k_{34})^{\mu_2} \mathcal{W}_{(12)}^{\mu_3\,\mu_2} \mathcal{W}_{(12)}^{\mu_4\mu_5} \mathcal{W}_{(34)\,\mu_4\mu_1} \mathcal{W}_{(34)\,\mu_5\mu_3}\Bigg]~.
\end{split}	
\end{equation}
The last two tensor structures appear only for four graviton amplitude. 
\begin{equation}
 \widehat{\mathcal T}_8 = (\mathcal{X}_{(1;2)} \cdot k_{34})  
 \qquad\text{and}\qquad  
 \mathcal T_{9} = (\mathcal{X}_{(1;2)}\cdot \mathcal{X}_{(3;4)})
 ~.
\end{equation}
These tensor structures appear for \texttt{psss}, \texttt{psps}, \texttt{gsss} and \texttt{gsgs}. 
Note that, 
\begin{enumerate}
	\item this choice of basis is NOT crossing symmetric invariant. 

	\item Tensor structures with $\widehat{}$ are note invariant under $(1,2)\longleftrightarrow (3,4)$ exchange. We define tensor sturctures with $\bar{}$ through $(1,2)\longleftrightarrow (3,4)$ exchange 
\begin{equation}
	(1,2)\longleftrightarrow (3,4)\quad \implies \quad \widehat{\mathcal{T}}\longleftrightarrow\overline{\mathcal{T}}~.
\end{equation}
\end{enumerate}

\section{The derivative method} 
\label{sec:ramderivativemethod}

 We have defined the numerator of a spinning amplitude due to spin $J$ exchange in eqn \eqref{hspinexchange14} and by comparing \eqref{hspinexchange13} and \eqref{hspinexchange14} we get
\begin{equation}
\texttt{A}^\texttt{(i|j)}_{\texttt{xyzw}}=	\Big(M_{\texttt{xyJ}}^\texttt{(i)}\Big)^{\mu_1\cdots\mu_J} ~\mathcal{P}^{(J)}_{\mu_1\cdots\mu_J,\nu_1\cdots\nu_J}~\Big(M_{\texttt{Jzw}}^\texttt{(j)}\Big)^{\nu_1\cdots\nu_J} ~.
\label{ramderivativemethod1}
\end{equation}
The meaning of this notation is explained below eqn \eqref{hspinexchange14}. Our modus operandi is the following observation
\begin{equation}
\texttt{A}^\texttt{(i|j)}_{\texttt{xyzw}}= 	\big(\mathcal{O}^\texttt{(i|j)}\big)^\texttt{ssss}_\texttt{xyzw}~~~ \texttt{A}_{\texttt{ssss}} \label{ramderivativemethod2}
\end{equation}
where $\big(\mathcal{O}^\texttt{(i|j)}\big)^\texttt{ssss}_\texttt{xyzw}$ are differential/multiplicative operators, which can raise the spin of any leg from a scalar to arbitrary massless or massive spin. Authors of \cite{Costa:2011dw} have used similar strategy to compute spinning CFT correlators. In this formulation, such operators can change the spin of the external leg only. It will be even more useful if one can find an operator that can change the spin and the mass of the exchange particle. In this article we focus only on the amplitudes with massless legs and their corresponding such operators. From the expression of the three point functions, it is possible to find the expression for these operators. In order to show this, we start from three point function of two scalars and one spin $J$,
\begin{equation}
\mathcal{A}^{\texttt{(0)}}_{\texttt{ssJ}}=	g(\epsilon_3 \odot k_{12})^J= g (\epsilon_3 \cdot k_{12})^J~.
\label{ramderivativemethod3}
\end{equation}
The three point function two photons and one spin $J$ has two structures: minimal coupling and non-minimal coupling ; see eqn \eqref{ramsetup32}. We rewrite the minimal coupling in the following way 
\begin{equation}
\Big[\mathcal{W}_{(12)}^{\mu \nu }(\epsilon_3)_{\mu }(\epsilon_3)_{\nu }\Big] (\epsilon_3 \cdot k_{12})^{J-2}= \Big[\mathcal{W}_{(12)}^{\tilde \mu \tilde \nu }{\Theta_{\tilde \mu }}^\mu {\Theta_{\tilde \nu }}^\nu (\epsilon_3)_{\mu }(\epsilon_3)_{\nu }\Big](\epsilon_3 \odot k_{12})^{J-2}~.
\label{ramderivativemethod4}
\end{equation}
Note that, we have converted back the $\eta$ dot products to  $\Theta$ dot products (see \eqref{hspinexchange12} for the definitions). Let us define the following differential operator
\begin{equation}
	\big(\mathcal{O}^{\texttt{(0)}}\big)^\texttt{ssJ}_\texttt{ppJ} =\frac{1}{J(J-1)} \mathcal{W}_{(12)}^{\tilde \mu \tilde \nu }\Theta_{\tilde \mu \mu} \Theta_{\tilde \nu \nu} \frac{\partial }{\partial (k_{12})_\mu} \frac{\partial }{\partial (k_{12})_\nu}~.
\label{ramderivativemethod5}
\end{equation}
Comparing the expressions in eqn \eqref{ramderivativemethod3} and in eqn \eqref{ramderivativemethod4}, we can verify that the spin 1-spin 1- spin $J$ three point function can written in terms of the action of the derivative operator \eqref{ramderivativemethod5} acting on scalar-scalar-spin $J$ three point function 
\begin{equation}
\mathcal{A}^{\texttt{(0)}}_{\texttt{ppJ}}=	\big(\mathcal{O}^{\texttt{(0)}}\big)^\texttt{ssJ}_\texttt{ppJ}~~ \mathcal{A}^{\texttt{(0)}}_{\texttt{ssJ}}~.
\label{ramderivativemethod11}
\end{equation}
Given this relation between three point functions, we can also relate the tree-level four point functions. For example, the numerator of four photon amplitude can be obtained in the following way 
\begin{align} 
\mathcal{A}^{\texttt{(i|j)}}_{\texttt{pppp}}= \big(\mathcal{O}^{\texttt{(j)}}\big)^\texttt{Jss}_\texttt{Jpp}~~ \mathcal{A}^{\texttt{(i|0)}}_{\texttt{ppss}} \qquad,\qquad
\mathcal{A}^{\texttt{(i|0)}}_{\texttt{ppss}}= \big(\mathcal{O}^{\texttt{(i)}}\big)^\texttt{ssJ}_\texttt{ppJ}~~ \mathcal{A}_{\texttt{ssss}}
\label{ramderivativemethod12}
%
\end{align} 
Similarly one can construct operators for non-minimal coupling between two photons and one spin $J$. In this case, the operator is simply a multiplicative operator
\begin{equation}
	 \big(\mathcal{O}^{\texttt{(1)}}\big)^\texttt{ssJ}_\texttt{ppJ} = \eta_{\mu\nu }\, \mathcal{W}_{(12)}^{\mu \nu }~.
\label{ramderivativemethod14}
\end{equation}
The greatest advantage of this formula is that this equation along with eqn \eqref{gegenpoly51} allows computing closed-form analytical results. One important aspect of this formulation is that, various dot products should be written with respect to $\Theta$. The conceptual reason behind this is that the propagator in \eqref{hspinexchange2} is defined with respect to $\Theta$ and as a result, if we evaluate the expression in \eqref{ramderivativemethod1} in the brute force way then various quantities will naturally come as theta dot product. For two arbitrary vectors, these two dot products do not necessarily agree. For example, consider the following two examples
\begin{equation}
(k_{12}\odot k_{12})=	(k_{12}\cdot k_{12})
\qquad\text{and}\qquad 
\Big[\mathcal{W}^{(12)}\Big]^{\mu\nu}\Theta_{\mu \nu }	=\frac{3}{4}\Big[\mathcal{W}^{(12)}\Big]^{\mu\nu}\eta_{\mu \nu }~.
\label{ramderivativemethod15}
\end{equation} 
From this we can clearly see that 
\begin{equation}
	\mathcal{W}_{(12)}^{\tilde \mu \tilde \nu }\Theta_{\tilde \mu \mu} \Theta_{\tilde \nu \nu} \frac{\partial }{\partial (k_{12})_\mu} \frac{\partial }{\partial (k_{12})_\nu}
(k_{12}\odot k_{12}) 
\ne 
\mathcal{W}_{(12)}^{ \mu \nu } \frac{\partial }{\partial (k_{12})_\mu} \frac{\partial }{\partial (k_{12})_\nu}	
(k_{12}\cdot k_{12})~.
\label{ramderivativemethod21}
\end{equation}
So, to implement the derivative method, we always write all the dot products in terms $\Theta$, and then we implement the derivative method.

Conversion from $\Theta$ dot product to $\eta$ product can be done only at the end; simplification at an intermediate step can lead to a wrong result. However, from \eqref{ramderivativemethod21} we can see that this $\Theta$ to $\eta$ can give rise to the wrong result only if the dot product involves a differential variable. For all other variables, we can convert $\Theta$ to $\eta$ at any step. So we implement the following strategy; for all the variables that don't appear in the differential operator, we convert $\Theta$ to $\eta$ at the beginning. In general, the differential operator involves derivatives with respect to many variables. Once all the derivatives of a particular variable are performed, we can convert the theta dot product to an eta product for that variable.

In the following section, we compute various tree level scattering amplitudes using the derivative method. We divide the amplitudes in two classes
\begin{enumerate}
	\item {\bf Class I:} When both the particles meeting at the vertex are the same
\begin{equation}
\textrm{(a) \texttt{ppss}}	
\qquad,\qquad 
\textrm{(b) \texttt{pppp}}	 
\qquad,\qquad  
\textrm{(c) \texttt{ggss}}	
\qquad,\qquad 
\textrm{(d) \texttt{ggpp}}	
\qquad,\qquad 
\textrm{(e) \texttt{gggg}}	
\label{classiexamples}
\end{equation}

\vspace{20pt}
	\item {\bf Class II:} When both the particles meeting at atleast one of the vertices are different 
\begin{equation}
\textrm{(a) \texttt{psss}}	
\qquad,\qquad 
\textrm{(b) \texttt{psps}}	
\qquad,\qquad 
\textrm{(c) \texttt{gsss}}	
\qquad,\qquad 
\textrm{(d) \texttt{gsgs}}	
\label{classiiexamples}
\end{equation}

\end{enumerate}  
For the above computations we have used Mathematica \cite{Mathematica}, especially two packages, the “xTensor” a package for abstract tensor algebra from the “xAct” bundle \cite{xAct} and the FeynCalc package \cite{Mertig:1990an,Shtabovenko:2016sxi,Shtabovenko:2020gxv}. We have used the xTensor package to apply the derivative operators on the corresponding tensors and the FeynCalc package to compute the angular distributions.

\section{Massless amplitudes: class I }
\label{sec:ramclassIamplitudes}

In this section, we evaluate class I amplitudes. For this class of amplitudes, the two particles that meet at a vertex are the same. A few examples are given in eqn \eqref{classiexamples}.
We mostly focus on computing the contribution from the minimal couplings, i.e. $\texttt{A}^\texttt{(0|0)}_{\texttt{xyzw}}$s. These are enough to determine the contributions due to the non-minimal couplings. We discuss how to compute the contributions due to the non-minimal coupling from the minimal couplings. 
To write down the amplitudes, we use various short-hand notations. We have summarised them in the appendix \ref{app:ramnotation}. 

\subsection{Photon amplitudes}

\subsubsection{\texttt{ppss}}
\label{subsec:ppssamplitude}

Let's start with the simplest case: two photons meeting at a vertex to give a spin $J$. In the other vertex, two scalars meet\footnote{In fact, this would be the only non-zero channel if the scalars have some equal and opposite charge.}. The corresponding Feynman diagram is being depicted in fig. \ref{fig:ppssfeynmandiagram}. 

\begin{figure}[h]
\begin{center}
\begin{tikzpicture}[line width=1.5 pt, scale=1.3]
 
\begin{scope}[shift={(0,0)}]	
		
\treefourpointexchange{photon}{photon}{realscalar}{realscalar}{spinj}{0}	

\labeltreefourpointexchange{ $k_1$ }{ $k_2$ }{ $k_3$ }{ $k_4$ }{}{0} 
\end{scope}

\end{tikzpicture} 
\end{center}
\caption{ $\texttt{ppss}$ amplitude due to exchange of a massive spinning particle }
\label{fig:ppssfeynmandiagram}
\end{figure}

The three-point function of two photons and one higher spin particle has two structures: we denoted them as minimal and non-minimal couplings, based on the number of momentum factors. The expressions for them are given in \eqref{ramsetup32}. The minimal coupling is non-zero only for $J\geq 2$.
The corresponding derivative operators can be found in \eqref{ramderivativemethod5}. The operator for minimal coupling $\big(\mathcal{O}^{\texttt{(0)}}\big)^\texttt{ssJ}_\texttt{ppJ}$ is a differential operator, but the operator for non-minimal coupling $\big(\mathcal{O}^{\texttt{(1)}}\big)^\texttt{ssJ}_\texttt{ppJ}$ is simply a multiplicative operator. The \texttt{ppss} four point function due to the minimal coupling is given by 
\begin{equation}
\texttt{A}_\texttt{ppss}^\texttt{(0|0)}= 	\big(\mathcal{O}^{\texttt{(0)}}\big)^\texttt{ssJ}_\texttt{ppJ}~~ \texttt{A}_\texttt{ssss} 
\label{ppsshighspinex4}~.
\end{equation} 
After the action of the derivative operators we can write in the form of \eqref{ramdecompositionI}
\begin{align}
\texttt{A}_\texttt{ppss}^\texttt{(0|0)}=	\mathcal{T}_{1(\textrm{\texttt{ppss}})}\, \mathcal{F}_{1(\textrm{\texttt{ppss}})} (s,t)+\mathcal{T}_{2(\textrm{\texttt{ppss}})}\, \mathcal{F}_{2(\textrm{\texttt{ppss}})} (s,t)
\end{align}
where $\mathcal{T}_{I(\textrm{\texttt{ppss}})}$ are the tensor factors and $\mathcal{F}_{I(\textrm{\texttt{ppss}})}$s are the form factors. In terms of tensor structures written in section \ref{subsec:basisfortenstr}, the \texttt{ppss} tensor structures are given by 
\begin{align}
\mathcal{T}_{1(\textrm{\texttt{ppss}})}= \widehat{\mathcal{T}}_{1} 
\qquad\text{and}\qquad	
\mathcal{T}_{2(\textrm{\texttt{ppss}})}=\widehat{\mathcal{T}}_{2}
\end{align} 
where the corresponding form factors are given by 
\begin{align} \mathcal F_{1\text{(\texttt{ppss})}} 
& = ~\frac{1}{4}\mathcal{N}_{J,D} (y_{s})^{J-2} \left[-3 \
\frac{d\gegen^{(\beta)}_{J-1}(z_{s})}{dz_{s}}+\frac{d^2\gegen^{(\beta)}_{J-2}(z_{s})}{dz_{s}^2}-2 z_{s} \
\frac{d^2\gegen^{(\beta)}_{J-1}(z_{s})}{dz_{s}^2}\right] \quad\text{and}
\\ 
\mathcal F_{2\text{(\texttt{ppss})}} 
 &= \mathcal{N}_{J,D}(y_{s})^{J-2} \frac{d^2\gegen^{(\beta)}_{J}(z_{s})}{dz_{s}^2}~.
 \end{align}
 In $3+1$ dimensions, $\mathcal F_{1\text{(\texttt{ppss})}}$ simplifies 
\begin{equation}
\, \mathcal{F}_{1(\textrm{\texttt{ppss}})} (s,t)=-\frac{1}{4}\mathcal{N}_{J,4}\,(y_{s})^{J-2}\frac{d^2}{dz_s^2}\legendre_J(z_s)=-\frac{1}{4}\mathcal{F}_{2(\textrm{\texttt{ppss}})} 	(s,t)~.
\label{ppsshighspinex11}
\end{equation}
Please note that this simplification does not happen in $D\geq 5$.

\paragraph{Angular distribution in 3+1 dimensions} Let's consider the amplitude for massless scalars, in the center of mass frame (defined in sec \ref{subsec:comconvention}) for the particular case of $3+1$ dimensions. In this frame, we evaluated the tensor structures; they are given in table \ref{tab:ppssangdist}. In this case, the amplitude reduces to 
\begin{equation}
	\mathcal{N}_{J,4}\, (s)^{J-2} \Bigg[	\Big(\mathcal{W}^{(12)}_{\mu\nu}(k_{34})^\mu(k_{34})^\nu\Big)-\frac{1}{4}\mathcal{W}^{(12)} (k_{34})^2\Bigg]\frac{d^2}{dz_s^2}\legendre_J(z_s)~.
\label{ppsshighspinex12}
\end{equation}
We have summarized the angular distributions for the tensor factors in table \ref{tab:ppssangdist}. We have explained the notation ($2^\pm\longrightarrow 0$ and $0\longrightarrow 0$) below equation (\ref{ram_equation_to_refer_later}). From this we can determine
\begin{equation}
 \Bigg[	\Big(\mathcal{W}^{(12)}_{\mu\nu}(k_{34})^\mu(k_{34})^\nu\Big)-\frac{1}{4}\mathcal{W}^{(12)} (k_{34})^2\Bigg]
 =	-\frac{1}{4}\delta^{h_1+h_2,2}\, s^2 \sin^2\theta\,  
 =\frac{s^2}{4}\delta^{h_1+h_2,2}(z_s^2-1)
\label{ppsshighspinex14}
\end{equation}
where, $h_1$ and $h_2$ are the helicities of the photons.
If we put this in eqn \eqref{ppsshighspinex12}, we obtain 
\begin{equation}
	\mathcal{N}_{J,4}\, \frac{s^2}{4}(z_s^2-1)\frac{\partial^2 \legendre_J(z_s)}{\partial z_s^2}~.
\label{ppsshighspinex15}
\end{equation}
Using the identity in eqn \eqref{eq:wignertolegendre}, for $h=2$ we get 
\begin{equation}
\mathcal{N}_{J,4}\,	\frac{s^J}{4}\sqrt{(J+2)(J+1)J(J-1)}\, d^{J}_{20}(\theta)~.
\label{ppsshighspinex21}
\end{equation}
The angular distribution of four point function of two scalar and two photon has the following form 
\begin{equation}
\begin{split}
\widetilde{\mathcal{N}}_{J;2,0} (\delta^{h_1+h_2,2})\frac{1}{s-m_J^2+\iimg \varepsilon } s^{J} \, d^{(J)}_{20}(\theta)  
\end{split}	
\label{ppsshighspinex22}
\end{equation}
where $\widetilde{\mathcal{N}}_{J;2,0}$ is given in eqn \eqref{angdist3}. This completes the verification of our answer for \texttt{ppss}.

\begin{table}[h]
 \begin{center}
\begin{tabular}{ ||p{4cm}||p{2.5cm}|p{2.5cm}|| }
   \hline
   \hline
   \multicolumn{3}{|c|}{{\bf Angular distributions for \texttt{ppss} tensor structures} } \\
   \hline
   \hline
   Tensor structures & $2^\pm\longrightarrow 0$ & $0\longrightarrow 0$ \\
   \hline
    $ \mathcal T_{1\text{(\texttt{ppss})}}$ & $0$ & $-s$ \\
   \hline
    $ \mathcal T_{2\text{(\texttt{ppss})}} $ & $-\frac{s^2}{4}\sin^2\theta$ & $-\frac{s^2}{4}$\\
   \hline
   \hline
  \end{tabular}
  \caption{Angular distributions for \texttt{ppss} tensor structures}
   \label{tab:ppssangdist} 
 \end{center}
\end{table}

Upto this point, we have done the amplitude, only for the minimal coupling. Let's now the consider the contribution due to the non-minimal coupling. The amplitude is simpler to evaluate in this case. It is given by 
\begin{align}
\texttt{A}_{\text{\texttt{ppss}}}^{\texttt{(1|0)}} &= \Big[\Tr\, \mathcal{W}_{(12)}\Big](k_{12})^{\mu_1}\cdots (k_{12})^{\mu_J} ~\mathcal{P}^{(J)}_{\mu_1\cdots\mu_J;\nu_1\cdots\nu_J}~(k_{34})^{\nu_1}\cdots (k_{34})^{\nu_J}
\nonumber\\
&= \Big[\Tr\, \mathcal{W}_{(12)}\Big] \texttt{A}_{\text{\texttt{ssss}}}
= \Big[\Tr\, \mathcal{W}_{(12)}\Big] \mathcal{N}_{J,D} (y_{s})^{J} \gegen^{(\beta)}_{J}(z_{s})~.
\end{align}

\subsubsection{\texttt{pppp} : The four photon amplitude }
\label{subsec:ppppamplitude}

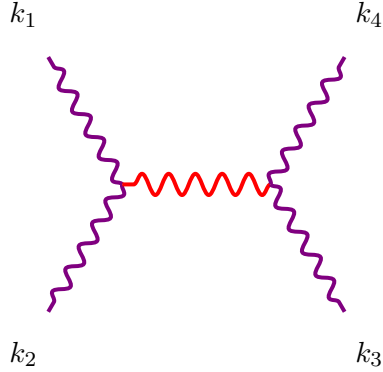
\begin{figure}[h]
\begin{center}
\begin{tikzpicture}[line width=1.5 pt, scale=1.3]
 
\begin{scope}[shift={(0,0)}]	
		
\treefourpointexchange{photon}{photon}{photon}{photon}{spinj}{0}	

\labeltreefourpointexchange{ $k_1$ }{ $k_2$ }{ $k_3$ }{ $k_4$ }{}{0} 
\end{scope}

\end{tikzpicture} 
\end{center}
\caption{Four photon amplitude due to exchange of a massive spinning particle }
\label{fig:fourphotonhspin}
\end{figure}

Now we want to derive the expression for tree-level four-photon amplitude due to higher spin exchange. The corresponding Feynman diagram can be found in fig. \ref{fig:fourphotonhspin}. The photon-photon spin $J$ three-point function has two different structures; however, it is easy to deal with non-minimal coupling. We considered that towards the end of this subsection. Currently we restrict the computation when we use minimal coupling at both the vertices.
In this case the four photon amplitude is given by 
\begin{equation}
\begin{split}
\texttt{A}_{\text{\texttt{pppp}}}^{\texttt{(0|0)}}
	&= \big(\mathcal{O}^{\texttt{(0)}}\big)^\texttt{Jss}_\texttt{Jpp}~~ \texttt{A}_{\text{\texttt{ppss}}}^{\texttt{(0|0)}}
= \big(\mathcal{O}^{\texttt{(0)}}\big)^\texttt{Jss}_\texttt{Jpp}~~\big(\mathcal{O}^{\texttt{(0)}}\big)^\texttt{ssJ}_\texttt{ppJ}~~ \texttt{A}_{\text{\texttt{ssss}}} ~.
\end{split}	
\label{pppphighspinex1}
\end{equation} 
The final answer for the four photon amplitude is given by 
\begin{equation}
\texttt{A}_{\text{\texttt{pppp}}}^{\texttt{(0|0)}} = \sum_{I=1}^5 \mathcal{T}_{I\textrm{(\texttt{pppp})}}\, \mathcal{F}_{I\textrm{(\texttt{pppp})}}(s,t)~.
\label{pppphighspinex2}
\end{equation}
$\mathcal{T}_{I\textrm{(\texttt{pppp})}}$s are the tensor factors. We write them in terms of the basis given in section \ref{subsec:basisfortenstr}. The reducible tensor structures\footnote{A tensor structure is reducible if it can be written as product of more than one Lorentz invariant tensor structures.} are
 \begin{align}
 \mathcal T_{1\text{(\texttt{\texttt{pppp}})}} = \widehat{\mathcal{T}}_{1}\, 	\overline{\mathcal{T}}_{1}\qquad,\qquad 
 \mathcal T_{2\text{(\texttt{\texttt{pppp}})}} = \widehat{\mathcal{T}}_{1} \overline{\mathcal{T}}_{2}
+\widehat{\mathcal{T}}_{2} \overline{\mathcal{T}}_{1}\qquad\text{and}\qquad 
 \mathcal T_{3\text{(\texttt{\texttt{pppp}})}} = \widehat{\mathcal{T}}_{2}\, 	\overline{\mathcal{T}}_{2}~.
 \end{align}
And there are two irreducible tensor structures 
 \begin{align}
 \mathcal T_{4\text{(\texttt{\texttt{pppp}})}} = \mathcal{T}_{3}\qquad\text{and}\qquad 
 \mathcal T_{5\text{(\texttt{\texttt{pppp}})}} = \mathcal{T}_{4}~.
 \end{align} 
Tensor factors $\mathcal{T}_{1\textrm{(\texttt{pppp})}} $ and $\mathcal{T}_{4\textrm{(\texttt{pppp})}} $ appear for all spin $J\geq 2$; $\mathcal{T}_{2\textrm{(\texttt{pppp})}} $ and $\mathcal{T}_{5\textrm{(\texttt{pppp})}} $ are there for $J\geq 3$. $\mathcal{T}_{3\textrm{(\texttt{pppp})}}$ appear only for spin $4$ or more. The corresponding form factors $\mathcal{F}_{I\textrm{(\texttt{pppp})}}(s,t)$ are given by 
\begin{align}
\mathcal{F}_{1\textrm{(\texttt{pppp})}}  
&= \frac{\mathcal{N}_{J,D}\,}{16} (y_{s})^{J-4}\left[-2  \
\frac{d^2\gegen^{(\beta)}_{J}(z_{s})}{dz_{s}^2} -18 \
\frac{d\gegen^{(\beta)}_{J-1}(z_{s})}{dz_{s}}+41 \frac{d^2\gegen^{(\beta)}_{J-2}(z_{s})}{dz_{s}^2}-14 \
\frac{d^3\gegen^{(\beta)}_{J-3}(z_{s})}{dz_{s}^3} \right.\nonumber\\
&~~~~~~~~~~\left.+\frac{d^4\gegen^{(\beta)}_{J-4}(z_{s})}{dz_{s}^4}-4 z_{s} \
\left(8 \frac{d^2\gegen^{(\beta)}_{J-1}(z_{s})}{dz_{s}^2}-8 \
\frac{d^3\gegen^{(\beta)}_{J-2}(z_{s})}{dz_{s}^3}+2 z_{s} \
\frac{d^3\gegen^{(\beta)}_{J-1}(z_{s})}{dz_{s}^3} \right.\right.\nonumber\\
&~~~~~~~~~~\left.\left.+\frac{d^4\gegen^{(\beta)}_{J-3}(z_{s})}{dz_{s}^4}-z_{s} \
\frac{d^4\gegen^{(\beta)}_{J-2}(z_{s})}{dz_{s}^4}\right)\right]
\quad ,\\
\mathcal{F}_{2\textrm{(\texttt{pppp})}}  
&=\frac{1}{4}\mathcal{N}_{J,D}\, (y_{s})^{J-4} \left[-7 \
\frac{d^3\gegen^{(\beta)}_{J-1}(z_{s})}{dz_{s}^3}+\frac{d^4\gegen^{(\beta)}_{J-2}(z_{s})}{dz_{s}^4}-2 z_{s} \
\frac{d^4\gegen^{(\beta)}_{J-1}(z_{s})}{dz_{s}^4}\right]
\quad ,\\ 
\mathcal{F}_{3\textrm{(\texttt{pppp})}}  
&=\mathcal{N}_{J,D}\,(y_{s})^{J-4} \frac{d^4\gegen^{(\beta)}_{J}(z_{s})}{dz_{s}^4}
\quad ,\\ 
\mathcal{F}_{4\textrm{(\texttt{pppp})}} &
=\mathcal{N}_{J,D}\,(y_{s})^{J-2} \frac{d^2\gegen^{(\beta)}_{J}(z_{s})}{dz_{s}^2}
\quad \text{and}\\ 
\mathcal{F}_{5\textrm{(\texttt{pppp})}} 
& = \mathcal{N}_{J,D}\, (y_{s})^{J-3} \frac{d^3\gegen^{(\beta)}_{J}(z_{s})}{dz_{s}^3}~.
\label{pppphighspinex4}
\end{align}
Note that the expression for $\mathcal{F}_{1\textrm{(\texttt{pppp})}}$ and $\mathcal{F}_{2\textrm{(\texttt{pppp})}}$ is very messy. The expressions simplify a lot for $D=4$. In this case, 
\begin{align}
\mathcal{F}_{1\textrm{(\texttt{pppp})}}&=\frac{1}{16}\mathcal{N}_{J,4}\, (y_{s})^{J-4}\Bigg[\frac{d^4\legendre_{J}(z_{s})}{dz_{s}^4}- 8\frac{d^2\legendre_{J}(z_{s})}{dz_{s}^2}-4z_s\frac{d^3\legendre_{J}(z_{s})}{dz_{s}^3}
\Bigg]
\quad\text{and}\\
	\mathcal{F}_{2\textrm{(\texttt{pppp})}}&= -\frac{\mathcal{N}_{J,4}}{4}\,(y_{s})^{J-4} \frac{d^4\legendre_{J}(z_{s})}{dz_{s}^4}~.
\end{align} 
Just like in the case of \texttt{ppss}, this simplification happens only in $D=4$.

\paragraph{Angular distribution in 3+1 dimensions} The angular distribution for various \texttt{pppp} tensor structure is given in the table \ref{tab:ppppangdist}. These expression are for center of mass frame which is defined in sec \ref{subsec:comconvention}. For the highest helicity exchange $\mathcal T_{1\textrm{(\texttt{pppp})}}=0 = \mathcal T_{3\textrm{(\texttt{pppp})}} $ and the rest are given by
\begin{align}
 \mathcal T_{2\textrm{(\texttt{pppp})}} &= \frac{s^2}{2}\cos^4\bigg(\frac\theta2\bigg)=\frac{s^2}{8}(z_s+1)^2
 \qquad,\qquad
 \mathcal T_{5\textrm{(\texttt{pppp})}} = \frac{s^4}{16}\sin^4\theta=\frac{s^4}{16}(z_s^2-1)^2 \quad\text{and} \\ 
 \mathcal T_{4\textrm{(\texttt{pppp})}} &= -\frac{s^3}{4}(\sin^2\theta) (\cos\theta+1)=\frac{s^3}{4}(z_s^2-1)(z_s+1)~.
\end{align}
Then the amplitude $\texttt{A}_{\text{\texttt{pppp}}}^{2^+\rightarrow 2^+}$ is 
\begin{align}
 = \mathcal{N}_{J,4}\,s^J\, \left[\frac{1}{J(J-1)}\right]^2 \frac{(1+z_s)^2}{16}\Bigg[\frac{(1-z_s)^2}{4}~ \frac{\partial^4 \legendre_J(z_s)}{\partial^4 z_s} - (1-z_s)~ \frac{\partial^3 \legendre_J(z_s)}{\partial z_s^3} + \frac{1}{2}~ \frac{\partial^2 \legendre_J(z_s)}{\partial z_s^2} \Bigg] ~.
 \end{align}
We use the identity \eqref{mathematicalidentityII11} to write it as 
\begin{equation}
\mathcal{N}_{J,4}\, s^J\, \frac{(J+2)(J+1)}{16\, J(J-1)}	d_{22}^{(J)}(\theta) = \Bigg[  \Big( \widetilde{\mathcal{N}}_{J;2,2}\Big)\, s^{J}\, d_{22}^{(J)}(\theta) \Bigg] 
\end{equation}
where $\widetilde{\mathcal{N}}_{J;2,2} $ is given in eqn \eqref{angdist3}. The other possibilities include $2^\pm\rightarrow 2^\mp$. This is simply exchange of $3\leftrightarrow 4$ which is nothing but $\theta \rightarrow \pi+\theta $. 
\begin{align}
\texttt{A}_{\text{\texttt{pppp}}}^{2^+\rightarrow 2^-} &= {(-1)^J}\Big(\texttt{A}_{\text{\texttt{pppp}}}^{2^+\rightarrow 2^+} \Big)_{z_{s}\rightarrow -z_{s}}\\
&=~ {(-1)^J}\Bigg[  \Big( \widetilde{\mathcal{N}}_{J;2,2}\Big)\, s^{J}\, d_{22}^{(J)}(\pi+\theta) \Bigg] =\Bigg[  \Big( \widetilde{\mathcal{N}}_{J;2,2}\Big)\, s^{J}\, d_{2-2}^{(J)}(\theta) \Bigg] ~.
\end{align}
Let's now consider the case $2^\pm \rightarrow 0$ (also $0\rightarrow 2^\pm $). In this case, only $\mathcal{T}_{2\text{(\texttt{pppp})}}$ and $\mathcal{T}_{3\text{(\texttt{pppp})}}$ is non-zero. And if we put the explicit expression for the form factors, we can check that 
\begin{align}
\texttt{A}_{(\text{\texttt{pppp}})}^{2^\pm \rightarrow 0} =0= \texttt{A}_{(\text{\texttt{pppp}})}^{0\rightarrow 2^\pm } ~.
\end{align}
The only remaining case is $0\rightarrow 0$. If we put all the expressions, we can verify that the amplitude vanishes in this case $\texttt{A}_{(\text{\texttt{pppp}})}^{0 \rightarrow 0} =0$.

\begin{table}[h!]
 \begin{center}

  \begin{tabular}{ ||p{3.5 cm}||p{3cm}|p{3cm}|p{2.5cm}|p{2.5cm}|| }
   \hline
   \hline
   \multicolumn{5}{|c|}{{\bf Angular distributions for \texttt{pppp} tensor structures} } \\
   \hline
   \hline
   Tensor structures & $2^\pm\rightarrow 2^\pm$ &$2^\pm\rightarrow 2^\mp $ & $2^\pm \rightarrow 0$ & $0\rightarrow 0$ \\
   \hline

   $ \mathcal{T}_{1\text{(\texttt{pppp})}}$ & $0$ & 0&$0$ & $s^2$ \\
   \hline
   $ \mathcal{T}_{2\text{(\texttt{pppp})}}$ & $0$ &0 & -$\frac{s^3}{4} \sin^2\theta$ & $\frac{s^3}{2}$ \\
   \hline
   $ \mathcal{T}_{3\text{(\texttt{pppp})}}$ & $\frac{s^4}{16} \sin^4\theta$ &$\frac{s^4}{16} \sin^4\theta$ &  -$\frac{s^4}{16} \sin^2\theta$ & $\frac{s^4}{16}$\\
   \hline
   $ \mathcal{T}_{4\text{(\texttt{pppp})}}$ & $\frac{s^2}{2} \cos^4(\frac\theta2)$ & $\frac{s^2}{2} \sin^4(\frac\theta2)$ & $0$ & $\frac{s^2}{2}$ \\
   \hline
   $ \mathcal{T}_{5\text{(\texttt{pppp})}}$ & $-\frac{s^3}{2} \sin^2\theta \cos^2(\frac\theta2)$ & $\frac{s^3}{2} \sin^2\theta \sin^2(\frac\theta2)$ & $0$ & $\frac{s^3}{4}\cos\theta$ \\
   \hline
    \hline
  \end{tabular}
  \caption{Angular distributions for the \texttt{pppp} tensor structures}
  \label{tab:ppppangdist} 
 \end{center}
\end{table}

\paragraph{Amplitude due to non-minimal coupling} Upto this point we focused only on $\texttt{A}_{\text{\texttt{\texttt{pppp}}}}^{\texttt{(0|0)}}$; this is the contribution when we use both the minimal coupling at both vertices. Since photon-photon-spin $J$ non-minimal coupling is non-zero, there are three more contributions: $\texttt{A}_{\text{\texttt{\texttt{pppp}}}}^{\texttt{(1|0)}}$, $\texttt{A}_{\text{\texttt{\texttt{pppp}}}}^{\texttt{(0|1)}}$ and $\texttt{A}_{\text{\texttt{\texttt{pppp}}}}^{\texttt{(1|1)}}$. 
\begin{align}
\texttt{A}_{\text{\texttt{\texttt{pppp}}}}^{\texttt{(1|0)}} &= \Big[\Tr\, \mathcal{W}_{(12)}\Big](k_{12})^{\mu_1}\cdots (k_{12})^{\mu_J} ~\mathcal{P}^{(J)}_{\mu_1\cdots\mu_J,\nu_1\cdots\nu_J}~(\mathcal{W}_{(34)})^{\nu_1\nu_2}(k_{34})^{\nu_3}\cdots (k_{34})^{\nu_J} 
\\
&= \Big[\Tr\, \mathcal{W}_{(12)}\Big] \Big(\texttt{A}_{\texttt{sspp}}^{(\texttt{0|0})}\Big)\nonumber
~.\\
\texttt{A}_{\text{\texttt{\texttt{pppp}}}}^{\texttt{(0|1)}} &= (\mathcal{W}_{(12)})^{\mu_1\mu_2}(k_{12})^{\mu_3}\cdots (k_{12})^{\mu_J} ~\mathcal{P}^{(J)}_{\mu_1\cdots\mu_J,\nu_1\cdots\nu_J}~ (k_{34})^{\nu_1}\cdots (k_{34})^{\nu_J} \Big[\Tr\mathcal{W}_{(34)}\Big]
\\
&= \Big[\Tr\,\mathcal{W}_{(34)}\Big] \Big(\texttt{A}_{\texttt{ppss}}^{(\texttt{0|0})}\Big)\nonumber
~.\\
\texttt{A}_{\text{\texttt{\texttt{pppp}}}}^{\texttt{(1|1)}} &= \Big[\Tr\, \mathcal{W}_{(12)}\, \Tr\, \mathcal{W}_{(34)}\Big](k_{12})^{\mu_1}\cdots (k_{12})^{\mu_J} ~\mathcal{P}^{(J)}_{\mu_1\cdots\mu_J,\nu_1\cdots\nu_J}~(k_{34})^{\nu_1}\cdots (k_{34})^{\nu_J}
\\
&= \Big[\Tr\, \mathcal{W}_{(12)}\, \Tr\, \mathcal{W}_{(34)}\Big] \Big(\texttt{A}_{\texttt{ssss}}\Big)~.\nonumber
\end{align}
We can see that these contributions can be written straightforwardly once we know the \texttt{ppss} numerator due to minimal coupling and \texttt{ssss} numerator.  

\subsection{Graviton amplitudes}

\subsubsection{\texttt{ggss}} 
\label{subsec:ggssamplitude}

Let's now focus on the class I amplitudes involving gravitons. The simplest of them is the \texttt{ggss} amplitude. The graviton-graviton-spin $J$ three point function has three independent structures (see \eqref{ramsetup33}) \cite{Chakraborty:2020rxf} . The expression in terms of double copy can be determined using \eqref{ramsetup31}. We use the expression in terms of double copy since those expressions are convenient to determine the contribution due to non-minimal coupling. Following our general strategy, we focus on the minimal coupling first. In this case, the numerator is given by  
\begin{align}
 \texttt{A}_{\text{\texttt{ggss}}}^{\texttt{(0|0)}} &= (\mathcal{W}_{(12)})^{\mu_1\mu_2}(\mathcal{W}_{(12)})^{\mu_3\mu_4}(k_{12})^{\mu_5}\cdots (k_{12})^{\mu_J} ~\mathcal{P}^{(J)}_{\mu_1\cdots\mu_J\,;\,\nu_1\cdots\nu_J}~(k_{34})^{\nu_1}\cdots (k_{34})^{\nu_J}
 \nonumber \\
 &=\left[\frac{1}{J(J-1)(J-2)(J-3)}\right] \mathcal{W}_{(12)}^{\tilde\mu_1 \tilde\mu_2}\mathcal{W}_{(12)}^{\tilde\mu_3 \tilde\mu_4} \Theta_{\tilde\mu_1 \mu_1} \Theta_{\tilde\mu_2 \mu_2}\Theta_{\tilde\mu_3 \mu_3} \Theta_{\tilde\mu_4 \mu_4}\nonumber\\
 &~~~~~~~~~~~~~~~~~~~~~~~~~~~~~~~~~~~~~~~~~~~~~~~ \frac{\partial}{\partial (k_{12})_{\mu_1}} \frac{\partial}{\partial (k_{12})_{\mu_2}} \frac{\partial}{\partial (k_{12})_{\mu_3}} \frac{\partial}{\partial (k_{12})_{\mu_4}} \Big( \texttt{A}_{\texttt{ssss}}\Big)~.\nonumber\\
 &=\big(\mathcal{O}^{\texttt{(0)}}\big)^\texttt{ssJ}_\texttt{ggJ}~~\texttt{A}_{\texttt{ssss}}
\end{align}
In this case, there are three independent tensor structures and all of the reducible $\Big[\textrm{(\texttt{ppss})}_I\Big]^2$ -
\begin{align}
 \mathcal T_{1\text{(\texttt{ggss})}} = \widehat{\mathcal{T}}_{1}\widehat{\mathcal{T}}_{1}
\qquad,\qquad  
 \mathcal T_{2\text{(\texttt{ggss})}} = \widehat{\mathcal{T}}_{1}\widehat{\mathcal{T}}_{2}
\qquad\text{and}\qquad   
 \mathcal T_{3\text{(\texttt{ggss})}} = \widehat{\mathcal{T}}_{2}\widehat{\mathcal{T}}_{2}~.
\end{align}
The corresponding form factors are given by 
\begin{equation}
\begin{split}
\mathcal F_{1\text{(\texttt{ggss})}}  
&= \mathcal{N}_{J,D}\, \frac{1}{16} (y_{s})^{J-4} \left[15 \
\frac{d^2\gegen^{(\beta)}_{J-2}(z_{s})}{dz_{s}^2}-10 \
\frac{d^3\gegen^{(\beta)}_{J-3}(z_{s})}{dz_{s}^3}+\frac{d^4\gegen^{(\beta)}_{J-4}(z_{s})}{dz_{s}^4}\right.\\
&~~~~~~~~~~\left.+4 z_{s} \
\left(5 \
\frac{d^3\gegen^{(\beta)}_{J-2}(z_{s})}{dz_{s}^3}-\frac{d^4\gegen^{(\beta)}_{J-3}(z_{s})}{dz_{s}^4}+z_{s} \
\frac{d^4\gegen^{(\beta)}_{J-2}(z_{s})}{dz_{s}^4}\right)\right]
\quad ,\\
\mathcal F_{2\text{(\texttt{ggss})}}  
&= \mathcal{N}_{J,D}\, \frac{1}{2} (y_{s})^{J-4} \left[-5 \
\frac{d^3\gegen^{(\beta)}_{J-1}(z_{s})}{dz_{s}^3}+\frac{d^4\gegen^{(\beta)}_{J-2}(z_{s})}{dz_{s}^4}-2 z_{s}\
\frac{d^4\gegen^{(\beta)}_{J-1}(z_{s})}{dz_{s}^4}\right]
\quad\text{and}\\
\mathcal F_{3\text{(\texttt{ggss})}}  
&=	\mathcal{N}_{J,D}\,(y_{s})^{J-4} \frac{d^4\gegen^{(\beta)}_{J}(z_{s})}{dz_{s}^4} ~.
\end{split}
\end{equation}

\paragraph{Angular distribution in 3+1 dimensions} For the case of $3+1$ dimensions and for massless scalars, we want to compute angular distribution. The angular distributions for various tensor structures are given in table \ref{tab:ggssangdist}.
\begin{table}[h!]
 \begin{center}
  \begin{tabular}{ ||p{3.5cm}||p{4cm}|p{4cm}||}
   \hline
   \hline
   \multicolumn{3}{|c|}{{\bf Angular distributions for \texttt{ggss} tensor structures} } \\
   \hline
   \hline
   Tensor structures & $4^\pm\longrightarrow 0$ & $0\longrightarrow 0$\\
   \hline
   $\mathcal T_{1\text{(\texttt{ggss})}}$ & $0$ & $s^2$\\
   \hline
   $\mathcal T_{2\text{(\texttt{ggss})}}$ & $0$& $\frac{s^3}{4}$ \\
   \hline
   $\mathcal T_{3\text{(\texttt{ggss})}}$ & $\frac{s^4}{16}\sin^4\theta$ & $\frac{s^4}{16}$\\
   \hline
   \hline
  \end{tabular}
  \caption{Angular distributions for different helicities for \texttt{ggss} amplitude}
  \label{tab:ggssangdist} 
 \end{center}
\end{table}
 When one of the gravitons (say 1) has positive helicity, and the other graviton has negative helicity, two of the tensor structures ($\mathcal T_{1\text{(\texttt{ggss})}},\mathcal T_{2\text{(\texttt{ggss})}}$) are zero.  The only non-zero one evaluates to 
\begin{align}
 \mathcal T_{3\text{(\texttt{ggss})}} &= \frac{s^4}{16}\sin^4\theta=\frac{s^4}{16}(1-z_s^2)^2 ~.
\end{align}
Then the full amplitude is given by 
\begin{align}
 \texttt{A}_{\text{\texttt{ggss}}}^{4^+\rightarrow 0} = &\mathcal{N}_{J,D}\,\frac{s^J}{16}(1-z_s^2)^2~ \frac{(J-4)!}{J!} \frac{\partial^4\legendre_J(z_s)}{\partial z_s^4}~ =(s)^J\Bigg[ \mathcal{N}_{J,D}\,~ \frac{(J-4)!}{J!}\sqrt{\frac{(J+4)!}{(J-4)!}} \Bigg] d^{(J)}_{40}(\theta)
 \nonumber \\
  =&  (s)^J \Big( \widetilde{\mathcal{N}}_{J;4,0}\Big)~d^{(J)}_{40}(\theta) ~.
\end{align}
Here we used \eqref{eq:wignertolegendre} for $m=4$. This matches with \eqref{angdist2}. One can check that 
\begin{equation}
 \widetilde{\mathcal{N}}_{J;4,0} = 	\Bigg[ \mathcal{N}_{J,D}\,~ \frac{(J-4)!}{J!}\sqrt{\frac{(J+4)!}{(J-4)!}} \Bigg] ~.
\end{equation}
The angular distribution for the choice $-4\rightarrow 0$ also follows in the same way since $\mathcal T_{3\text{(\texttt{ggss})}}$ takes same value. One can also check $d^{(J)}_{40}(\theta)=d^{(J)}_{-40}(\theta)$. So for $-4\rightarrow 0$ also follows from the same analysis.

\paragraph{Amplitudes due to the non-minimal couplings} Graviton-graviton-spin $J$ three point function has three different structures. Let's now compute the amplitudes due to the two non-minimal couplings. First we compute the amplitude due to the non-minimal coupling
\begin{align}
\texttt{A}_{\text{\texttt{ggss}}}^{\texttt{(1|0)}} &= \mathcal{W}_{(12)}\, (\mathcal{W}_{(12)})^{\mu_1\mu_2}(k_{12})^{\mu_3}\cdots (k_{12})^{\mu_J} ~\mathcal{P}^{(J)}_{\mu_1\cdots\mu_J,\nu_1\cdots\nu_J}~(k_{34})^{\nu_1}\cdots (k_{34})^{\nu_J}
\nonumber
\\
&=\left(\frac{1}{J(J-1)}\right) \mathcal{W}_{(12)} (\mathcal{W}_{(12)})^{\tilde\mu_1 \tilde\mu_2} \Theta_{\tilde\mu_1 \mu_1} \Theta_{\tilde\mu_2 \mu_2} \frac{\partial}{\partial (k_{12})_{\mu_1}} \frac{\partial}{\partial (k_{12})_{\mu_2}} \Big(\texttt{A}_{\texttt{ssss}}\Big)
\nonumber
\\
&= \Big[\Tr\, \mathcal{W}_{(12)}\Big] \Big(\texttt{A}_{\texttt{ppss}}^{\texttt{(0|0)}}\Big)~.
\end{align}
Let's now compute the contribution to the non-non-minimal coupling. It is given by 
\begin{align}
\texttt{A}_{\text{\texttt{ggss}}}^{\texttt{(2|0)}} &=\Big[\Tr\, \mathcal{W}_{(12)}\,\Tr\, \mathcal{W}_{(12)}\Big](k_{12})^{\mu_1}\cdots (k_{12})^{\mu_J} ~\mathcal{P}^{(J)}_{\mu_1\cdots\mu_J,\nu_1\cdots\nu_J}~(k_{34})^{\nu_1}\cdots (k_{34})^{\nu_J}
\nonumber
\\
&= \Big[\Tr\, \mathcal{W}_{(12)}\, \Tr\, \mathcal{W}_{(12)}\Big] \Big(\texttt{A}_{\texttt{ssss}}\Big)~.
\end{align}

 \subsubsection{\texttt{gggg} : The four graviton amplitude}
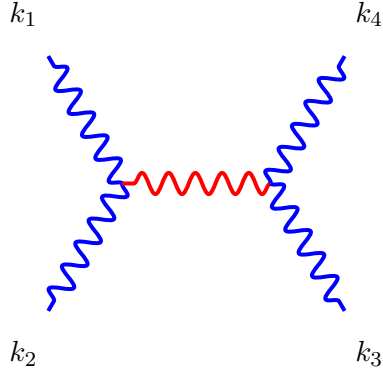
\begin{figure}[h]
\begin{center}
\begin{tikzpicture}[line width=1.5 pt, scale=1.3]
 
\begin{scope}[shift={(0,0)}]	
		
\treefourpointexchange{graviton}{graviton}{graviton}{graviton}{spinj}{0}	

\labeltreefourpointexchange{ $k_1$ }{ $k_2$ }{ $k_3$ }{ $k_4$ }{}{0} 
\end{scope}

\end{tikzpicture} 
\end{center}
\caption{Four graviton amplitude due to exchange of a massive spinning particle }
\label{fig:fourgravitonhspin}
\end{figure}

Now we focus on the most interesting one: the four graviton amplitude due to exchange of a massive $J$ particle. The corresponding Feynman diagram can be found in fig. \ref{fig:fourgravitonhspin}. 
 Before we dive into the spin $J$ expression, we first focus on a simple case. The graviton-graviton-spin $J$ minimal coupling is non-zero for spins $\geq 4$. Let's consider the expression for the four graviton amplitude due to the spin $4$ exchange; it is given by  
\begin{equation}
	\mathcal{M}_s+(1\leftrightarrow 3)+(1\leftrightarrow 4)
\end{equation}
where $(s-M^2+\iimg \varepsilon)\, \mathcal{M}_s$ is given by 
\begin{equation}
\begin{split}
 \frac{\iimg }{672}\Bigg[&+11\,(\mathcal{W}_{(12)})^2(\mathcal{W}_{(34)})^2 
-64\,(\mathcal{W}_{(12)})(\mathcal{W}_{(34)})(\mathcal{W}_{(12)})^{\mu_1\mu_2}\Big[ (\mathcal{W}_{(34)})_{\,\mu_1\mu_2}+ (\mathcal{W}_{(34)})_{\,\mu_2\mu_1}\Big] 
\\
& +56 \Big[ (\mathcal{W}_{(12)})^{\mu_1\mu_2}(\mathcal{W}_{(34)})_{\,\mu_1\mu_2} +(\mathcal{W}_{(12)})^{\mu_1\mu_2}(\mathcal{W}_{(34)})_{\,\mu_2\mu_1} 
\Big]^2
\\
&+ 56(\mathcal{W}_{(12)})^{\mu_1\mu_2}(\mathcal{W}_{(12)})^{\mu_3\mu_4}\Big[(\mathcal{W}_{(34)})_{\,\mu_1\mu_4}(\mathcal{W}_{(34)})_{\,\mu_3\mu_2}
+(\mathcal{W}_{(34)})_{\,\mu_4\mu_1}(\mathcal{W}_{(34)})_{\,\mu_2\mu_3}
\\
 &+
2 (\mathcal{W}_{(34)})_{\,\mu_1\mu_4}(\mathcal{W}_{(34)})_{\,\mu_2\mu_3}+2
(\mathcal{W}_{(34)})_{\,\mu_1\mu_3}(\mathcal{W}_{(34)})_{\,\mu_2\mu_4}
+2
(\mathcal{W}_{(34)})_{\,\mu_1\mu_3}(\mathcal{W}_{(34)})_{\,\mu_4\mu_2}
\Big] \Bigg]~.
\end{split}
\end{equation}
We see that the expression is bulky just for spin 4 itself. The expression for spin $J$ is significantly more lengthy than this.  For a general spin, the four graviton amplitude is given by 
\begin{align}
\mathcal{A}_{\texttt{gggg}} = \sum_{\texttt{i,j}}\,g^\texttt{(i)}_{\texttt{ggJ}} \, g^\texttt{(j)}_{\texttt{Jgg}}\frac{\iimg}{s - m_\text{J}^2} ~\texttt{A}^\texttt{(i|j)}_{\texttt{gggg}}+\textrm{other two channels}~.
\end{align}
First we focus on the one which is most difficult to compute; when both vertices are the minimal coupling. In this case, $\texttt{A}_{\text{\texttt{gggg}}}^{\texttt{(0|0)}}$ for the $s$ channel exchange is given by 
\begin{align}
(\mathcal{W}_{(12)})^{\mu_1\mu_2}(\mathcal{W}_{(12)})^{\mu_3\mu_4}(k_{12})^{\mu_5}\cdots (k_{12})^{\mu_J} ~\mathcal{P}^{(J)}_{\mu_1\cdots\mu_J,\nu_1\cdots\nu_J}~ (\mathcal{W}_{(34)})^{\nu_1\nu_2}(\mathcal{W}_{(34)})^{\nu_3\nu_4}(k_{34})^{\nu_5}\cdots (k_{34})^{\nu_J}~.
 \nonumber  
\end{align}
We can compute it using the derivative method as
\begin{align}
 \texttt{A}_{\text{\texttt{gggg}}}^{\texttt{(0|0)}} =&\left[\frac{1}{J(J-1)(J-2)(J-3)}\right]^2 \mathcal{W}_{(12)}^{\tilde\mu_1 \tilde\mu_2}\mathcal{W}_{(12)}^{\tilde\mu_3 \tilde\mu_4} \mathcal{W}_{(34)}^{\tilde\nu_1 \tilde\nu_2}\mathcal{W}_{(34)}^{\tilde\nu_3 \tilde\nu_4}\nonumber\\
 &~~~~~~~~~~\Theta_{\tilde\mu_1 \mu_1} \Theta_{\tilde\mu_2 \mu_2}\Theta_{\tilde\mu_3 \mu_3} \Theta_{\tilde\mu_4 \mu_4} \Theta_{\tilde\nu_1 \nu_1} \Theta_{\tilde\nu_2 \nu_2}\Theta_{\tilde\nu_3 \nu_3} \Theta_{\tilde\nu_4 \nu_4} \nonumber\\
 &~~~~~~~~~~ \frac{\partial}{\partial (k_{12})_{\mu_1}} \frac{\partial}{\partial (k_{12})_{\mu_2}} \frac{\partial}{\partial (k_{12})_{\mu_3}} \frac{\partial}{\partial (k_{12})_{\mu_4}} \frac{\partial}{\partial (k_{34})_{\nu_1}} \frac{\partial}{\partial (k_{34})_{\nu_2}} \frac{\partial}{\partial (k_{34})_{\nu_3}} \frac{\partial}{\partial (k_{34})_{\nu_4}} \Big(\texttt{A}_{\texttt{ssss}}\Big)~\nonumber\\
 &= \big(\mathcal{O}^{\texttt{(0)}}\big)^\texttt{ssJ}_\texttt{ggJ}~~\big(\mathcal{O}^{\texttt{(0)}}\big)^\texttt{Jss}_\texttt{Jgg}~~ \texttt{A}_{\texttt{ssss}}~.
\end{align}
The amplitude can be written in terms of 19 tensor structures. 17 of these tensor structures are reducible, i.e. they can be product of two Lorentz scalars. 2 of these tensor structures are irreducible. We decompose the 19 structures in 5 different classes depending on their irreducibility. The tensor structures for four graviton amplitude is given by 
\begin{enumerate}
\item $\Big[\textrm{(\texttt{ppss})}_I\otimes\textrm{(\texttt{sspp})}_I\Big]^2$ -  
 \begin{align}
  \mathcal T_{1\text{(\texttt{gggg})}} &=\widehat{\mathcal{T}}_{1}^2\, 	\overline{\mathcal{T}}_{1}^2
 \qquad,\qquad      
 \mathcal T_{2\text{(\texttt{gggg})}} = \widehat{\mathcal{T}}_{1}\widehat{\mathcal{T}}_{2}\, 	\overline{\mathcal{T}}_{1}^2 + \overline{\mathcal{T}}_{1}\overline{\mathcal{T}}_{2}   \widehat{\mathcal{T}}_{1}^2
 \qquad,\qquad   
   \mathcal T_{3\text{(\texttt{gggg})}} = \widehat{\mathcal{T}}_{1} \, \widehat{\mathcal{T}}_{2}	\overline{\mathcal{T}}_{1}\overline{\mathcal{T}}_{2}
  \qquad, \\  
  \mathcal T_{4\text{(\texttt{gggg})}} &= \widehat{\mathcal{T}}_{2}^2\, 	\overline{\mathcal{T}}_{1}^2 + \widehat{\mathcal{T}}_{1}^2\, 	\overline{\mathcal{T}}_{2}^2    
 \qquad,\qquad  
  \mathcal T_{5\text{(\texttt{gggg})}} =  \widehat{\mathcal{T}}_{2}^2\, 	\overline{\mathcal{T}}_{1}\overline{\mathcal{T}}_{2} + \widehat{\mathcal{T}}_{1}\widehat{\mathcal{T}}_{2}\, 	\overline{\mathcal{T}}_{2}^2
 \qquad,\qquad    
   \mathcal T_{6\text{(\texttt{gggg})}} =  \widehat{\mathcal{T}}_{2}^2\, 	\overline{\mathcal{T}}_{2}^2 \qquad,
\nonumber    
 \end{align} 

\item $\Big[\textrm{(\texttt{ppss})}_I\otimes\textrm{(\texttt{sspp})}_I\otimes\textrm{(\texttt{\texttt{pppp}})}_I\Big]$
 \begin{align}
  \mathcal T_{7\text{(\texttt{gggg})}} =&\widehat{\mathcal{T}}_{1}\, 	\overline{\mathcal{T}}_{1} \mathcal{T}_{3}
 \qquad,\qquad   
  \mathcal T_{8\text{(\texttt{gggg})}} =\widehat{\mathcal{T}}_{1}\, 	\overline{\mathcal{T}}_{1} \mathcal{T}_{4}
 \qquad,\qquad  
  \mathcal T_{9\text{(\texttt{gggg})}} =(\widehat{\mathcal{T}}_{1} \overline{\mathcal{T}}_{2} 
+\widehat{\mathcal{T}}_{2} \overline{\mathcal{T}}_{1})\mathcal{T}_{3} \qquad,
\\
  \mathcal T_{10\text{(\texttt{gggg})}} =&(\widehat{\mathcal{T}}_{1} \overline{\mathcal{T}}_{2}
  +\widehat{\mathcal{T}}_{2} \overline{\mathcal{T}}_{1})\mathcal{T}_{4}
 \qquad,\qquad  
  \mathcal T_{11\text{(\texttt{gggg})}} =\widehat{\mathcal{T}}_{2}\, 	\overline{\mathcal{T}}_{2} \mathcal{T}_{3}
 \qquad,\qquad   
  \mathcal T_{12\text{(\texttt{gggg})}} = \widehat{\mathcal{T}}_{2}\, 	\overline{\mathcal{T}}_{2} 	\mathcal{T}_{4} \qquad,
 \nonumber  
 \end{align}
  
\item $\Big[\textrm{(\texttt{\texttt{pppp}})}_I\Big]^2$ 
\begin{align}
\mathcal T_{13\text{(\texttt{gggg})}}=\mathcal{T}_{3}\mathcal{T}_{3}
\qquad,\qquad
\mathcal T_{14\text{(\texttt{gggg})}}=\mathcal{T}_{3}\mathcal{T}_{4}
\qquad,\qquad
\mathcal T_{15\text{(\texttt{gggg})}}=\mathcal{T}_{4}\mathcal{T}_{4}
 \qquad,
\end{align}

\item $\Big[\textrm{(\texttt{sspp})}_I\otimes \textrm{(\texttt{ggpp})}_I\Big]\oplus \Big[\textrm{(\texttt{ppss})}_I\otimes \textrm{(\texttt{ppgg})}_I\Big]$ - 
\begin{align}
 \mathcal T_{16\text{(\texttt{gggg})}}=\widehat{\mathcal{T}}_{5}	\overline{\mathcal{T}}_{1} + \widehat{\mathcal{T}}_{1}\overline{\mathcal{T}}_{5}
\qquad,\qquad 
 \mathcal T_{17\text{(\texttt{gggg})}}=\widehat{\mathcal{T}}_{5}	\overline{\mathcal{T}}_{2} + \widehat{\mathcal{T}}_{2}\overline{\mathcal{T}}_{5}	 \qquad,
\end{align}

\item $\Big[ \textrm{(\texttt{gggg})}_I\Big]$
 \begin{align}
  \mathcal T_{18\text{(\texttt{gggg})}}= \mathcal{T}_{6}\qquad\text{and}\qquad
  \mathcal T_{19\text{(\texttt{gggg})}}= \mathcal{T}_{7}~.
 \end{align}
\end{enumerate}
Let's now write down the corresponding form factors. In this case, 6 form factors have long expression in terms of Gegenbauer polynomial. We write those expressions in appendix \ref{app:ggggformfactor}. The form factor corresponding to $\Big[\textrm{(\texttt{ppss})}_I\otimes\textrm{(\texttt{sspp})}_I\otimes\textrm{(\texttt{\texttt{pppp}})}_I\Big]$ tensor structures are given by 
\begin{align} 
\mathcal F_{1\text{(\texttt{gggg})}} &= \frac{1}{128}\sum_{a=0}^{\lfloor \frac{J}{2} \rfloor} A(J,a,D) 
\big[ (k_{12})^2  (k_{34})^2 \big]^{a-4} \big( k_{12}\cdot k_{34} \big)^{J-2 a-8} \nonumber\\
 &~~~~~~~~~~\Big[-4 \
 \big( k_{12}\cdot k_{34} \big)^4 \big[(k_{12})^2  (k_{34})^2 \big]^2 (J-2 a-3) (J-2 a-1) (J-2 a) \
(J-2 (a+1))\nonumber\\
 &~~~~~~~~~~-16 \big( k_{12}\cdot k_{34} \big)^6 \big[ (k_{12})^2 (k_{34})^2 \big] a^2 (J-2 a-1) \
(J-2 a) (-2 J+2 a+1)^2\nonumber\\
 &~~~~~~~~~~+8 \big( k_{12}\cdot k_{34} \big)^8 (a-1)^2 a^2 \left(1-4 \
(J-a)^2\right)^2\Big]\quad,
\\
\mathcal F_{2\text{(\texttt{gggg})}} &= \frac{1}{4}\sum_{a=0}^{\lfloor \frac{J}{2} \rfloor} A(J,a,D) a \
(J-2 a)(J-2a-1) \big[ (k_{12})^2 (k_{34})^2 \big]^{a-3} \big( k_{12}\cdot k_{34} \big)^{J-2
a-4} \nonumber\\
 &~~~~~~~~~~\Big[\big( k_{12}\cdot k_{34} \big)^2 (a-1) a (2 J-2 a+1) (-2 J+2 a+1)^2\nonumber\\
 &~~~~~~~~~~-\big[ (k_{12})^2 (k_{34})^2 \big] (J-2 a-3) (2 J-2 a-1) (J-2 a-2)\Big]\quad,
\end{align}

\begin{align} 
\mathcal F_{3\text{(\texttt{gggg})}} &= \frac{1}{2} \,\sum_{a=0}^{\lfloor \frac{J}{2} \rfloor} A(J,a,D)
 (J-2 a)(J-2 a-1) (J-2 a-2)(J-2 a-3) \big[ (k_{12})^2 (k_{34})^2 \big]^{a-2} \big( k_{12}\cdot k_{34} \big)^{J-2 a-6} \nonumber\\
 &~~~~~~~~~~\left[2 \big( k_{12}\cdot k_{34} \big)^2 a^2 (-2 
J+2 a+1)^2-\big[ (k_{12})^2 (k_{34})^2 \big] (J-2 a-4)(J-2 a-5) \right] \quad,
\\
\mathcal F_{4\text{(\texttt{gggg})}} &= \sum_{a=0}^{\lfloor \frac{J}{2} \rfloor} A(J,a,D)\frac{1}{4} \
 (J-2 a)(J-2 a-1) (J-2 a-2)(J-2 a-3) \nonumber\\
 &~~~~~~~~~~ a(a-1) \left(4 (J-a)^2-1\right)
\big[ (k_{12})^2  (k_{34})^2 \big]^{a-2} \big( k_{12}\cdot k_{34} \big)^{J-2 a-4}\quad,
\\
\mathcal F_{5\text{(\texttt{gggg})}}  
&= \frac{1}{2}\mathcal{N}_{J,D}\,(y_{s})^{J-8} \left[-13 \
\frac{d^7\mathcal{G}^{(\beta)}_{J-1}(z_{s})}{dz_{s}^7}+\frac{d^8\mathcal{G}^{(\beta)}_{J-2}(z_{s})}{dz_{s}^8}-2 z_{s} \
\frac{d^8\mathcal{G}^{(\beta)}_{J-1}(z_{s})}{dz_{s}^8}\right] \quad\text{and}
\\
\mathcal F_{6\text{(\texttt{gggg})}} & = \mathcal{N}_{J,D}\,(y_{s})^{J-8} \frac{d^8\mathcal{G}^{(\beta)}_{J}(z_{s})}{dz_{s}^8} ~.
\end{align}
Let's now write down the form factors corresponding to $\Big[\textrm{(\texttt{ppss})}_I\otimes\textrm{(\texttt{sspp})}_I\Big]^2$ tensor structures. Those 6 form factors are given by 
\begin{align}
 \mathcal F_{7\text{(\texttt{gggg})}} &=\frac{1}{2} \sum_{a=0}^{\lfloor \frac{J}{2} \rfloor} A(J,a,D) \
 (J-2 a)(J-2 a-1) \big[ (k_{12})^2 (k_{34})^2 \big]^{a-2} \big( k_{12}\cdot k_{34} \big)^{J-2a-4} \nonumber\\
 &~~~~~~~~~~\left[2 \big( k_{12}\cdot k_{34} \big)^2 a^2 (-2 J+2 a+1)^2-\big[ (k_{12})^2 k_{34})^2 \big] \
(J-2 a-3) (J-2a-2)\right] \quad,
\\ 
\mathcal F_{8\text{(\texttt{gggg})}} &= \frac{1}{2} \sum_{a=0}^{\lfloor \frac{J}{2} \rfloor} A(J,a,D)\
(J-2 a)(J-2a-1) (J-2a-2) \big[ (k_{12})^2 (k_{34})^2 \big]^{a-2} \
\big( k_{12}\cdot k_{34} \big)^{J-2 a-5} \nonumber\\
 &~~~~~~~~~~\left[2 \big( k_{12}\cdot k_{34} \big)^2 a^2 (-2 J+2 \
a+1)^2-\big[ (k_{12})^2 (k_{34})^2 \big] (J-2 a-3) (J-2a-4)\right]\quad,
\\
\mathcal F_{9\text{(\texttt{gggg})}} & = \mathcal{N}_{J,D}\, (y_{s})^{J-6} \left[-9 \
\frac{d^5\mathcal{G}^{(\beta)}_{J-1}(z_{s})}{dz_{s}^5}+\frac{d^6\mathcal{G}^{(\beta)}_{J-2}(z_{s})}{dz_{s}^6}-2 z_{s} \
\frac{d^6\mathcal{G}^{(\beta)}_{J-1}(z_{s})}{dz_{s}^6}\right] \quad,
\\
\mathcal F_{10\text{(\texttt{gggg})}} &=\mathcal{N}_{J,D}\, (y_{s})^{J-7} \left[-11 \
\frac{d^6\mathcal{G}^{(\beta)}_{J-1}(z_{s})}{dz_{s}^6}+\frac{d^7\mathcal{G}^{(\beta)}_{J-2}(z_{s})}{dz_{s}^7}-2 z_{s} \
\frac{d^7\mathcal{G}^{(\beta)}_{J-1}(z_{s})}{dz_{s}^7}\right] \quad,
\\
\mathcal F_{11\text{(\texttt{gggg})}}  
& = 4\, \mathcal{N}_{J,D}\, (y_{s})^{J-6} \frac{d^6\mathcal{G}^{(\beta)}_{J}(z_{s})}{dz_{s}^6}\quad\text{and}
\\ 
\mathcal F_{12\text{(\texttt{gggg})}}  
&= 4\,\mathcal{N}_{J,D}\, (y_{s})^{J-7} \frac{d^7\mathcal{G}^{(\beta)}_{J}(z_{s})}{dz_{s}^7} ~.
\end{align}
The expressions for $ \mathcal F_{7\text{(\texttt{gggg})}} $ and $ \mathcal F_{8\text{(\texttt{gggg})}} $ in terms of Gegenbauer polynomial can be found in appendix \ref{app:ggggformfactor}. The form factor corresponding to $\Big[\textrm{(\texttt{\texttt{pppp}})}_I\Big]^2$ tensor structures are given by 
\begin{align} 
\mathcal F_{13\text{(\texttt{gggg})}}  
& = 2\,\mathcal{N}_{J,D}\, (y_{s})^{J-4} \frac{d^4\mathcal{G}^{(\beta)}_{J}(z_{s})}{dz_{s}^4} \quad,
\\
\mathcal F_{14\text{(\texttt{gggg})}}  
&= 4\, \mathcal{N}_{J,D}\, (y_{s})^{J-5} \frac{d^5\mathcal{G}^{(\beta)}_{J}(z_{s})}{dz_{s}^5} \quad\text{and}
\\ 
\mathcal F_{15\text{(\texttt{gggg})}}  
&= 2\, \mathcal{N}_{J,D}\, (y_{s})^{J-6} \frac{d^6\mathcal{G}^{(\beta)}_{J}(z_{s})}{dz_{s}^6}~.
\end{align}
These three form factors have very simple expression in terms of the derivative of the Gegenbauer polynomial. The form factor corresponding to $\Big[\textrm{(\texttt{sspp})}_I\otimes \textrm{(\texttt{ggpp})}_I\Big]\oplus \Big[\textrm{(\texttt{ppss})}_I\otimes \textrm{(\texttt{ppgg})}_I\Big]$ tensor structures are as follows 
\begin{align} 
\mathcal F_{16\text{(\texttt{gggg})}}  
& 
=\mathcal{N}_{J,D}\, (y_{s})^{J-6} \left[-9 \
\frac{d^5\mathcal{G}^{(\beta)}_{J-1}(z_{s})}{dz_{s}^5}+\frac{d^6\mathcal{G}^{(\beta)}_{J-2}(z_{s})}{dz_{s}^6}-2 z_{s} \
\frac{d^6\mathcal{G}^{(\beta)}_{J-1}(z_{s})}{dz_{s}^6}\right]\quad\text{and} 
\\
\mathcal F_{17\text{(\texttt{gggg})}}  
& = 4\,\mathcal{N}_{J,D}\, (y_{s})^{J-6} \frac{d^6\mathcal{G}^{(\beta)}_{J}(z_{s})}{dz_{s}^6} ~.
\end{align}
We are left the two Form factor corresponding to the irreducible tensor structures $\Big[ \textrm{(\texttt{gggg})}_I\Big]$. Their expressions are extremely simple and they are given below 
\begin{align} 
\mathcal F_{18\text{(\texttt{gggg})}}  
& = 2\,\mathcal{N}_{J,D}\, (y_{s})^{J-4} \frac{d^4\mathcal{G}^{(\beta)}_{J}(z_{s})}{dz_{s}^4} \quad\text{and}
\\
\mathcal F_{19\text{(\texttt{gggg})}}  
& = 4\,\mathcal{N}_{J,D}\, (y_{s})^{J-5} \frac{d^5\mathcal{G}^{(\beta)}_{J}(z_{s})}{dz_{s}^5}~.
\end{align}

\paragraph{Angular distribution in 3+1 dimensions} Now, we verify the answer by computing angular distribution in $3+1$ dimensions. The angular distributions for the tensor structures are written in table \ref{tab:ggggangdist}. From the structure of the minimal coupling, we expect the answer to be zero for $4^\pm\rightarrow 0,0\rightarrow 4^\pm $ and $0\rightarrow 0$ helicity configuration. Furthermore, it should reproduce the correct Wigner matrix for $4^\pm\rightarrow 4^\pm $ and $4^\pm\rightarrow 4^\mp $ helicity configurations. 
\begin{table}[h!]
	\begin{center}
		\begin{tabular}{ ||p{3.25 cm}||p{3.5 cm}|p{3.5 cm}|p{2cm}|p{2.5 cm}||}
			\hline
			\hline
			\multicolumn{5}{|c|}{{\bf Angular distributions for \texttt{gggg} tensor structures} } \\
			\hline
			\hline
			\multicolumn{5}{|c|}{{$\Big((\textrm{\texttt{ppss}})_I\otimes  (\textrm{\texttt{sspp}})_I\Big)^2$ }} \\
			\hline
			\hline
			Tensor structures & $4^\pm\longrightarrow 4^\pm$ &$4^\pm\longrightarrow 4^\mp $& $4^\pm\longrightarrow 0$ & $0\longrightarrow 0$ \\
			\hline
			
			$\mathcal T_{1\text{(\texttt{gggg})}}$ & $0$ &$0$& $0$ & $s^{8}$ \\
			\hline
			$\mathcal T_{2\text{(\texttt{gggg})}}$ & $0$ &$0$& $0$ & $\frac{s^{8}}{2}$ \\
			\hline
			$\mathcal T_{3\text{(\texttt{gggg})}}$ & $0$ &$0$& $0$ & $\frac{s^{8}}{16}$ \\
			\hline
			$\mathcal T_{4\text{(\texttt{gggg})}}$ & $0$ &$0$& $\frac{s^{8}}{16}\sin^4\theta$ & $\frac{s^{8}}{8}$ \\
			\hline
			$\mathcal T_{5\text{(\texttt{gggg})}}$ & $0$ &$0$& $\frac{s^9}{64}\sin^4\theta$ & $\frac{s^8}{32}$ \\
			\hline
			$\mathcal T_{6\text{(\texttt{gggg})}}$ & $\frac{s^8}{256}\sin^8\theta$ &$\frac{s^8}{256}\sin^8\theta$& $\frac{s^8}{256}\sin^4\theta$ & $\frac{s^8}{256}$ \\
			\hline
			\hline
			\multicolumn{5}{|c|}{{$\Big((\textrm{\texttt{ppss}})_I \otimes (\textrm{\texttt{sspp}})_I \otimes (\textrm{\texttt{\texttt{pppp}}})_I \Big)$ }}\\
			\hline
			\hline    
			$\mathcal T_{7\text{(\texttt{gggg})}}$ & $0$ &$0$& $0$ & $\frac{s^6}{2}$ \\
			\hline
			$\mathcal T_{8\text{(\texttt{gggg})}}$ & $0$ &$0$& $0$ & $\frac{s^7}{4}\cos\theta$ \\	
			\hline
			$\mathcal T_{9\text{(\texttt{gggg})}}$ & $0$ &$0$& $0$ & $\frac{s^6}{4}$ \\
			\hline
			$\mathcal T_{10\text{(\texttt{gggg})}}$ & $0$ &$0$& $0$ & $\frac{s^7}{8}\cos\theta$ \\
			\hline
			$\mathcal T_{11\text{(\texttt{gggg})}}$ & $\frac{s^6}{32}\sin^4\theta\cos^4\big(\frac{\theta}{2}\big)$ &$\frac{s^6}{32}\sin^4\theta\sin^4\big(\frac{\theta}{2}\big)$& $0$ & $\frac{s^6}{32}$ \\
			\hline
			$\mathcal T_{12\text{(\texttt{gggg})}}$ & $-\frac{s^7}{32}\sin^2\theta \cos^6\left(\frac{\theta}{2}\right) $ & $\frac{s^7}{32}\sin^2\theta \sin^6\left(\frac{\theta}{2}\right) $& $0$ & $\frac{s^7}{64}\cos\theta$ \\
			\hline
			\hline
			\multicolumn{5}{|c|}{{$\Big((\textrm{\texttt{\texttt{pppp}}})_I \otimes (\textrm{\texttt{\texttt{pppp}}})_I \Big)$ }}\\
			\hline
			\hline
			$\mathcal T_{13\text{(\texttt{gggg})}}$ & $\frac{s^4}{4}\cos^8\big(\frac{\theta}{2}\big)$ &$\frac{s^4}{4}\sin^8\big(\frac{\theta}{2}\big)$& $0$ & $\frac{s^4}{4}$ \\
			\hline
			$\mathcal T_{14\text{(\texttt{gggg})}}$ & $-\frac{s^5}{4}\sin^2\theta\cos^6\big(\frac{\theta}{2}\big)$ &$\frac{s^5}{4}\sin^2\theta\sin^6\big(\frac{\theta}{2}\big)$& $0$ & $\frac{s^5}{8}\cos\theta$ \\
			\hline
			$\mathcal T_{15\text{(\texttt{gggg})}}$ & $\frac{s^6}{4}\sin^4\theta\cos^4\big(\frac{\theta}{2}\big)$ &$\frac{s^6}{4}\sin^4\theta\sin^4\big(\frac{\theta}{2}\big)$& $0$ & $\frac{s^6}{16}\cos^2\theta$ \\
			\hline
			\hline
			\multicolumn{5}{|c|}{{ $\Big((\textrm{\texttt{sspp}})_I \otimes (\textrm{\texttt{ggpp}})_I \Big) \oplus \Big((\textrm{\texttt{ppss}})_I \times (\textrm{\texttt{ppgg}})_I \Big)$ }}\\
			\hline
			\hline  
			$\mathcal T_{16\text{(\texttt{gggg})}}$ & $0$ &$0$& $0$ & $\frac{s^6}{8}$ \\
			\hline
			$\mathcal T_{17\text{(\texttt{gggg})}}$ & $2s^6 \sin^4\left(\frac{\theta}{2}\right) \cos^8\left(\frac{\theta}{2}\right)$ &$2s^6 \sin^8\left(\frac{\theta}{2}\right) \cos^4\left(\frac{\theta}{2}\right)$& $0$ & $\frac{s^6}{32}$ \\
			\hline
			\hline
			\multicolumn{5}{|c|}{{ $ (\textrm{\texttt{gggg}})_I $ }}\\
			\hline
			\hline  
			$\mathcal T_{18\text{(\texttt{gggg})}}$ & $\frac{s^4}{2} \cos^8\left(\frac{\theta}{2}\right)$ & $\frac{s^4}{2} \sin^8\left(\frac{\theta}{2}\right)$& $0$ & $\frac{s^4}{8}$
			\\
			\hline
			$\mathcal T_{19\text{(\texttt{gggg})}}$ & $-\frac{s^5}{2} \sin^2\theta \cos^6\left(\frac{\theta}{2}\right)$ & $\frac{s^5}{2} \sin^2\theta \sin^6\left(\frac{\theta}{2}\right)$ & $0$ & $\frac{s^5}{16} \cos\theta$ \\
			\hline       
			\hline       
			
		\end{tabular}
	\end{center}
	\caption{Angular distributions for different \texttt{gggg} tensor structures.}
 	\label{tab:ggggangdist} 
\end{table}

Let's first consider the helicity configuration $4^\pm\rightarrow 4^\pm $. In this case only 9 out of 19 tensor structures is non-zero. 
Putting the expression of the tensor structures and form factors in $\texttt{A}_{\text{\texttt{gggg}}}$ we obtain 
\begin{align}
 \texttt{A}_{\text{\texttt{gggg}}}^{4^+\rightarrow 4^+} 
 &= \mathcal{N}_{J,D} (s)^{J}\left[\frac{1}{J(J-1)(J-2)(J-3)}\right]^2 \frac{(z_{s}+1)^4}{16}\nonumber\\
 &~~~~~~~\Bigg[\frac{1}{16} (1-z_{s})^4 \frac{\partial^8 P_J(z_{s})}{\partial z_{s}^8} + \frac{1}{2} (1-z_{s})^2 \frac{\partial^6 P_J(z_{s})}{\partial z_{s}^6} - (1-z_{s})^3 \frac{\partial^7 P_J(z_{s})}{\partial z_{s}^7} +\frac{1}{2} \frac{\partial^4 P_J(z_{s})}{\partial z_{s}^4} \nonumber\\
 &~~~~~~~-2 (1-z_{s}) \frac{\partial^5 P_J(z_{s})}{\partial z_{s}^5} + 2 (1-z_{s})^2 \frac{\partial^6 P_J(z_{s})}{\partial z_{s}^6} + (1-z_{s})^2 \frac{\partial^6 P_J(z_{s})}{\partial z_{s}^6} \nonumber\\
 &~~~~~~~+ (1-z_{s})^2 \frac{\partial^6 P_J(z_{s})}{\partial z_{s}^6}+ \frac{\partial^4 P_J(z_{s})}{\partial z_{s}^4} - 4 (1-z_{s}) \frac{\partial^5 P_J(z_{s})}{\partial z_{s}^5} \Bigg]~.
\end{align}
Using the mathematical identity in eqn \eqref{mathematicalidentityIII51} we get
\begin{align}
 \texttt{A}_{\text{\texttt{gggg}}}^{4^+\rightarrow 4^+} &= \mathcal{N}_{J,D} (s)^{J} \frac{1}{256}\frac{\Gamma(J+5)}{\Gamma(J-3)} \frac{(z_{s}+1)^4}{16}\jacobi_{J-4}^{0,8}(z_{s}) 
 \nonumber \\
  &= \Bigg[ (s)^J \Big( \widetilde{\mathcal{N}}_{J;4,4}\Big) d_{44}^{(J)}(\theta)\Bigg] 
\end{align}
where $\widetilde{\mathcal{N}}_{J;4,4}$ is defined in \eqref{angdist3}. This is in agreement with \eqref{angdist2}. By following same set of equations we can show that
\begin{align}
\texttt{A}_{\text{\texttt{gggg}}}^{4^-\rightarrow 4^-} = \Bigg[(s)^J \Big( \widetilde{\mathcal{N}}_{J;4,4}\Big) d_{-4-4}^{(J)}(\theta)\Bigg] ~.
\end{align}
Let's now look at the helicity configuration $4^\pm\rightarrow 4^\mp$. This is simply exchange of $3$ and $4$. 
\begin{align}
\texttt{A}_{\text{\texttt{gggg}}}^{4^+\rightarrow 4^-} &= {(-1)^J}\Big(\texttt{A}_{\text{\texttt{gggg}}}^{4^+\rightarrow 4^+} \Big)_{z_{s}\rightarrow -z_{s}}
=(s)^J \Big( \widetilde{\mathcal{N}}_{J;4,4}\Big) d_{4-4}^{(J)}(\theta)~.
\end{align}
Now we consider, the helicity configurations $4^\pm\rightarrow 0 $ and $0\rightarrow 4^\pm $. In this case only three of tensor structures are non-zero. Putting the corresponding form factors we verified that the amplitude vanishes in this case. The last case to consider is the helicity configuration $0\rightarrow 0 $. In this case, all the tensor structures are non-zero. Putting the tensor structures and form factors we verified using Mathematica that the amplitude vanishes for this case.

\paragraph{Amplitudes due to the non-minimal coupling} The graviton-graviton-spin $J$ three point function has 3 different structures. As a result, the four graviton amplitude due massive spin $J$ exchange has 9 different contribution. Upto this point, we considered the one which comes solely from the minimal coupling. The other eight are easy to evaluate. We write them here.
\begin{align}
\texttt{A}_{\text{\texttt{gggg}}}^\texttt{(1|0)} &= \Big[\Tr\, \mathcal{W}_{(12)}\,\Big] \Big(\texttt{A}_{\texttt{ppgg}}^{\texttt{(0|0)}}\Big)~.
\\
\texttt{A}_{\text{\texttt{gggg}}}^\texttt{(0|1)} &= \Big[ \Tr\, \mathcal{W}_{(34)}\Big] \Big(\texttt{A}_{\texttt{ggpp}}^{\texttt{(0|0)}}\Big)~.
\\
\texttt{A}_{\text{\texttt{gggg}}}^\texttt{(2|0)} &= \Big[\Tr\, \mathcal{W}_{(12)}\, \Tr\, \mathcal{W}_{(12)}\Big] \Big(\texttt{A}_{\texttt{ssgg}}^{\texttt{(0|0)}}\Big)~.
\\
\texttt{A}_{\text{\texttt{gggg}}}^\texttt{(0|2)} &= \Big[\Tr\, \mathcal{W}_{(34)}\, \Tr\, \mathcal{W}_{(34)}\Big] \Big(\texttt{A}_{\texttt{ggss}}^{\texttt{(0|0)}}\Big)~.
\\
\texttt{A}_{\text{\texttt{gggg}}}^\texttt{(1|1)} &= \Big[\Tr\, \mathcal{W}_{(12)}\, \Tr\, \mathcal{W}_{(34)}\Big] \Big(\texttt{A}_{\texttt{pppp}}^{\texttt{(0|0)}}\Big)~.
\\
\texttt{A}_{\text{\texttt{gggg}}}^\texttt{(2|1)} &= \Big[\Tr\, \mathcal{W}_{(12)}\, \Tr\, \mathcal{W}_{(12)}\, \Tr\, \mathcal{W}_{(34)}\Big] \Big(\texttt{A}_{\texttt{sspp}}^{\texttt{(0|0)}}\Big)~.
\\
\texttt{A}_{\text{\texttt{gggg}}}^\texttt{(1|2)} &= \Big[\Tr\, \mathcal{W}_{(12)}\, \Tr\, \mathcal{W}_{(34)}\Tr\, \mathcal{W}_{(34)}\Big] \Big(\texttt{A}_{\texttt{ppss}}^{\texttt{(0|0)}}\Big)~.
\\
\texttt{A}_{\text{\texttt{gggg}}}^\texttt{(2|2)} &= \Big[\Tr\, \mathcal{W}_{(12)}\, \Tr\, \mathcal{W}_{(12)}\, \Tr\, \mathcal{W}_{(34)} \Tr\, \mathcal{W}_{(34)}\Big] \Big(\texttt{A}_{\texttt{ssss}} \Big)~.
\end{align}
The expression for $(\texttt{A}_{\text{\texttt{gggg}}}^\texttt{(2|2)})$ is also given in L.21 of \cite{Chowdhury:2019kaq}.

\section{Massless amplitudes: class II}
\label{sec:ramclassIIamplitudes}

In this section, we compute class II amplitude with massless external states. Class II amplitudes have at least one vertex where two  particles of different spin meet. In this section, we compute the following four class II amplitudes
\begin{equation}
\textrm{(a) \texttt{psss}}	
\qquad,\qquad 
\textrm{(b) \texttt{psps}}	
\qquad,\qquad 
\textrm{(c) \texttt{gsss}}	
\qquad,\qquad 
\textrm{(d) \texttt{gsgs}}	
\end{equation}
One important feature is that all the three-point functions are unique, and hence all the class II amplitudes are unique.

\subsection{Photon amplitudes}

\subsubsection{\texttt{psss}}
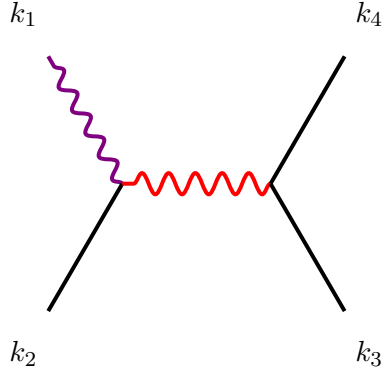
\begin{figure}[h]
\begin{center}
\begin{tikzpicture}[line width=1.5 pt, scale=1.3]
 
\begin{scope}[shift={(0,0)}]	
		
\treefourpointexchange{photon}{realscalar}{realscalar}{realscalar}{spinj}{0}	

\labeltreefourpointexchange{ $k_1$ }{ $k_2$ }{ $k_3$ }{ $k_4$ }{}{0} 
\end{scope}

\end{tikzpicture} 
\end{center}
\caption{\texttt{psss} amplitude due to exchange of a massive spinning particle }
\label{fig:psssfeynmandiag}
\end{figure}

We start with the simplest class II amplitude: \texttt{psss}. In this both the three point functions are unique. Let's consider the $s$-channel amplitude. The corresponding Feynman diagram is depicted in fig. \ref{fig:psssfeynmandiag}. The numerator is given by 
\begin{align}
 \texttt{A}_{\texttt{psss}}=  & (\mathcal{X}_{(1;2)})^{\mu_1}(k_{12})^{\mu_2}\cdots (k_{12})^{\mu_J} ~\mathcal{P}^{(J)}_{\mu_1\cdots\mu_J\,;\,\nu_1\cdots\nu_J}~(k_{34})^{\nu_1}\cdots (k_{34})^{\nu_J}~.
\end{align}
This numerator can be obtained from the four scalar amplitude using the derivative method 
\begin{align}
 \texttt{A}_{\texttt{psss}}&= \left(\frac{1}{J}\right) \mathcal{X}_{(1;2)}^{\tilde\mu} \Theta_{\tilde\mu \mu} \frac{\partial}{\partial (k_{12})_\mu} \Big( \texttt{A}_{\texttt{ssss}}\Big)~\nonumber\\
 &=\big(\mathcal{O}^{\texttt{(0)}}\big)^\texttt{ssJ}_\texttt{psJ}~~ \texttt{A}_{\texttt{ssss}}~.
\end{align}
Using the expression for $\texttt{A}_{\texttt{ssss}}$ in terms of the Gegenbauer polynomial and the identity \eqref{gegenpoly51}, this derivative is straightforward to evaluate. The final answer can be written as a product of form factor and tensor factor as in \eqref{ramdecompositionI}. In this the tensor structure is unique and it given by 
\begin{align}
 \mathcal T_{\text{(\texttt{psss})}} = \widehat{\mathcal T}_8
\end{align}
and the corresponding form factor is given by 
\begin{align} \mathcal F_{\text{(\texttt{psss})}}  
 &= \mathcal{N}_{J,D}\, (y_{s})^{J-1}\frac{d\mathcal{G}^{(\beta)}_J(z_{s})}{dz_{s}}~.
\end{align}
The expression for other channels can simply be obtained $2\longleftrightarrow 3$ and $2\longleftrightarrow 4$ exchange. 

\paragraph{Angular distribution in $3+1$ dimensions} Let's now consider the angular distribution for this process when the scalar is also massless. The angular distribution for the tensor structures can be found in table \ref{tab:psssangdist}. First we restrict to the case when the photon has a positive helicity .
\begin{table}[h]
 \begin{center}
  \begin{tabular}{ ||c||c|c|| }
   \hline
   \hline
   \multicolumn{3}{|c|}{{\bf Angular distributions for different helicities} } \\
   \hline
      \hline
   Tensor structures & $1^+\rightarrow 0$ & $1^-\rightarrow 0$ \\
   \hline
   $ \mathcal T_{\text{(\texttt{psss})}} $ & $\frac{-s\sqrt{s}}{2\sqrt{2}}\sin\theta$& $\frac{s\sqrt{s}}{2\sqrt{2}}\sin\theta$ \\
   \hline
   \hline

  \end{tabular}
  \caption{Angular distributions for the \texttt{psss} tensor structures}
  \label{tab:psssangdist} 
 \end{center}
\end{table}
For positive helicity of the photon, the tensor factor evaluates to  
\begin{align}
 \mathcal T_{\text{(\texttt{psss})}} &= \frac{-s\sqrt{s}}{2\sqrt{2}}\sin\theta = \frac{-s\sqrt{s}}{2\sqrt{2}}\sqrt{1-z^2}~.
\end{align}
Then the full amplitude is given by 
\begin{align}
 \mathcal T_{\text{(\texttt{psss})}}\, \mathcal F_{\text{(\texttt{psss})}} 
 =\frac{-s\sqrt{s}}{2\sqrt{2}}(\sqrt{1-z^2})~ \mathcal{N}_{J,4}\, s^{J-1} \frac{1}{J} \frac{\partial \legendre_J(z)}{\partial z}~.
\end{align}
Now we use eqn \eqref{eq:wignertolegendre} for $h=1$ to get 
\begin{equation}
	\frac{\sqrt{s}}{2\sqrt{2}}~ \mathcal{N}_{J,4}\, s^{J} \sqrt{\frac{(J+1)}{J}} d^{(J)}_{01}(\theta) 
= \sqrt{\frac{s}{2}} \Big(\widetilde{\mathcal{N}}_{J;1,0} \Big) s^{J} d^{(J)}_{01}(\theta) ~.
\end{equation}
The expression including the normalisation agrees with \eqref{angdist2}. 
It is straightforward to check that 
\begin{equation}
	\frac{1}{2}\mathcal{N}_{J,4}\sqrt{\frac{(J+1)}{J}}=\widetilde{\mathcal{N}}_{J;1,0} ~.
\end{equation}
If we flip the helicity of the photon then $ \mathcal T_{\text{(\texttt{psss})}}$ flips sign; from the properties of the Wigner matrix we know $d^{(J)}_{0-1}(\theta) =-d^{(J)}_{01}(\theta) $. Hence our answer gives the correct angular distribution for the other helicity of the photon.

\subsubsection{\texttt{psps}} 

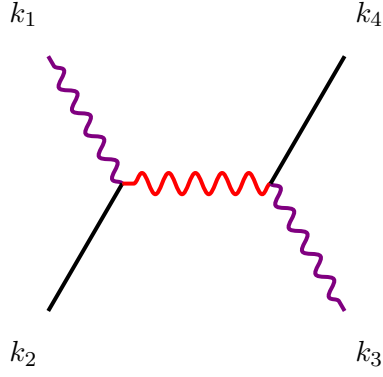
\begin{figure}[h]
\begin{center}
\begin{tikzpicture}[line width=1.5 pt, scale=1.3]
 
\begin{scope}[shift={(0,0)}]	
		
\treefourpointexchange{photon}{realscalar}{photon}{realscalar}{spinj}{0}	

\labeltreefourpointexchange{ $k_1$ }{ $k_2$ }{ $k_3$ }{ $k_4$ }{}{0} 
\end{scope}

\end{tikzpicture} 
\end{center}
\caption{\texttt{psps} amplitude due to exchange of a massive higher particle }
\label{fig:pspsfeynmandiag}
\end{figure}

Now we compute the \texttt{psps} amplitude. An important feature of this amplitude is that since the two vertices are the same and hence the coupling constants are the same, the residue at the pole is a positive definite quantity. The $s$-channel Feynman diagram is being depicted in fig. \ref{fig:pspsfeynmandiag}. 

In the derivative method, the amplitude is given by 
\begin{align}
 \texttt{A}_{\texttt{psps}}&=   \mathcal{X}_{(1;2)}^{\tilde\mu} \Theta_{\tilde\mu \mu} \frac{\partial}{\partial (k_{12})_\mu} \mathcal{X}_{(3;4)}^{\tilde\nu} \Theta_{\tilde\nu \nu} \frac{\partial}{\partial (k_{34})_\nu}  
 \texttt{A}_{\texttt{ssss}} \nonumber\\
 &=\big(\mathcal{O}^{\texttt{(0)}}\big)^\texttt{ssJ}_\texttt{psJ}~~\big(\mathcal{O}^{\texttt{(0)}}\big)^\texttt{Jss}_\texttt{Jps}~~ \texttt{A}_{\texttt{ssss}}~.
\end{align}
It is straightforward to evaluate the amplitude in this way. For this simple case, we present the details here. 
\begin{eqnarray}
 && \mathcal{X}_{(1;2)}^{\tilde\mu} \Theta_{\tilde\mu \mu} \frac{\partial}{\partial (k_{12})_\mu} \Bigg[ \sum_{a=0}^{\lfloor\frac{J}{2}\rfloor} A(J,a,D) ~\Big[(k_{12}\odot k_{12})\, (k_{34}\odot k_{34})\Big]^{a} (J-2a)\Big(k_{12}\odot k_{34}\Big)^{J-2a-1} (\mathcal{X}_{(3;4)} \odot k_{12}) \Bigg] 
 \nonumber  \\
 &=& \sum_{a=0}^{\lfloor\frac{J}{2}\rfloor} A(J,a,D)(J-2a) ~\Big[(k_{12}\odot k_{12})\, (k_{34}\odot k_{34})\Big]^{a} \Big(k_{12}\odot k_{34}\Big)^{J-2a-2} 
 \nonumber  \\
 &&\Bigg[ (J-2a-1) (\mathcal{X}_{(3;4)} \odot k_{12}) ~(\mathcal{X}_{(1;2)} \odot k_{34}) 
 +  \Big(k_{12}\odot k_{34}\Big) (\mathcal{X}_{(3;4)}\odot \mathcal{X}_{(1;2)}) \Bigg]~.
\end{eqnarray}
We write the final answer as a linear sum of two independent tensor structures. 
\begin{equation}
 \texttt{A}_{\text{\texttt{psps}}}=	\mathcal T_{1\text{(\texttt{psps})}}\, \mathcal F_{1\text{(\texttt{psps})}}(s,t) + \mathcal T_{2\text{(\texttt{psps})}} \mathcal F_{2\text{(\texttt{psps})}} (s,t)
\end{equation}
where $\mathcal T_{I\text{(\texttt{psps})}}$ are the tensor factors and $\mathcal F_{I\text{(\texttt{psps})}}(s,t)$ are the form factors. The tensor factors are given by 
\begin{equation}
 \mathcal T_{1\text{(\texttt{psps})}} = \widehat{\mathcal T}_8 \bar{\mathcal T}_8 
 \qquad\text{and}\qquad  
 \mathcal T_{2\text{(\texttt{psps})}} = \mathcal{T}_9
\end{equation}
where the corresponding form factors are given by 
\begin{equation}
\begin{split}
 \mathcal{F}_{1\text{(\texttt{psps})}} 
&=\mathcal{N}_{J,D}\,(y_{s})^{J-2}\frac{d^2\mathcal{G}^{(\beta)}_J(z_{s})}{dz_{s}^2}  
 \qquad\text{and}\qquad 
 \mathcal{F}_{2\text{(\texttt{psps})}} 
 =\mathcal{N}_{J,D}\,(y_{s})^{J-1}\frac{d\mathcal{G}^{(\beta)}_J(z_{s})}{dz_{s}}~.
\end{split}
\end{equation}
We can see the form factors are extremely simple.

\paragraph{Angular distribution} In the center of mass frame, the angular distribution of the tensor structures for different choice of helicity is given in table \ref{tab:pspsangdist}. Let's first consider the case $1^+\rightarrow 1^+$. For this choice, our answer is given by 
\begin{equation}
 \texttt{A}_{\texttt{psps}}^{1^+\rightarrow 1^+}=  \mathcal{N}_{J,4}\frac{s^{J+1}}{8J^2}(z_s+1)\Bigg[  \frac{\partial \legendre_J(z_s)}{\partial z_s} + (z_s-1) \frac{\partial^2 \legendre_J(z_s)}{\partial z_s^2} \Bigg]~.
\end{equation}
Using the mathematical identity, given in eqn \eqref{mathematicalidentityI5}, 
we get the angular distribution in 3+1 dimensions to be 
\begin{equation}
 s^{J+1} \mathcal{N}_{J,4}\frac{(J+1)}{8J} d_{11}^{(J)}(\theta)~.
\end{equation}
Putting the expression for $\mathcal{N}_{J,4}$, we get 
\begin{equation}
\mathcal{N}_{J,4}\frac{(J+1)}{8J}=	 \frac{1}{2} 	\widetilde{\mathcal{N}}_{J;1,1} ~.
\end{equation}
\begin{table}[h]
 \begin{center}
  \begin{tabular}{ ||c||c|c|| }
   \hline
   \hline
   \multicolumn{3}{|c|}{{\bf Angular distributions for different helicities} } \\
   \hline
   \hline
   Tensor structures & $1^\pm\rightarrow 1^\pm $& $1^\pm\rightarrow 1^\mp $ \\
   \hline
   $ \mathcal T_{1\text{(\texttt{psps})}} $ & $-\frac{s^3}{8}\sin^2\theta$ & $\frac{s^3}{8}\sin^2\theta$ \\
   \hline
   $ \mathcal T_{2\text{(\texttt{psps})}} $ & $\frac{s^2}{8} (\cos\theta+1)$ &$-\frac{s^2}{8} (\cos\theta-1)$\\
   \hline
    \hline
  \end{tabular}
  \caption{Angular distributions for different \texttt{psps} tensor structures}
  \label{tab:pspsangdist} 
 \end{center}
\end{table}
The computation is exactly the same if we choose negative helicity for both the photons (note that $d_{11}^{(J)}(\theta)=d_{-1-1}^{(J)}(\theta)$). Another choice of helicity ($1^+\rightarrow 1^- $) corresponds to the exchange of $3$ and $4$, which is same as $\theta \rightarrow \pi+\theta$. We see that there is an extra minus for the tensor structures for that case. The minus sign along with the property of Wigner matrix \eqref{eq:wignerprop2} produces the correct angular distribution  for that helicity configuration  
\begin{align}
 \texttt{A}_{\text{\texttt{psps}}}^{1^+\rightarrow 1^-} &={(-1)^J} \Big( \texttt{A}_{\text{\texttt{psps}}}^{1^+\rightarrow 1^+} \Big)_{z_{s}\rightarrow -z_{s}}= (-1)^Js^{J+1}\Bigg( (-1) { \frac{1}{2}}	\widetilde{\mathcal{N}}_{J;1,1} ~d_{11}^{(J)}(\pi+\theta) \Bigg)\\
&=  (s)^{J} { \frac{s}{ 2}}	\widetilde{\mathcal{N}}_{J;1,1} ~d_{1-1}^{(J)}(\theta)~.
\end{align}

\subsection{Graviton amplitudes}

\subsubsection{\texttt{gsss}}
Let's now consider the $\texttt{gsss}$ amplitude.  Since the graviton cannot carry any other charge, the amplitude gets contribution only from even spins. The $\texttt{gsJ}$ three-point is unique and it is the square of $\texttt{ps}\frac{\texttt{J}}{2}$ three-point function. The numerator of $\texttt{gsss}$ is given by 
\begin{align}
\texttt{A}_{\texttt{gsss}}= (\mathcal{X}_{(1;2)})^{\mu_1} (\mathcal{X}_{(1;2)})^{\mu_2}(k_{12})^{\mu_3}\cdots (k_{12})^{\mu_J} ~\mathcal{P}^{(J)}_{\mu_1\cdots\mu_J;\nu_1\cdots\nu_J}~(k_{34})^{\nu_1}\cdots (k_{34})^{\nu_J}~.
\end{align}
We can (again !) use the derivative method to compute this numerator
\begin{align}
\texttt{A}_{\texttt{gsss}}&=  \left[\frac{1}{J(J-1)}\right] \mathcal{X}_{(1;2)}^{\tilde\mu_1} \mathcal{X}_{(1;2)}^{\tilde\mu_2} \Theta_{\tilde\mu_1 \mu_1} \Theta_{\tilde\mu_2 \mu_2} \frac{\partial}{\partial (k_{12})_{\mu_1}} \frac{\partial}{\partial (k_{12})_{\mu_2}} \Big( \texttt{A}_{\texttt{ssss}}\Big)~\nonumber\\
&=\big(\mathcal{O}^{\texttt{(0)}}\big)^\texttt{ssJ}_\texttt{gsJ}~~\texttt{A}_{\texttt{ssss}}~.
\end{align}
Just like for the three point function, the derivative operator is also the square apart from the $J$ dependent factor outside. The $s$-channel answer is given by  
\begin{align}
\mathcal T_{\text{(\texttt{gsss})}}\mathcal F_{\text{(\texttt{gsss})}}\qquad\text{where}\qquad \mathcal T_{\text{(\texttt{gsss})}} &=  \widehat{\mathcal T}_8^2
\end{align}
and the corresponding form factor is 
\begin{equation}
\begin{split}
\mathcal F_{\text{(\texttt{gsss})}} 
&=\mathcal{N}_{J,D}\, (y_{s})^{J-2}\frac{d^2\mathcal{G}^{(\beta)}_J(z_{s})}{dz_{s}^2}~.
\end{split}
\end{equation}

\paragraph{Angular distribution in $3+1$ dimensions} Now, we restrict to the case of $3+1$ dimensions, and we further assume that the scalars are massless. In the center of mass frame (given in \eqref{subsec:comconvention}) the angular distribution for the tensor structures are given in Table \ref{tab:gsssangdist}. For example, consider $+2$ helicity for the graviton.
\begin{table}[h]
	\begin{center}
		\begin{tabular}{ ||c||c|| }
			\hline
			\hline
			\multicolumn{2}{|c|}{{\bf Angular distributions for different helicities} } \\
			\hline
			\hline
			Tensor structure & $2^\pm \rightarrow 0$ \\
			\hline
			$ \mathcal T_{\text{(\texttt{gsss})}} $ & $-\frac{s^3}{8}\sin^2\theta$ \\
			\hline
			\hline
		\end{tabular}
		\caption{Angular distributions for the \texttt{gsss} tensor factor.}
		\label{tab:gsssangdist} 
	\end{center}
\end{table}
In this case the tensor structure is given by 
\begin{align}
\mathcal T_{\text{(\texttt{gsss})}} &= -\frac{s^3}{8}\sin^2(\theta) = -\frac{s^3}{8}(1-z_s^2)
\end{align}
and hence the full amplitude is given by 
\begin{align}
\texttt{A}_{\text{\texttt{gsss}}}^{2^+\rightarrow 0} &=\mathcal T_{\text{(\texttt{gsss})}}\, \mathcal F_{\text{(\texttt{gsss})}} =-\frac{s^{J+1}}{8}(1-z_s^2)~ \mathcal{N}_{J,D}\, ~ \frac{1}{J(J-1)}  \frac{\partial^2 \legendre_J(z_s)}{\partial z_s^2}~.
\end{align}
From \eqref{eq:wignertolegendre} with $h=2$ we obtain 
\begin{equation}
\frac{\sqrt{s}}{2\sqrt{2}}~ \mathcal{N}_{J,4}\, s^{J} \sqrt{\frac{(J+1)(J+2)}{J(J-1)}} d^{(J)}_{20}(\theta) 
=  \frac{s}{2} \Big(\widetilde{\mathcal{N}}_{J;2,0} \Big) s^{J} d^{(J)}_{20}(\theta) ~.
\end{equation}
It is straightforward to check that 
\begin{equation}
\frac{1}{4}\mathcal{N}_{J,4}\sqrt{\frac{(J+1)(J+2)}{J(J-1)}}=\widetilde{\mathcal{N}}_{J;2,0} ~.
\end{equation}
The computation for $-2$ helicity of the graviton is exact the same.

\subsubsection{\texttt{gsgs}}
We go ahead to compute the last class II amplitude: the $\texttt{gsgs}$ amplitude. The $s$ channel numerator for the $\texttt{gsgs}$ amplitude is given by 
\begin{align}
 \texttt{A}_{\texttt{gsgs}}= (\mathcal{X}_{(1;2)})^{\mu_1}(\mathcal{X}_{(1;2)})^{\mu_2}(k_{12})^{\mu_3}\cdots (k_{12})^{\mu_J} ~\mathcal{P}^{(J)}_{\mu_1\cdots\mu_J;\nu_1\cdots\nu_J}~(\mathcal{X}_{(3;4)})^{\nu_1}(\mathcal{X}_{(3;4)})^{\nu_2}(k_{34})^{\nu_3}\cdots (k_{34})^{\nu_J}~.
\end{align}
In this case the corresponding derivative operator has the following expression  
\begin{align} 
 \texttt{A}_{\texttt{gsgs}}= &\left[\frac{1}{J(J-1)}\right]^2 \mathcal{X}_{(1;2)}^{\tilde\mu_1}\mathcal{X}_{(1;2)}^{\tilde\mu_2} \Theta_{\tilde\mu_1 \mu_1}\Theta_{\tilde\mu_2 \mu_2} \frac{\partial}{\partial (k_{12})_{\mu_1}}\frac{\partial}{\partial (k_{12})_{\mu_2}}
 \nonumber \\ 
 &~~~~~~~~~~~~~~\mathcal{X}_{(3;4)}^{\tilde\nu_1}\mathcal{X}_{(3;4)}^{\tilde\nu_2} \Theta_{\tilde\nu_1 \nu_1}\Theta_{\tilde\nu_2 \nu_2} \frac{\partial}{\partial (k_{34})_{\nu_1}}\frac{\partial}{\partial (k_{34})_{\nu_2}} \Big( \texttt{A}_{\text{ssss}}\Big) \nonumber\\
 =&\big(\mathcal{O}^{\texttt{(0)}}\big)^\texttt{ssJ}_\texttt{gsJ}~~\big(\mathcal{O}^{\texttt{(0)}}\big)^\texttt{Jss}_\texttt{Jgs}~~\texttt{A}_{\texttt{ssss}}.
 ~.
\end{align}
The final amplitude can be written in terms three independent (reducible) tensor structures 
\begin{align}
 \mathcal T_{1\text{(\texttt{gsgs})}} = (\widehat{\mathcal T}_8)^2(\overline{\mathcal T}_8)^2 \qquad,\qquad
 \mathcal T_{2\text{(\texttt{gsgs})}} = \widehat{\mathcal T}_8\, \overline{\mathcal T}_8 \mathcal T_9\qquad\text{and}\qquad
 \mathcal T_{3\text{(\texttt{gsgs})}} = (\mathcal T_9)^2~.
\end{align}
The corresponding form factors are given by 
 \begin{align}
\mathcal F_{1\text{(\texttt{gsgs})}}  
& = \mathcal{N}_{J,D}\,(y_{s})^{J-4} \frac{d^4\mathcal{G}^{(\beta)}_{J}(z_{s})}{dz_{s}^4}
\qquad,\qquad 
\mathcal F_{2\text{(\texttt{gsgs})}}  
 = 4\, \mathcal{N}_{J,D}\, (y_{s})^{J-3} \frac{d^3\mathcal{G}^{(\beta)}_{J}(z_{s})}{dz_{s}^3}
\quad\text{and}\\
\mathcal F_{3\text{(\texttt{gsgs})}}  
&= 2\, \mathcal{N}_{J,D}\,(y_{s})^{J-2} \frac{d^2\mathcal{G}^{(\beta)}_{J}(z_{s})}{dz_{s}^2}~.
\end{align}

\paragraph{Angular distribution in $3+1$ dimensions} For the computation of the angular distribution in $3+1$ dimensions, we restrict to the case when the scalar is also massless. If we consider positive helicity for both the gravitons then the tensor structures evaluates to 
\begin{table}[h]
 \begin{center}
  \begin{tabular}{ ||c||c|c|| }
   \hline
   \hline
   \multicolumn{3}{|c|}{{\bf Angular distributions for different helicities} } \\
   \hline
   \hline
   Tensor structures & $2^\pm\longrightarrow 2^\pm$& $2^\pm\longrightarrow 2^\mp $ \\
   \hline
   $ \mathcal T_{1\text{(\texttt{gsgs})}} $ & $\frac{s^6}{64}\sin^4\theta$& $\frac{s^6}{64}\sin^4\theta$ \\
   \hline
   $ \mathcal T_{2\text{(\texttt{gsgs})}} $ & $-\frac{s^5}{64} \sin^2\theta(\cos\theta+1)$ & $-\frac{s^5}{64} \sin^2\theta(\cos\theta-1)$\\
   \hline
   $ \mathcal T_{3\text{(\texttt{gsgs})}} $ & $\frac{s^4}{64} (\cos\theta+1)^2$& $\frac{s^4}{64} (\cos\theta-1)^2$ \\
   \hline
    \hline
  \end{tabular}
  \caption{Angular distributions for different \texttt{gsgs} tensor structures}
  \label{tab:gsgsangdist}
 \end{center} 
\end{table}
\begin{align}
 \mathcal T_{1\text{(\texttt{gsgs})}} &= \frac{s^6}{64}\sin^4\theta=\frac{s^6}{64}(1-z_s)^2(1+z_s)^2\quad,\\
 \mathcal T_{2\text{(\texttt{gsgs})}} &= -\frac{s^5}{64}\sin^2\theta\, (\cos\theta+1)=-\frac{s^5}{64}(1-z_s) (1+z_s)^2\quad,\\
 \mathcal T_{3\text{(\texttt{gsgs})}} &= \frac{s^4}{64} (\cos\theta+1)^2=\frac{s^4}{64} (z_s+1)^2~.
\end{align}
For the above choice of helicities, the full amplitude is given by 
\begin{align}
 \texttt{A}_{\text{\texttt{gsgs}}}^{2^+\rightarrow 2^+} &=\mathcal T_{1\text{(\texttt{gsgs})}}\, \mathcal F_{1\text{(\texttt{gsgs})}} + \mathcal T_{2\text{(\texttt{gsgs})}}\, \mathcal F_{2\text{(\texttt{gsgs})}} + \mathcal T_{3\text{(\texttt{gsgs})}}\, \mathcal F_{3\text{(\texttt{gsgs})}}
 \\
 &= {\frac{s^2}{4}}\mathcal{N}_{J,4}\,(s)^J\, \left[\frac{1}{J(J-1)}\right]^2 \frac{(1+z_s)^2}{16}\Bigg[\frac{(1-z_s)^2}{4}~ \frac{\partial^4 \legendre_J(z_s)}{\partial^4 z_s} - (1-z)~ \frac{\partial^3 \legendre_J(z_s)}{\partial z_s^3} + \frac{1}{2}~ \frac{\partial^2 \legendre_J(z_s)}{\partial z_s^2} \Bigg] ~.
 \nonumber  
\end{align}
Using \textbf{Mathematical identity II} given in \eqref{mathematicalidentityII11}
\begin{equation}
 {\frac{s^2}{4}}\mathcal{N}_{J,4}\, s^J\, \frac{(J+2)(J+1)}{16\, J(J-1)}	d_{22}^{(J)}(\theta) = {\frac{s^2}{4}}\Bigg[  \Big( \widetilde{\mathcal{N}}_{J;2,2}\Big)\, s^{J}\, d_{22}^{(J)}(\theta) \Bigg] ~.
\end{equation}
Now we consider other choices of helicities. If we flip both the helicities then  we obtain $\texttt{A}_{\text{\texttt{gsgs}}}^{2^-\rightarrow 2^-}~$. 
\begin{equation}
 \texttt{A}_{\text{\texttt{gsgs}}}^{2^+\rightarrow 2^+}=  \texttt{A}_{\text{\texttt{gsgs}}}^{2^-\rightarrow 2^-}~.
\end{equation}
This is consistent with angular distribution since 
\begin{equation}
	d_{22}^{(J)}(\theta) =d_{-2-2}^{(J)}(\theta) ~.
\end{equation}
Let's now consider the case $2^+\longrightarrow 2^-$; Physically this case correspond to exchange of $3$ and $4$ in the previous case. From the expressions, given in table \ref{tab:gsgsangdist} (and property of Legendre polynomial given in \eqref{gegenpoly5}) we can see that 
\begin{align}
 \texttt{A}_{\text{\texttt{gsgs}}}^{2^+\rightarrow 2^-} &= {(-1)^J}\Big[ \texttt{A}_{\text{\texttt{gsgs}}}^{2^+\rightarrow 2^+} \Big]_{z_{s}\rightarrow -z_{s}}~.
\end{align}
The same set equations hold here with ${z_{s}\rightarrow -z_{s}}$. Hence the final answer is 
\begin{equation}
(-1)^J {\frac{s^2}{4}}\Bigg[  \Big( \widetilde{\mathcal{N}}_{J;2,2}\Big)\, s^{J}\, d_{22}^{(J)}(\theta+\pi ) \Bigg] 
=
 {\frac{s^2}{4}}\Bigg[  \Big( \widetilde{\mathcal{N}}_{J;2,2}\Big)\, s^{J}\, d_{2-2}^{(J)}(\theta ) \Bigg] ~.
\end{equation}
In the same way, we can check the angular distribution for the case $2^-\rightarrow 2^+$ (This is simply the $1\leftrightarrow2$ exchange of the first case). 

\section{Conclusion and future directions}
\label{sec:ramconclusion}

This paper computes the amplitude of spinning massless external states due to massive spin $J$ exchange in arbitrary space-time dimensions. The massive states can be classified by the representation of the little group . In $D$ spacetime dimensions the little group for massive particles is $SO(D-1)$. We have considered only the completely symmetric traceless representation of $SO(D-1)$. All the irreducible representations (i.e. spherical tensors) of  $SO(3)$ are completely symmetric and traceless. As a result, in $3+1$ dimensions, these are the only kind of higher spin particles. In higher dimensions, more general irreducible representations (mixed-symmetric, completely anti-symmetric) are allowed. States, transforming under mixed-symmetric representations, also appear in various string theories. One would like to know the expression for the tree level exchanges due to these states too. In this paper, we considered only the bosonic states. A generic theory of massive higher spins would also have fermionic states, and one would like to derive similar results for fermionic external states and fermionic exchanges.

One of the important tools in this paper is differential/multiplicative operators to construct the spinning amplitude from scalar amplitudes. We would like to investigate whether it is possible to extend this to loop amplitudes in QFT and (tree/loop) amplitudes in string theory. At this point, the operators are useful analytical gadgets. It is not clear whether these operators have any physical meaning. 

In this work, we entirely focus on the massless external states. There are more general amplitudes with massive external states. One example of such amplitude is the Compton amplitude (with photon, graviton). Expression for Compton amplitude is known using massive spinor-helicity formalism \cite{Arkani-Hamed:2017jhn}, and this amplitude is extremely useful in the black hole literature \cite{Guevara:2018wpp, Guevara:2019fsj}. More generally, amplitudes with all massive external states can help us to understand black hole scattering. It is possible to compute amplitudes with massive spinning external states using the differential/multiplicative operators.   

String theory has a plethora of higher spin particles. The mass and the spin obey are very particular relations in string theory. One would like to explore whether it is possible to extend this derivative method to string theory. In string theory, the spectrum of single-particle states is generated by a set of oscillators; for example, in bosonic string theory, they act on the Tachyonic vacuum to create all the higher spin states. It is worth investigating whether the spectrum generating algebra can be promoted to an amplitude generating algebra and what it has to do with the (broken) higher spin symmetries of string theory. We suspect that the extension of this derivative method to string theory would help us to answer those questions.

We cross-checked our answers by computing angular distribution in $3+1$ dimensions. A similar analysis should be possible in higher dimensions, too, and for that, one would like to know the partial wave expansion of spinning amplitudes in higher dimensions. To our knowledge, the expression for partial wave expansion for spinning amplitudes in higher dimensions is not known.

Our key motivation behind this work is to understand the question posed by Juan Maldacena\footnote{\href{https://physics.princeton.edu/strings2014/slides/Strominger.pdf}{Talk} by Andrew Strominger in strings 2015.} {\it ``What is the general theory of weakly coupled, interacting higher spin particles? Is string theory the only solution, in the same way, that GR is the only solution of a similar question involving massless spin particles and leading order in derivatives ?"} The work of CMEZ \cite{Camanho:2014apa} has shown us that a theory with a finite number of massive higher spin particles and/or higher derivative interaction terms are inconsistent with causality. A possible way to restore causality is to have an infinite number of higher spin particles. CMEZ also showed that string theory, which has an infinite number of higher spin particles, is casual; however, in string theory, the spectrum and coupling constants are of a very particular form.

Given that the expression for the four-graviton and the four-photon amplitude due to massive spin $J$ exchange is now known, one would like to understand the constraint of causality and unitarity on the spectrum and the coupling constants. The authors of \cite{Caron-Huot:2016icg, Sever:2017ylk, Nayak:2017qru} pursued this question for four scalar amplitudes. Using the result of this paper, one can pursue this question for spinning amplitudes too. The expression for the four graviton amplitude is bulky. It  drastically simplies in a supersymmetric theory (especially for theories with 32 supercharges). To do this, we need to know the massive super-multiplets and their coupling to gravitons. We hope to address them in the future.  

\paragraph{Acknowledgement} We are grateful to Shreyansh Agrawal, Arindam Bhattacharyya, Chandan Jana, R Loganayagam, Arpita Mitra, Manoj K Mandal, Shiraz Minwalla, Kumar Tanay, and Sarvesh Upadhyay for many discussions. We are thankful to Kushal Chakraborty, Chandan Jana, Manoj K Mandal, Arpita Mitra, Ashoke Sen, and Sarvesh Upadhyay for their comments on a preliminary version the draft.  {We also thank the referee for making important suggestions for improving the scope of this paper. We would like to thank Nabamita Banerjee, Suvankar Dutta, and Manoj K Mandal for their constant support and encouragement. M KNB is grateful to DST, India, for the KVPY fellowship. RP is grateful to IISER Bhopal for the hospitality and honorarium during his visit. AR would like to thank IISER Bhopal for the generous support towards a newly joined faculty and for the Institute start-up grant. Finally, we are grateful to the people of India for their generous funding for research in basic sciences.

\appendix 

\section{Notation and convention}
\label{app:ramnotation}

\begin{subequations}   
\begin{eqnarray} 
\textrm{Spacetime metric}\quad \quad && \eta_{\mu \nu }=\diag(-1,1,\cdots,1)
\\
\textrm{Lorentz indices}\quad \quad && \mu,\nu 
\\
\textrm{Momentum}\quad \quad && k_\mu 
\\
\textrm{Particle labels }\quad \quad && a,b,c,d
\\
\textrm{Momentum difference}\quad \quad && k_{ab}=k_a-k_b
\\
\textrm{Mandestam variables}\quad \quad && s,t,u
\\
\textrm{Form factor}\quad \quad && \mathcal{F} (s,t,u)
\\
\textrm{Tensor factor}\quad \quad && \mathcal{T} 
\\
\textrm{Wigner (small-)$d$ matrix}\quad \quad && d_{h^\prime h}^{(j)}( \beta ) 
\\
\textrm{Legendre polynomial}\quad \quad && \legendre_J(z)
\\
\textrm{Gegenbauer polynomial}\quad \quad && \gegen^{(\beta)}_n(z)
\\
\textrm{Jacobi polynomial}\quad \quad && \jacobi^{(\alpha,\beta)}_{n}(z)
\\
\textrm{Masses of particles}\quad \quad && m, m_a, m_J 
\\
\textrm{Spin}\quad \quad && J
\\
\textrm{Helicity}\quad \quad && h
\\
\textrm{Linearized Maxwell field strength}\quad \quad && \mathcal{B}_{\mu \nu}
\\
\textrm{Linearized Riemann tensor}\quad \quad && \mathcal{R}_{\mu \nu \rho \sigma}
\end{eqnarray} 
\end{subequations} 
We follow the following convention for the Mandelstam variables 
\begin{eqnarray}
s=-(k_1+k_2)^2
&\quad,\quad &
t=-(k_1+k_4)^2\quad \text{and}
\nonumber\\
&u=-(k_1+k_3)^2&~.
\label{hscomp22} 
\end{eqnarray}
This is same as convention in Green-Schwarz-Witten \cite{Green:2012oqa} \footnote{vol.1 page 373, 378} but different from Polchinski \cite{Polchinski:1998rq}. We also follow the convention such that all the {\it external particles} are {\it outgoing}.

\paragraph{A few more definition for scattering amplitudes} 

For our the purpose scattering amplitude, it is convenient to define following quantities 
\begin{equation}
\begin{split}
	&x_s=(k_{12}\odot k_{34})\qquad,\qquad y_s=\sqrt{(k_{12}\odot k_{12}) (k_{34}\odot k_{34})}
\qquad,\qquad z_s= \frac{x_s}{y_s}\quad;
\\	
	&x_t=(k_{14}\odot k_{23})\qquad,\qquad y_t=\sqrt{(k_{14}\odot k_{14}) (k_{23}\odot k_{23} )} \qquad,\qquad z_t= \frac{x_t}{y_t}\quad;
\\		
&x_u=(k_{13}\odot k_{24})\qquad,\qquad y_u=\sqrt{(k_{13}\odot k_{13}) (k_{24}\odot k_{24})} \qquad,\qquad z_u= \frac{x_u}{y_u}~.
\
\end{split}	
\end{equation}

\subsection{Convention for the center of mass frame}
\label{subsec:comconvention}
We have computed angular distribution in the center of mass frame at various places. Our choice for the center of mass frame is the following. The momenta for the incoming particles are  
\begin{equation}
	\begin{split}
		k_1= E(-1,0,0,1)
\qquad\text{and}\qquad		
		k_2= E(-1,0,0,-1)~.
	\end{split}
\label{fourmasslessconv1.1}
\end{equation}
The momenta of the out-going particles are 
\begin{equation}
	\begin{split}
		k_3= E(1,\sin\theta,0,\cos\theta )
\qquad\text{and}\qquad		
		k_4= E(1,-\sin\theta,0,-\cos\theta )~.
	\end{split}
\label{fourmasslessconv1.2} 
\end{equation}
For this choice of Mandelstam variables are
\begin{equation}
	s=4E^2\qquad,\qquad t= -2E^2(1-\cos\theta) 
	\qquad\text{and}\qquad u= -2E^2(1+ \cos\theta) ~.
\end{equation}
In 3+1 dimensions, massless particles have two polarizations. The polarizations take the following form
\begin{equation}
\begin{split} 
	\epsilon^{(\pm )}_1=&\frac{1}{\sqrt{2}}\left(0,\mp 1, - \iimg,0 \right) 
\qquad\qquad,\qquad  
	\epsilon^{(\pm)}_2=\frac{1}{\sqrt{2}}\left(0,\pm 1,- \iimg ,0 \right) \quad,	
\\
	\epsilon^{(\pm)}_3 =& \frac{ 1}{\sqrt{2}}\left(0,\pm\cos\theta ,-\iimg , \mp\sin\theta \right) 
\qquad\text{and}\qquad 
	\epsilon^{(\pm)}_4 =  \frac{ 1}{\sqrt{2}}\left(0,\mp\cos\theta ,\iimg , \pm\sin\theta \right) ~.
	\end{split}
\end{equation}

\section{Partial wave analysis for tree-level amplitude with massless external state}
\label{app:rampartialwave}

Here we review the partial wave analysis for spinning amplitude in 3+1 dimensions. In \cite{Hebbar:2020ukp}, there is a very nice review of partial wave expansion for general spinning amplitudes in $3+1$ space-time dimensions. For this paper, we restrict our analysis to {\it tree-level amplitudes with massless external states} \footnote{There is a similar analysis in section 6.3 of \cite{Arkani-Hamed:2020blm}. The authors explained in simple terms the appearance of the Wigner matrix.}. The partial wave decomposition doesn't depend on the perturbative approximation. However, in that case, the co-efficient of the partial waves are functions of momentum. The tree-level approximation allows us to write the precise dependence on the momenta. Also, for this restrictive case, it is possible to determine everything (including normalization) just from the partial wave analysis. The $S$-matrix is usually decomposed into a trivial part and the interaction part
\begin{equation}
	S=1+\iimg\, T~.
\label{rampartialwave1}
\end{equation}
We are interested in the tree level contribution to the quantity 
\begin{eqnarray}
	\langle \outvc|T|\invc\rangle ~.
\label{rampartialwave2}
\end{eqnarray}
We do this analysis in the frame given in Eqn \eqref{fourmasslessconv1.1} \& \eqref{fourmasslessconv1.2}. The massless states can be denoted by energy, the direction of motion, and helicity. For this paper, the motion is always restricted in the $x-z$ plane, and hence direction motion is uniquely determined by a single angle $\theta$. So a massless state is given by\footnote{More generally, a massless state is denoted by 
\begin{equation}
	|E, \theta, \phi, h\rangle 
\nonumber	
\end{equation}
} 
\begin{equation}
	|E, \theta, h\rangle~. 
\label{rampartialwave3}
\end{equation}
The incoming states are denoted by 
\begin{equation}
	|E, 0, h_1\rangle 
\otimes |E,\pi , h_2\rangle 	~.
\label{rampartialwave4}
\end{equation}
In terms of Poincare representation theory, this is simply the product of two massless representations; it can be written in terms of irreducible representations. The composite state has energy $2E$ and no spatial momentum. Hence it can be written in terms of massive representations of the Poincare group. A state in the massive representation can be denoted by ``mass"($=p_\mu p^\mu$), components of the spatial momentum, ``spin $J$"  and $z$ component of the spin (in the rest frame)
\begin{equation}
	|p_\mu p^\mu, \vec p,J,J_z \rangle 
\qquad\text{and}\qquad 	
\mathcal{W}_\mu \mathcal{W}^\mu=J(J+1) (p_\mu p^\mu) ~.
\label{rampartialwave5}
\end{equation}
Here $\mathcal{W}_\mu $ is the Pauli-Lubanski vector. In this particular case, we want to write 
\begin{equation}
	|E, 0, h_1\rangle 
\otimes |E, \pi , h_2\rangle 
=\sum_{J}\sum_{J_z} |4E^2, \vec 0,J,J_z \rangle 	~.
\label{rampartialwave11}
\end{equation}
By acting with $\mathcal{J}_z$ operator on the both side, we can determine
\begin{equation}
	J_z =h_1-h_2~.
\label{rampartialwave12}
\end{equation}
But the total spin $J$ cannot be fixed; it is unbounded above. There are at least two ways to understand it 
\begin{enumerate}
	\item (Intuitively) The two-particle have relative motion because of the non-zero relative momenta $k_{12}=k_1-k_2$. So depending on their separation, they can get arbitrary orbital angular momentum. This orbital angular momentum contributes to the spin of the composite system. Since the relative motion is restricted to the $x-z$ plane (in our convention), it cannot contribute to $J_z$. 

	\item (Algebratically) We can construct an operator $\mathcal{T}_0^{(J)}(k_{12})=(k_{12})^{\mu_1}\cdots (k_{12})^{\mu_J}$.  This operator can be constructed from momenta of the two states in the LHS of eqn \eqref{rampartialwave11} . This operator has spin $J$ and $J_z=0$ (since $k_{12}$ is along $z$ axis). By acting this operator on the state in RHS, we can always increase its spin without changing $J_z$ eigenvalue.  
\end{enumerate}
We follow the second way. We write the RHS in terms of minimal value for $J$ that is consistent with $J_z$ and we create any higher spin state by action of $\mathcal{T}_0^{(J)}(k_{12})$. In this convention,  
 \begin{equation}
	|E, 0, h_1\rangle 
\otimes |E, \pi , h_2\rangle 
= |4E^2, \vec 0,h_{12},h_{12} \rangle 	
\qquad\text{where}\qquad h_{ab}=h_a-h_b~.
\label{rampartialwave13}
\end{equation}
For out going states the LHS is 
\begin{equation}
	|E, \theta, h_3\rangle  
\otimes |E,\pi+\theta , h_4\rangle 	~.
\label{rampartialwave14}
\end{equation}
It is very similar to the incoming state except from the fact that it is rotated by an angle $\theta $. Let's now go back to our original expression \eqref{rampartialwave2}. $T$ can be written in terms of Dyson series. We restrict to the term that contribute to the tree level exchange. We also insert a complete of states inside. Then, it can be written as 
\begin{eqnarray}
	\sum_{n=0}^\infty \langle \outvc|H_{I}(\infty)| n ; \infty\rangle\langle n ; \infty | n ; -\infty\rangle\langle n ; -\infty|H_{I}(-\infty) |\invc\rangle  
\label{rampartialwave15}
\end{eqnarray}
where $| n ,t\rangle$ is the $n$-particle state of the interacting theory at time $t$ and $H_{I}(t)$ is the interaction Hamiltonian at time $t$. For the purpose this paper, we restrict to the tree-level amplitudes; i.e. the $n=1$ term \footnote{We assume that the vacuum is stable and it rules out $n=0$ term in the summation. }
\begin{eqnarray}
	 \langle \outvc|H_{I}(\infty)| 1; \infty\rangle\langle 1 ; \infty | 1 ; -\infty\rangle\langle 1;-\infty|H_{I}(-\infty) |\invc\rangle~.  
\label{rampartialwave21}
\end{eqnarray}
We write all the states in the momentum space. In a generic theory there are more than one $1$-particle states. Let's now focus on the formula for a particular $1$-particle state with spin $J$ and mass $m$. In this case, it becomes
\begin{eqnarray}
	 \langle \outvc|H_{I}(\infty)|m_j^2,\vec p, J, J_z ; \infty\rangle\langle m_j^2,\vec p, J, J_z ; \infty | m_j^2,\vec p, J, J_z ; -\infty\rangle\langle m_j^2,\vec p, J, J_z ; -\infty|H_{I}(-\infty) |\invc\rangle ~.
\nonumber\\
\label{rampartialwave22}
\end{eqnarray}
The middle piece is easy to evaluate. It is simply the momentum space propagator 
\begin{equation}
	\langle m_j^2,\vec p, J, J_z ; \infty | m_j^2,\vec p, J, J_z ; -\infty\rangle=\frac{\iimg }{p^2-m_J^2+\iimg \varepsilon}~.
\label{rampartialwave23}
\end{equation}
Then let's focus on the last term. We use eqn \eqref{rampartialwave13} 
\begin{eqnarray}
\langle m_j^2,\vec p, J, J_z ; -\infty|H_{I}(-\infty) |\invc\rangle  
=
\langle m_j^2,\vec p, J, J_z ; -\infty|H_{I}(-\infty)|4E^2, \vec 0,h_{12},h_{12} \rangle 	~.
\label{rampartialwave24}
\end{eqnarray}
From the Wigner-Eckart theorem, the above quantity gets contribution of an operator with spin $J-h_{12}$: $\mathcal{T}^{J-h_{12}}_{h}$. We denote an orthonormal state in a $SO(3)$ representation as 
\begin{equation}
	|j,j_z\rangle \quad:\quad \mathcal{J}^2|j,j_z\rangle=j(j+1)|j,j_z\rangle
\quad\text{and}\quad 	
\mathcal{J}_z|j,j_z\rangle=j_z|j,j_z\rangle~.
\end{equation}
The operator can only be function of $k_{12}$s and hence $h=0$. The over-lap can be figured out using Clebsch-Gordon coefficient. Here
\begin{equation}
k_{12}= (|k_{12}|)	|1,0\rangle \langle 1,0| ~.
\label{rampartialwave25}
\end{equation}
Then 
\begin{equation}
	\mathcal{T}^{J}_{0}(k_{12})= (|k_{12}|)^{J} 	|1,0\rangle \langle 1,0| \otimes\textrm{$J$ -times}\otimes 	|1,0\rangle \langle 1,0| 
	= (|k_{12}|)^{J} 	\Big[ \mathcal{C}_J|J,0\rangle \langle J,0| + \cdots\Big]~.
\label{rampartialwave31}
\end{equation}
We can evaluate $\mathcal{C}_J$ from Clebsch-Gordon coefficients. Let's start from the following Clebsch-Gordon coefficient  
\begin{equation}
\langle j+1,0|j,0;1,0\rangle = \sqrt{\frac{j+1}{2j+1}} ~.
\label{rampartialwave32}
\end{equation}
Then the Clebsch-Gordon to add $J$ number of $|1,0\rangle$s to get $|J,0\rangle $ is 
\begin{equation}
 \mathcal{C}_J	=\prod_{j=0}^{J-1} \langle j+1,0|j,0;1,0\rangle = \frac{\pi^{1/4} 2^{-\frac{J}{2}} \sqrt{\Gamma (J+1)}}{\sqrt{\Gamma \left(J+\frac{1}{2}\right)}}~.
\label{rampartialwave33}
\end{equation}  
Now we want to compute 
\begin{equation}
	\langle h,h|\mathcal{T}^{(J-h)}_0|J,h\rangle~.
\label{rampartialwave34}
\end{equation}
Let's first add $J-h$ number of $|1,0\rangle$s to get $|J-h,0\rangle$ 
\begin{equation}
	\langle J-h,0| 	\Bigg[\prod_{i=1}^{J-h}\otimes |1_i,0\rangle \Bigg]= \mathcal{C}_{J-h} ~.
\label{rampartialwave35}
\end{equation}
Then add $|J-h,0\rangle$ with $|h,h\rangle $ to get $|J,h\rangle$. The corresonding Clebsch-Gordon coefficient is given by 
\begin{equation}
	\langle J,h|J-h,0;h,h\rangle = \frac{(-1)^{2 (J-h)} }{\sqrt[4]{\pi }} 	
 \sqrt{\frac{  4^{J-h}  (J+h)! \Gamma \left(-h+J+\frac{1}{2}\right)}{\Gamma (2 J+1)   }} 	~.
\label{rampartialwave41}
\end{equation}
Then the quantity in \eqref{rampartialwave34} is given by 
\begin{equation}
	\langle h,h|\mathcal{T}^{(J-h)}_0|J,h\rangle= (|k_{12}|)^{J} \mathcal{C}_{J,h}
\label{rampartialwave42}
\end{equation}
where $\mathcal{C}_{J,h}$ is given by 
\begin{equation}
	\mathcal{C}_{J,h}=\mathcal{C}_{J-h} \langle J,h|J-h,0;h,h\rangle
\label{rampartialwave43}
\end{equation}
We put this back in eqn \eqref{rampartialwave24} to get 
\begin{eqnarray}
\langle m_j^2,\vec p, J, J_z ; -\infty|H_{I}(-\infty)|4E^2, \vec 0,h_{12},h_{12} \rangle 	
= 
(|k_{12}|)^{J} \mathcal{C}_{J,h_{12}}~.
\end{eqnarray}
Similarly, one can do the analysis for the out going states. The only difference is that the outgoing states are rotated with respect to the incoming states. So for the out-going states we get 
\begin{equation}
	 \langle \outvc|H_{I}(\infty)|m_j^2,\vec p, J, J_z ; \infty\rangle = d^{(J)}_{hh^\prime }(\theta) (|k_{34}|)^{J} \mathcal{C}_{J,h_{34}}~.
\end{equation}
Then in $3+1$ dimensions, and when all the external particles are massless, the scattering amplitude in \eqref{rampartialwave22} takes the following form 
\begin{equation}
	\widetilde{\mathcal{N}}_{J;h,h^\prime} \frac{1}{s-m_J^2+\iimg \varepsilon }\frac{s^{J}}{2}\, d^{(J)}_{hh^\prime }(\theta)
\label{apppartialwavereview1}
\end{equation} 
where $\widetilde{\mathcal{N}}_{J;h,h^\prime}$ is 
\begin{equation}
\widetilde{\mathcal{N}}_{J;h,h^\prime}=	\mathcal{C}_{J,s}\, \mathcal{C}_{J,h^\prime }
\qquad\text{and}\qquad
 \mathcal{C}_{J,h}=	2^{\frac{J}{2}-h} \sqrt{ \frac{ \Gamma (J-h+1)\, \Gamma (J+h+1)}{ \Gamma (2 J+1)} }~.
\label{apppartialwavereview2}
\end{equation}

\section{Orthogonal polynomials }
\label{app:orthogonalpoly}

Orthogonal polynomials denote a family of polynomials such that any two distinct members of the family are orthogonal to each other with respect to the inner product. We discuss some key features of Gegenbauer polynomials, Jacobi polynomials, and Wigner matrices for our purposes. 
\subsection{Spinning polynomial }

In $3+1$, the Legendre polynomial provides a basis to write down any amplitude with external scalar particles. In other dimensions, the Gegenbauer polynomial plays the same role. In \cite{Arkani-Hamed:2017jhn, Arkani-Hamed:2020blm, Liu:2020fgu}, the analogous basis for spinning external particles in $3+1$ were investigated. The claim is that such amplitudes can be written in terms of Jacobi polynomials. In our work, we investigated the same questions in various dimensions. In $3+1$ dimensions, we can express our answer in terms of Jacobi polynomials. In higher dimensions, any spinning amplitude, due to the exchange of completely symmetric higher spin particles, can be written in terms of the derivative of the Gegenbauer polynomial.

\subsection{Short summary on Gegenbauer polynomials}
\label{subsec:gegenpoly}

Gegenbauer polynomial is the higher dimensional analogue of Legendre polynomial. Gegenbauer polynomial appears in the multipole expansion of newtonian potential in higher dimensional spacetime
\begin{eqnarray}
\frac{1}{(1-2\,z\, t+t^2)^{\beta }}	=\sum_{n=0}^\infty \gegen^{(\beta)}_n(z)\, t^n~.
\label{gegenpoly1}	
\end{eqnarray}
In standard literature, the variable in the superscript is usually denoted by the letter $\alpha$. We choose to use $\beta$ to avoid confusion with string length squared $\alpha^\prime $. This reduces to Legendre polynomial for $\beta=1/2$. 
For $D$ spacetime dimension\footnote{In 3 space-dimension, the multipole expansion of coulomb potential gives Legendre polynomial. In $D-1$ space dimensions, if we do analogous expansion of Columb potential then we obtain Gegenbauer polynomial. } 
\begin{eqnarray}
\beta=\frac{(D-3)}{2}
\label{gegenpoly2}	
\end{eqnarray}
An explicit expression of Gegenbauer polynomial is 
\begin{equation}
	\gegen^{(\beta)}_n(z)= \sum_{a=0}^{\lfloor \frac{n}{2}\rfloor }(-1)^a\frac{\Gamma (n-a+\beta) }{a! \Gamma(\beta) (n-2a)!} (2z)^{n-2a}~.
\label{gegenpoly3}	
\end{equation}
Another useful formula for Gegenbauer polynomial is the Rodrigues formula
\begin{equation}
\gegen^{(\beta)}_n(z)= \frac{1}{2^n n !}\frac{\Gamma(\beta+\frac{1}{2})\Gamma(n+2\beta)}{\Gamma(2\beta)\Gamma(\beta+n+\frac{1}{2})}(z^2-1)^{-\beta+\frac{1}{2}} \frac{d^n}{dz^n } \Big[ (z^2-1)^{n+\beta-\frac{1}{2}}\Big] ~.
\label{gegenpoly4}	
\end{equation}
Gegenbauer polynomial (and hence Legendre polynomial) satisfies the following property
\begin{equation}
\gegen^{(\beta)}_n(-z)=(-1)^n\, \gegen^{(\beta)}_n(z)	~.
\label{gegenpoly5}	
\end{equation}

\paragraph{Connection to the four scalar amplitude}
The numerator of the four scalar amplitude is given by 
\begin{equation}
	 \sum_{a=0}^{\lfloor\frac{J}{2}\rfloor}~A(J,a,D) \Big[(k_{12}\cdot k_{12})( k_{34}\cdot k_{34} )\Big]^{a}(k_{12}\cdot k_{34} )^{J-2a}
\label{gegenpoly61}	
\end{equation}
here $A(J,a,D)$ is given by 
\begin{equation}
 A(J,a,D) = \Bigg[\frac{(-1)^a J!(2J+D-2a-5)!!}{2^a a!(J-2a)!(2J+D-5)!!}\Bigg]~.
\label{gegenpoly62}	
\end{equation}
Comparing this with eqn \eqref{gegenpoly3} we can see that it is same as
\begin{equation}
	s^J\frac{\Gamma(J+1)\, \Gamma \left(\beta\right)}{2^J\Gamma \left( \beta+J\right)} \gegen^{(\beta)}_J(\hat k_{12}\cdot \hat k_{34})
\qquad\text{where}\qquad
\beta= \frac{D-3}{2}	~.
\label{gegenpoly63}	
\end{equation}
In order to find normalization consider $a=0$
\begin{equation}
 A(J,0,D) = \Bigg[\frac{ J!(2J+D-5)!!}{ J!(2J+D-5)!!}\Bigg]=1
\label{gegenpoly64}	
\end{equation}
whereas the coefficient of the $a=0$ term in eqn \eqref{gegenpoly3} is 
\begin{equation}
	2^J\frac{\Gamma (J+\beta) }{\Gamma(\beta) J!} =2^J\frac{\Gamma (J+\beta) }{\Gamma(\beta)\, \Gamma(J+1)} ~.
\label{gegenpoly65}	
\end{equation}
This determines the normalization; the factor of $s^J$ can be fixed by dimensional analysis. 


\paragraph{Vector derivative of Gegenbauer polynomial}
Let's consider two space-like unit vector $\widehat n$ and $\widehat q$ such that 
\begin{equation}
	\widehat n\cdot \widehat q=\cos \theta =z ~.
\label{gegenpoly41}	
\end{equation}
Consider the Gegenbauer polynomial of the form 
\begin{equation}
	\gegen^{(\beta)}_\ell(\widehat n\cdot \widehat q)\equiv \gegen^{(\beta)}_\ell(z)~.
\label{gegenpoly42}	
\end{equation}
Then we can show that 
\begin{equation}
		\widehat r \cdot 
	\frac{\partial}{\partial {\widehat n}}\Big[
	\gegen^{(\beta)}_\ell(z) \Big]= (\hat r\cdot \hat q)	\gegen^{(\beta)\prime}_\ell(z) -(\hat r\cdot \hat n)	\gegen^{(\beta)\prime}_{\ell-1}(z)
\label{gegenpoly51}	
\end{equation}
where $\hat r$ is some other unit vector and $^\prime$ denotes derivative of Gegenbauer polynomial w.r.t $z$.

\subsection{Jacobi polynomial}
\label{subsec:jacobipoly}

Derivatives of the Jacobi polynomial $\jacobi^{(\alpha,\beta)}_{n}(z)$ satisfy the following relations 
\begin{equation}
\label{eq:jacobiderivative1}
	\frac{d^k}{dz^k}\jacobi^{(\alpha,\beta)}_{n}(z) = \frac{1}{2^k}\frac{\Gamma(\alpha+\beta+n+k+1)}{\Gamma(\alpha+\beta+n+1)}\jacobi^{(\alpha+k,\beta+k)}_{n-k}(z)~.
\end{equation}
Jacobi polynomial also obeys the following recursion relation 
\begin{equation}
	(z-1)\frac{d}{dz}\jacobi^{(\alpha,\beta)}_{n}(z) = (\alpha + n)\jacobi^{(\alpha-1,\beta+1)}_{n}(z)-\alpha\, \jacobi^{(\alpha,\beta)}_{n}(z)~.
\label{eq:jacobirecurrence1}	
\end{equation}
Legendre polynomial is simply one particular Jacobi polynomial 
\begin{equation}
	\legendre_J(z)\equiv \jacobi^{(0,0)}_{J}(z)~.
\label{eq:jacobitolegendre}	
\end{equation}
Then from this equation and eqn \eqref{eq:jacobiderivative1} it follows 
\begin{equation}
\label{eq:jacobiderivative}
	\frac{d^k}{dz^k}\legendre_J(z) = \frac{1}{2^k}\frac{\Gamma(J+k+1)}{\Gamma(J+1)}\jacobi^{(k,\, k)}_{J-k}(z)~.
\end{equation}
We use this equation many times to show the angular distributions. More generally Gegenbauer polynomial is a special case of Jacobi polynomial 
\begin{equation}
\gegen^{(\beta)}_n(z)= \frac{(2\alpha)_n}{(\alpha+1/2)_n}	 \jacobi^{(\beta-1/2,\, \beta-1/2)}_{n}(z)
\label{eq:jacobitogegen}	
\end{equation}
where $(a)_n$ is Pochhammer function. Jacobi polynomials satisfy the following symmetry property
\begin{align}
\jacobi^{(\alpha,\beta)}_J(-z) = (-1)^J \jacobi^{(\beta,\alpha)}_{J}(z)~.
\label{eq:jacobiparity}	
\end{align}

\subsection{Wigner matrix}
\label{subsec:wignermatrix}

Wigner (small-)$d$ matrix is defined as 
\begin{eqnarray}
  d_{h^\prime h}^{(j)}( \beta ) 
&=&   \left\langle j,h^\prime\left| \exp\left[-\frac{\iimg \amom_y }{\hbar }\beta 
  \right ] \right|j,h\right\rangle  ~.
\label{eq:wignermatrixdefn}
\end{eqnarray}
It is quantum rotation matrix in $x$-$z$ plane; it rotates the states belonging to $2j+1$ dimensional irreducible representation by angle $\theta$. A few important property of the Wigner matrices are the following 
\begin{align}
d_{h',h}^{(j)}(\pi+\theta) &= (-1)^{j-h} d_{h',-h}^{(j)}(\theta)\quad\text{and}
\label{eq:wignerprop1}
\\
d_{h',h}^{(j)}(\theta) &= (-1)^{h-h'} d_{h, h'}^{(j)}(\theta)= d_{-h,-h'}^{(j)}(\theta)~.
\label{eq:wignerprop2}
\end{align}
For the purpose of this paper, the following mathematical identities are useful
\begin{enumerate}
	\item Wigner matrix with one argument being zero can be written as derivative of Legendre polynomial
\begin{equation}
\label{eq:wignertolegendre}
	d^{(j)}_{h0}(\theta)= \sqrt{\frac{(j-h)!}{(j+h)!}}(-1)^h (1-z^2)^{h/2}\frac{d^h}{dz^h} P_{j}(z)~.
\end{equation}
	\item Wigner matrices can written in terms of Jacobi polynomial in the following way 
\begin{equation}
\label{eq:wignertojacobi}
	d^{(j)}_{h,h^\prime}(\theta) = \sqrt{\frac{(j-h)!(j+h)!}{(j-h^\prime)!(j+h^\prime)!}}\,\Big(\cos\frac{\theta}{2}\,\Big)^{h+h^\prime}\Big(\sin\frac{\theta}{2}\,\Big)^{h-h^\prime}\jacobi^{(h-h^\prime,h+h^\prime)}_{j-h}(\cos\theta) ~.
\end{equation}
We list a few cases which will be relevant for us (here $z=\cos \theta$)
\begin{enumerate}

	\item 

\begin{equation}
	d^{(j)}_{1,1}(\theta) = \Big(\cos\frac{\theta}{2}\,\Big)^{2}\jacobi^{(0,2)}_{j-1}(\cos\theta) = \bigg[\frac{z+1}{2}\bigg]\jacobi^{(0,2)}_{j-1}(z)\quad,
	\label{jacobitowigner4}	
\end{equation}

	\item 

\begin{equation}
	d^{(j)}_{2,2}(\theta) = \Big(\cos\frac{\theta}{2}\,\Big)^{4}\jacobi^{(0,4)}_{j-2}(\cos\theta) = \bigg[\frac{z+1}{2}\bigg]^2\jacobi^{(0,4)}_{j-2}(z)\quad\text{and}
\label{jacobitowigner5}	
\end{equation}

%

	\item 

\begin{equation}
	d^{(j)}_{4,4}(\theta) = \Big(\cos\frac{\theta}{2}\,\Big)^{8}\jacobi^{(0,8)}_{j-4}(\cos\theta) = \bigg[\frac{z+1}{2}\bigg]^4\jacobi^{(0,8)}_{j-4}(z)~.
\label{jacobitowigner7}	
\end{equation} 
\end{enumerate}
\end{enumerate}

\subsection{Mathematical identities}
\label{subsec:mathidentity}
Now we derive various mathematical identities which have been used in the paper. 
\paragraph{Mathematical identity I}

In this section we show that 
\begin{equation}
\Bigg[  \frac{\partial \legendre_j(z)}{\partial z} + (z-1) \frac{\partial^2 \legendre_j(z)}{\partial z^2} \Bigg]=\frac{j(j+1)}{2}\jacobi^{(0,2)}_{j-1}(z)~.
\label{mathematicalidentityI1}
\end{equation}
We start from the LHS
\begin{equation}
	 \bigg[\frac{d}{dz}\jacobi^{(0,0)}_{j}(z)+(z-1)\frac{d^2}{dz^2}\jacobi^{(0,0)}_{j}(z)\bigg] ~.
\label{mathematicalidentityI2}
\end{equation}
Using the expression for derivatives of Jacobi polynomial in eqn \eqref{eq:jacobiderivative} we get 
\begin{equation}
	\frac{(j+1)}{2}\bigg(\jacobi^{(1,1)}_{j-1}(z)+(z-1)\frac{d}{dz}\jacobi^{(1,1)}_{j-1}(z)\bigg)~.
\label{mathematicalidentityI3}
\end{equation}
Now using the recursion relation \eqref{eq:jacobirecurrence1} we get 
\begin{equation}
\frac{(j+1)}{2}\bigg(\cancel{\jacobi^{(1,1)}_{j-1}(z)}-\cancel{\jacobi^{(1,1)}_{j-1}(z)} + j\jacobi^{(0,2)}_{j-1}(z)\bigg)
	=\frac{j(j+1)}{2}\jacobi^{(0,2)}_{j-1}(z) ~.
\label{mathematicalidentityI4}
\end{equation}
As a consequence of \eqref{mathematicalidentityI1} and \eqref{jacobitowigner4} we obtain 
\begin{equation}
d_{11}^{(j)}(\theta)= \bigg[\frac{z+1}{2}\bigg]\jacobi^{(0,2)}_{j-1}(z) =  \frac{(z+1)}{j(j+1)}\Bigg[  \frac{\partial \legendre_j(z)}{\partial z} + (z-1) \frac{\partial^2 \legendre_j(z)}{\partial z^2} \Bigg]~.
\label{mathematicalidentityI5}
\end{equation}
We use this identity for \texttt{psps} amplitude.

\paragraph{Mathematical identity II}
In this section we show that 
\begin{equation}
	 \Bigg[2\, \frac{\partial^2 \legendre_J(z)}{\partial z^2} + (z-1)^2\,  \frac{\partial^4 \legendre_J(z)}{\partial z^4} + 4(z-1)  \frac{\partial^3 \legendre_J(z)}{\partial z^3} \Bigg]
=\frac{1}{4} (J+2)(J+1)J(J-1)\jacobi^{(0,4)}_{J-2}(z)	 ~.
\label{mathematicalidentityII1}
\end{equation}
We start from the LHS and using \eqref{eq:jacobiderivative} we can write it as 
\begin{equation}
	\frac{1}{2} (J+2)(J+1)\Bigg[ \jacobi^{(2,2)}_{J-2}(z)+ \frac{(z-1)^2\,(J+3)}{4}\Big[\frac{d}{dz}\jacobi^{(3,3)}_{J-3}(z)\Big] + (z-1)(J+3)\jacobi^{(3,3)}_{J-3}(z)\Bigg] ~.
\label{mathematicalidentityII2}
\end{equation}
We use the definition of derivative of Jacobi polynomial to get 
\begin{equation}
\frac{1}{2} (J+2)(J+1)\Bigg[\frac{(z-1)J(J+3)}{4}\jacobi^{(2,4)}_{J-3}(z) + \frac{J}{2}\jacobi^{(1,3)}_{J-2}(z)\Bigg]~.
\label{mathematicalidentityII3}
\end{equation}
We rewrite the first term as a derivative and then use identity \eqref{eq:jacobirecurrence1} to get the answer,
\begin{equation}
\frac{J}{2}\Bigg[(J-1)\jacobi^{(0,4)}_{J-2}(z)-\cancel{\jacobi^{(1,3)}_{J-2}(z)}+\cancel{\jacobi^{(1,3)}_{J-2}(z)}\Bigg]
\label{mathematicalidentityII4}
\end{equation}
From \eqref{jacobitowigner5} and \eqref{mathematicalidentityII1} we get 
\begin{equation}
	(z+1)^2\Bigg[2\, \frac{\partial^2 \legendre_J(z)}{\partial z^2} + (z-1)^2\,  \frac{\partial^4 \legendre_J(z)}{\partial z^4} + 4(z-1)  \frac{\partial^3 \legendre_J(z)}{\partial z^3} \Bigg]
=4(J+2)(J+1)J(J-1)\, d^{(j)}_{2,2}(\theta) 	~.
\label{mathematicalidentityII11}
\end{equation}
We use this identity for \texttt{pppp} and \texttt{gsgs} amplitudes.

\paragraph{Mathematical identity III}

In this section, we show that
\begin{equation}
	\begin{split}
\frac{1}{256}\frac{\Gamma(J+5)}{\Gamma(J-3)}\jacobi^{(0,8)}_{J-4}
=&\Big[\frac{3}{2}\frac{d^4}{dz^4}\legendre_J(z) + 6(z-1)\frac{d^5}{dz^5}\legendre_J(z)+\frac{9(z-1)^2}{2}\frac{d^6}{dz^6}\legendre_J(z)\\
		&+(z-1)^3\frac{d^7}{dz^7}\legendre_J(z)+\frac{(z-1)^4}{16}\frac{d^8}{dz^8}\legendre_J(z)\Big]
	\end{split}
\label{mathematicalidentityIII51}
\end{equation}
We start from RHS. First we write
\begin{equation}
	\frac{d^{4+n}}{dz^{4+n }}\legendre_J(z) 
=
\frac{\Gamma (J+5)}{16 \Gamma(J+1)}	\frac{d^{n}}{dz^{n }} \jacobi_{J-4}^{(4,4)}(z)
\end{equation}
Now we successively use \eqref{eq:jacobirecurrence1} for the right-hand side of above equation
\begin{equation}
\begin{split}
	\frac{d^{4}}{dz^{4}}\legendre_J(z) 
&=
\frac{\Gamma (J+5)}{16 \Gamma(J+1)}	 \jacobi_{J-4}^{(4,4)}(z)
\\
	(z-1)\frac{d^{5}}{dz^{5}}\legendre_J(z) 
&= \frac{\Gamma (J+5)}{16 \Gamma(J+1)}	 \Big[ J \jacobi_{J-4}^{(3,5)}(z)-4 \jacobi_{J-4}^{(4,4)}(z)\Big]	
\\
	(z-1)^2\frac{d^{6}}{dz^{6}}\legendre_J(z) 
&= \frac{\Gamma (J+5)}{16 \Gamma(J+1)}	 \Bigg[\frac{\Gamma(J+1)}{\Gamma(J-1)}  \jacobi_{J-4}^{(2,6)}(z)-8 J \jacobi_{J-4}^{(3,5)}(z)+20 \jacobi_{J-4}^{(4,4)}(z)
\Bigg]	
\\
	(z-1)^3\frac{d^7}{dz^7}\legendre_J(z) 
&= \frac{\Gamma (J+5)}{16 \Gamma(J+1)}	 \Bigg[
\frac{\Gamma(J+1)}{\Gamma(J-2)} \jacobi_{J-4}^{(1,7)}(z)-12\frac{\Gamma(J+1)}{\Gamma(J-1)}  \jacobi_{J-4}^{(2,6)}(z) \
\\
&\qquad\qquad \qquad +60 J \jacobi_{J-4}^{(3,5)}(z)-120 \jacobi_{J-4}^{(4,4)}(z) 
\Bigg]
\\
	(z-1)^4\frac{d^8}{dz^8}\legendre_J(z)   
&= \frac{\Gamma (J+5)}{16 \Gamma(J+1)}	 \Bigg[ \frac{\Gamma(J+1)}{\Gamma(J-3)} \jacobi_{J-4}^{(0,8)}(z)-16 \frac{\Gamma(J+1)}{\Gamma(J-2)}\jacobi_{J-4}^{(1,7)}(z)  
\\
&\qquad\qquad \qquad
+120 \frac{\Gamma(J+1)}{\Gamma(J-1)} \jacobi_{J-4}^{(2,6)}(z)
-480 J \jacobi_{J-4}^{(3,5)}(z)+840 \jacobi_{J-4}^{(4,4)}(z)
\Bigg]
\end{split}
\end{equation}
Putting these in the RHS of \eqref{mathematicalidentityIII51} we get the LHS. We use this identity for \texttt{gggg} amplitude.

%
%

\section{Form factors for the \texttt{gggg} amplitude }
\label{app:ggggformfactor}

The expressions for 6 of the form factors ($\mathcal F_{1\text{(\texttt{gggg})}}, \mathcal F_{2\text{(\texttt{gggg})}}, \mathcal F_{3\text{(\texttt{gggg})}}, \mathcal F_{4\text{(\texttt{gggg})}}, \mathcal F_{7\text{(\texttt{gggg})}}, \mathcal F_{8\text{(\texttt{gggg})}}$) are very long in terms of Gegenbauer polynomial. We write those expressions here.
\begin{align} 
\mathcal F_{1\text{(\texttt{gggg})}}  
& = -\left(\frac{1}{32} (y_{s})^{J-8} \
\frac{d^4\mathcal{G}^{(\beta)}_{J}(z_{s})}{dz_{s}^4}\right)-\frac{1}{32} 
(y_{s})^{J-8} \left(-50 \
\frac{d^3\mathcal{G}^{(\beta)}_{J-1}(z_{s})}{dz_{s}^3}+73 \frac{d^4\mathcal{G}^{(\beta)}_{J-2}(z_{s})}{dz_{s}^4}\right.\nonumber\\
&~~~~~~~~~~\left.-18 \
\frac{d^5\mathcal{G}^{(\beta)}_{J-3}(z_{s})}{dz_{s}^5}+\frac{d^6\mathcal{G}^{(\beta)}_{J-4}(z_{s})}{dz_{s}^6}-4 z_{s} \
\left(12 \frac{d^4\mathcal{G}^{(\beta)}_{J-1}(z_{s})}{dz_{s}^4} -10 \
\frac{d^5\mathcal{G}^{(\beta)}_{J-2}(z_{s})}{dz_{s}^5}\right.\right.\nonumber\\
&~~~~~~~~~~\left.\left. +2 z_{s} \
\frac{d^5\mathcal{G}^{(\beta)}_{J-1}(z_{s})}{dz_{s}^5} +\frac{d^6\mathcal{G}^{(\beta)}_{J-3}(z_{s})}{dz_{s}^6}-z_{s} \
\frac{d^6\mathcal{G}^{(\beta)}_{J-2}(z_{s})}{dz_{s}^6}\right)\right)\nonumber\\
&~~~~~~~~~~+\frac{1}{256} (y_{s})^{J-8} \
\left(1800 \frac{d^2\mathcal{G}^{(\beta)}_{J-2}(z_{s})}{dz_{s}^2}-13800 \
\frac{d^3\mathcal{G}^{(\beta)}_{J-3}(z_{s})}{dz_{s}^3}+16689 \
\frac{d^4\mathcal{G}^{(\beta)}_{J-4}(z_{s})}{dz_{s}^4}\right.\nonumber\\
&~~~~~~~~~~\left.-6228 \
\frac{d^5\mathcal{G}^{(\beta)}_{J-5}(z_{s})}{dz_{s}^5}+898 \frac{d^6\mathcal{G}^{(\beta)}_{J-6}(z_{s})}{dz_{s}^6}-52 \
\frac{d^7\mathcal{G}^{(\beta)}_{J-7}(z_{s})}{dz_{s}^7}+\frac{d^8\mathcal{G}^{(\beta)}_{J-8}(z_{s})}{dz_{s}^8}\right.\nonumber\\
&~~~~~~~~~~\left.+8 z_{s} \
\left(1000 \frac{d^3\mathcal{G}^{(\beta)}_{J-2}(z_{s})}{dz_{s}^3}-3616 \
\frac{d^4\mathcal{G}^{(\beta)}_{J-3}(z_{s})}{dz_{s}^4}+2441 \
\frac{d^5\mathcal{G}^{(\beta)}_{J-4}(z_{s})}{dz_{s}^5}-525 \frac{d^6\mathcal{G}^{(\beta)}_{J-5}(z_{s})}{dz_{s}^6}\right.\right.\nonumber\\
&~~~~~~~~~~\left.\left.+41 \
\frac{d^7\mathcal{G}^{(\beta)}_{J-6}(z_{s})}{dz_{s}^7}-\frac{d^8\mathcal{G}^{(\beta)}_{J-7}(z_{s})}{dz_{s}^8}+z_{s} \
\left(872 \frac{d^4\mathcal{G}^{(\beta)}_{J-2}(z_{s})}{dz_{s}^4}-1880 \
\frac{d^5\mathcal{G}^{(\beta)}_{J-3}(z_{s})}{dz_{s}^5}\right.\right.\right.\nonumber\\
&~~~~~~~~~~\left.\left.\left.+757 \frac{d^6\mathcal{G}^{(\beta)}_{J-4}(z_{s})}{dz_{s}^6}-90 \
\frac{d^7\mathcal{G}^{(\beta)}_{J-5}(z_{s})}{dz_{s}^7}+3 \frac{d^8\mathcal{G}^{(\beta)}_{J-6}(z_{s})}{dz_{s}^8}+2 z_{s} \
\left(112 \frac{d^5\mathcal{G}^{(\beta)}_{J-2}(z_{s})}{dz_{s}^5}\right.\right.\right.\right.\nonumber\\
&~~~~~~~~~~\left.\left.\left.\left.-160 \
\frac{d^6\mathcal{G}^{(\beta)}_{J-3}(z_{s})}{dz_{s}^6}+38 \frac{d^7\mathcal{G}^{(\beta)}_{J-4}(z_{s})}{dz_{s}^7}-2 \
\frac{d^8\mathcal{G}^{(\beta)}_{J-5}(z_{s})}{dz_{s}^8}+z_{s} \left(8 \
\frac{d^6\mathcal{G}^{(\beta)}_{J-2}(z_{s})}{dz_{s}^6}\right.\right.\right.\right.\right.\nonumber\\
&~~~~~~~~~~\left.\left.\left.\left.\left.-8 \
\frac{d^7\mathcal{G}^{(\beta)}_{J-3}(z_{s})}{dz_{s}^7}+\frac{d^8\mathcal{G}^{(\beta)}_{J-4}(z_{s})}{dz_{s}^8}\right)\right)\right)\right)\right) ~.
 \end{align}
 
\begin{align} 
\mathcal F_{2\text{(\texttt{gggg})}}  
& = \frac{1}{32} (y_{s})^{J-8} \left[1764 \
\frac{d^4\mathcal{G}^{(\beta)}_{J-2}(z_{s})}{dz_{s}^4}-1791 \
\frac{d^5\mathcal{G}^{(\beta)}_{J-3}(z_{s})}{dz_{s}^5}+457 \frac{d^6\mathcal{G}^{(\beta)}_{J-4}(z_{s})}{dz_{s}^6}-39 \
\frac{d^7\mathcal{G}^{(\beta)}_{J-5}(z_{s})}{dz_{s}^7}\right.\nonumber\\
&~~~~~~~~~~\left.+\frac{d^8\mathcal{G}^{(\beta)}_{J-6}(z_{s})}{dz_{s}^8}+2 z_{s} \
\left\{900 \frac{d^5\mathcal{G}^{(\beta)}_{J-2}(z_{s})}{dz_{s}^5}-573 \
\frac{d^6\mathcal{G}^{(\beta)}_{J-3}(z_{s})}{dz_{s}^6}+82 \frac{d^7\mathcal{G}^{(\beta)}_{J-4}(z_{s})}{dz_{s}^7}\right.\right.\nonumber\\
&~~~~~~~~~~\left.\left.-3 \
\frac{d^8\mathcal{G}^{(\beta)}_{J-5}(z_{s})}{dz_{s}^8}+2 z_{s} \left(116 \
\frac{d^6\mathcal{G}^{(\beta)}_{J-2}(z_{s})}{dz_{s}^6}-47 \frac{d^7\mathcal{G}^{(\beta)}_{J-3}(z_{s})}{dz_{s}^7}+8 z_{s} \
\frac{d^7\mathcal{G}^{(\beta)}_{J-2}(z_{s})}{dz_{s}^7}\right.\right.\right.\nonumber\\
&~~~~~~~~~~\left.\left.+3 \frac{d^8\mathcal{G}^{(\beta)}_{J-4}(z_{s})}{dz_{s}^8}-2 z_{s} \
\frac{d^8\mathcal{G}^{(\beta)}_{J-3}(z_{s})}{dz_{s}^8}\right)\right\}\nonumber\\
&~~~~~~~~~~-\left.4 \
\left(-9 \
\frac{d^5\mathcal{G}^{(\beta)}_{J-1}(z_{s})}{dz_{s}^5}+\frac{d^6\mathcal{G}^{(\beta)}_{J-2}(z_{s})}{dz_{s}^6}-2 z_{s} \
\frac{d^6\mathcal{G}^{(\beta)}_{J-1}(z_{s})}{dz_{s}^6}\right)\right]~.
\\
\mathcal F_{3\text{(\texttt{gggg})}} 
 =& \frac{1}{4} (y_{s})^{J-8} \left[-162 \
\frac{d^5\mathcal{G}^{(\beta)}_{J-1}(z_{s})}{dz_{s}^5}+161 \frac{d^6\mathcal{G}^{(\beta)}_{J-2}(z_{s})}{dz_{s}^6}-26 \
\frac{d^7\mathcal{G}^{(\beta)}_{J-3}(z_{s})}{dz_{s}^7}+\frac{d^8\mathcal{G}^{(\beta)}_{J-4}(z_{s})}{dz_{s}^8}\right.
\nonumber\\
&~~~\left.-4 z_{s} \
\left(20 \frac{d^6\mathcal{G}^{(\beta)}_{J-1}(z_{s})}{dz_{s}^6}-14 \
\frac{d^7\mathcal{G}^{(\beta)}_{J-2}(z_{s})}{dz_{s}^7}+2 z_{s} \
\frac{d^7\mathcal{G}^{(\beta)}_{J-1}(z_{s})}{dz_{s}^7}+\frac{d^8\mathcal{G}^{(\beta)}_{J-3}(z_{s})}{dz_{s}^8}\right.\right.
\nonumber\\
&~~~\left.\left.-z_{s} \
\frac{d^8\mathcal{G}^{(\beta)}_{J-2}(z_{s})}{dz_{s}^8}\right)-2\
\frac{d^6\mathcal{G}^{(\beta)}_{J}(z_{s})}{dz_{s}^6}~.\right]
\\
\mathcal F_{4\text{(\texttt{gggg})}} 
& =\frac{1}{16} (y_{s})^{J-8} \left[143 \
\frac{d^6\mathcal{G}^{(\beta)}_{J-2}(z_{s})}{dz_{s}^6}-26 \
\frac{d^7\mathcal{G}^{(\beta)}_{J-3}(z_{s})}{dz_{s}^7}+\frac{d^8\mathcal{G}^{(\beta)}_{J-4}(z_{s})}{dz_{s}^8}\right.\nonumber\\
&~~~~~~~~~~\left.+4 z_{s} \
\left(13 \
\frac{d^7\mathcal{G}^{(\beta)}_{J-2}(z_{s})}{dz_{s}^7}-\frac{d^8\mathcal{G}^{(\beta)}_{J-3}(z_{s})}{dz_{s}^8}+z_{s} \
\frac{d^8\mathcal{G}^{(\beta)}_{J-2}(z_{s})}{dz_{s}^8}\right)\right] ~.
\\
\mathcal F_{7\text{(\texttt{gggg})}} & = \frac{1}{4} (y_{s})^{J-6} \left[-50 \
\frac{d^3\mathcal{G}^{(\beta)}_{J-1}(z_{s})}{dz_{s}^3}+73 \frac{d^4\mathcal{G}^{(\beta)}_{J-2}(z_{s})}{dz_{s}^4}-18 \
\frac{d^5\mathcal{G}^{(\beta)}_{J-3}(z_{s})}{dz_{s}^5}+\frac{d^6\mathcal{G}^{(\beta)}_{J-4}(z_{s})}{dz_{s}^6}\right.\nonumber\\
&~~~\left.-4 z_{s} \
\left(12 \frac{d^4\mathcal{G}^{(\beta)}_{J-1}(z_{s})}{dz_{s}^4}-10 \
\frac{d^5\mathcal{G}^{(\beta)}_{J-2}(z_{s})}{dz_{s}^5}+2 z_{s} \
\frac{d^5\mathcal{G}^{(\beta)}_{J-1}(z_{s})}{dz_{s}^5}+\frac{d^6\mathcal{G}^{(\beta)}_{J-3}(z_{s})}{dz_{s}^6}\right.\right.
\nonumber\\
&~~~\left.\left.-z_{s} \
\frac{d^6\mathcal{G}^{(\beta)}_{J-2}(z_{s})}{dz_{s}^6}\right) -2
\frac{d^4\mathcal{G}^{(\beta)}_{J}(z_{s})}{dz_{s}^4} \right] ~.
\\
\mathcal F_{8\text{(\texttt{gggg})}}  
& = \frac{1}{4} (y_{s})^{J-7} \left[-98 \
\frac{d^4\mathcal{G}^{(\beta)}_{J-1}(z_{s})}{dz_{s}^4}+113 \frac{d^5\mathcal{G}^{(\beta)}_{J-2}(z_{s})}{dz_{s}^5}-22 \
\frac{d^6\mathcal{G}^{(\beta)}_{J-3}(z_{s})}{dz_{s}^6}+\frac{d^7\mathcal{G}^{(\beta)}_{J-4}(z_{s})}{dz_{s}^7}\right.\nonumber\\
&~~~\left.-4 z_{s} \
\left(16 \frac{d^5\mathcal{G}^{(\beta)}_{J-1}(z_{s})}{dz_{s}^5}-12 \
\frac{d^6\mathcal{G}^{(\beta)}_{J-2}(z_{s})}{dz_{s}^6}+2 z_{s} \
\frac{d^6\mathcal{G}^{(\beta)}_{J-1}(z_{s})}{dz_{s}^6}+\frac{d^7\mathcal{G}^{(\beta)}_{J-3}(z_{s})}{dz_{s}^7}\right.\right.
\nonumber\\
&~~~\left.\left.-z_{s} \
\frac{d^7\mathcal{G}^{(\beta)}_{J-2}(z_{s})}{dz_{s}^7}\right)-2 \
\frac{d^5\mathcal{G}^{(\beta)}_{J}(z_{s})}{dz_{s}^5} ~.\right]
\end{align}

\bibliographystyle{utphys}
\bibliography{hspinsmatrix} 

\end{document}